\documentclass[final,onefignum,onetabnum]{siamart190516}

\usepackage{lipsum}
\usepackage{amsfonts}
\usepackage{graphicx}
\usepackage{epstopdf}
\usepackage{algorithmic}
\ifpdf
\DeclareGraphicsExtensions{.eps,.pdf,.png,.jpg}
\else
 \DeclareGraphicsExtensions{.eps}
\fi

\usepackage{dsfont}
\usepackage[caption=false]{subfig}

\DeclareMathOperator{\tr}{tr}

\newsiamremark{remark}{Remark}
\newsiamremark{hypothesis}{Hypothesis}
\crefname{hypothesis}{Hypothesis}{Hypotheses}
\newsiamthm{claim}{Claim}

\headers{3D Genome Organization in Diploid Organisms}{A. Belyaeva, K. Kubjas, L. J. Sun and C. Uhler}

\title{Identifying 3D Genome Organization in Diploid Organisms via Euclidean Distance Geometry\thanks{Submitted.
\funding{Anastasiya Belyaeva was supported by an NSF Graduate Research Fellowship
(1122374), the Abdul Latif Jameel World Water and Food Security Lab (J-WAFS) at MIT and the MIT J-Clinic for Machine Learning and Health. Kaie Kubjas was supported by the European Union's Horizon 2020 research and innovation programme: Marie Sk\l{}odowska-Curie grant agreement No. 748354, research carried out at LIDS, MIT and Team PolSys, LIP6, Sorbonne University. Caroline Uhler was partially supported by NSF (DMS-1651995), ONR (N00014-17-1-2147 and N00014-18-1-2765), IBM, and a Simons Investigator Award. 
}}}

\author{Anastasiya Belyaeva\thanks{Laboratory for Information and Decision Systems, Department of Electrical Engineering and Computer Science, and Institute for Data, Systems and Society, Massachusetts Institute of Technology, Cambridge, MA 
  (\email{belyaeva@mit.edu}, \email{sunl@mit.edu}, \email{cuhler@mit.edu}).}
\and Kaie Kubjas\thanks{Department of Mathematics and Systems Analysis, Aalto University, Espoo, Finland
  (\email{kaie.kubjas@aalto.fi}).}
\and Lawrence J. Sun\footnotemark[2]
\and Caroline Uhler\footnotemark[2].}

\usepackage{amsopn}

\ifpdf
\hypersetup{
  pdftitle={Identifying 3D Genome Organization in Diploid Organisms via Euclidean Distance Geometry},
  pdfauthor={A. Belyaeva, K. Kubjas, L. J. Sun and C. Uhler}
}
\fi

\makeatletter
\newcommand*{\addFileDependency}[1]{% argument=file name and extension
  \typeout{(#1)}
  \@addtofilelist{#1}
  \IfFileExists{#1}{}{\typeout{No file #1.}}
}
\makeatother

\begin{document}

\maketitle

% REQUIRED
\begin{abstract}
The spatial organization of the DNA in the cell nucleus plays an important role for gene regulation, DNA replication, and genomic integrity. Through the development of chromosome conformation capture experiments (such as 3C, 4C, Hi-C) it is now possible to obtain the contact frequencies of the DNA at the whole-genome level. In this paper, we study the problem of reconstructing the 3D organization of the genome from such whole-genome contact frequencies. A standard approach is to transform the contact frequencies into noisy distance measurements and then apply semidefinite programming (SDP) formulations to obtain the 3D configuration. However, neglected in such reconstructions is the fact that most eukaryotes including humans are diploid and therefore contain two copies of each genomic locus. We prove that the 3D organization of the DNA is not identifiable from distance measurements derived from contact frequencies in diploid organisms. In fact, there are infinitely many solutions even in the noise-free setting. We then discuss various additional biologically relevant and experimentally measurable constraints (including distances between neighboring genomic loci and higher-order interactions) and prove identifiability under these conditions. Furthermore, we provide SDP formulations for computing the 3D embedding of the DNA with these additional constraints and show that we can recover the true 3D embedding with high accuracy from both noiseless and noisy measurements. Finally, we apply our algorithm to real pairwise and higher-order contact frequency data and show that we can recover known genome organization patterns.
\end{abstract}

% REQUIRED

% REQUIRED

\section{Introduction}
It is now well established that the spatial organization of the genome in the cell nucleus plays an important role for cellular processes including gene regulation, DNA replication, and the maintenance of genomic integrity~\cite{gene, Uhler_Trends, Uhler_Nat_Rev}. Notably, a recent study~\cite{wang2018crispr} showed a causal link between three-dimensional (3D) genome organization
%genome architecture 
and gene regulation, where gene repositioning was induced and subsequent changes in gene expression were observed. This motivates the development of methods to reconstruct the 3D structure of the genome to study its functions.
%Therefore, reconstructing the three-dimensional (3D) structure of the genome is critical for studying the functions of the genome. 

The genetic information in cells is contained in the DNA, which is organized into chromosomes and packed into the cell nucleus.
%Humans, like most eukaryotes, are diploid meaning they have two copies of the genome (coming from each parent). In human cells the genome consists of an approximately 3 meter long DNA sequence, which is divided into two copies of 23 chromosomes.
%(coming from each parent). 
Chromosome confirmation capture techniques (such as 3C, 4C, Hi-C, Capture-C) have enabled the interrogation of the contact frequencies between pairs of genomic loci at the whole-genome scale~\cite{Dekker1306,Simonis,genedist,Hughes}. In Hi-C, for example, interacting chromosome regions are crosslinked (i.e., frozen), the DNA is then fragmented,  the crosslinked fragments are ligated, and paired-end sequencing is applied to the ligation products and mapped to a reference genome~\cite{genedist}. By binning the genome and ascribing each read pair into the corresponding bin, one obtains a contact frequency matrix between genomic loci that is commonly of the size $10^6\times 10^6$. 
%textcolor{green}{Maybe a better division would be distance-based, physics-based (MD simulations) and probabilitstic (MCMC sampling and Poisson modeling of contact frequency)?}

Different computational approaches for reconstructing the 3D genome organization from contact frequency data have been considered. 
%There are three major types of computational approaches for reconstruction of the 3D genome organization from contact frequency data: distance-based, ensemble, and statistical approaches. 
Distance-based approaches convert contact frequencies $F_{ij}$ into spatial distances $D_{ij}$ and find a Euclidean embedding of the points in 3D~\cite{duan2010three, chromsde, lesne20143d, rieber2017miniMDS}. Ensemble methods such as MCMC5C and BACH~\cite{rousseau2011three, hu2013bayesian} learn a set of possible 3D structures by defining a probabilistic model for contact frequencies and generating an ensemble of structures via MCMC sampling.
%, while methods such as
%Similarly, MOGEN~\cite{trieu2016mogen} MOGEN~\cite{trieu2016mogen} obtain a set of 3D structures via gradient ascent with different initializations.
%that can explain the data while MOGEN~\cite{trieu2016mogen} learns multiple 3D structures by starting gradient ascent at different initializations
%set of structures such that either the average distances or the contact probability between every two loci are consistent with the observed interaction frequencies
%GEM~\cite{zhu2018GEM} learns an ensemble, learn a networks of interactions based on Hi-C data, ensemble comes from the different mizing proportions, model data as a mixture of conformations and minimize the KL divergence between observed and actual contact frequencies
%by maximizing the scoring function
%\cite{wang2015bayesian} use bayesian EM to learn ensemble of strcutres
%\textcolor{red}{Add another method to say how this works; does this include the physics simulations?}. 
Other ensemble methods include molecular dynamics simulations that model DNA as a polymer and output an ensemble of 3D structures ~\cite{lieberman2009comprehensive, mirny2011fractal, dipierro2016transferablemodel, Qi2019MD}.
%beads on a string: Monte Carlo simulations to construct ensembles of fractal globules (lieberman-aiden) mechanism-based modeling physics modeling
%data-driven physics modeling (Pierro, Bin Zhang)
%physics-models where force-field parameters are learned from Hi-C data
Finally, statistical methods have also been proposed that directly model contact counts instead of distances, using for example the Poisson distribution~\cite{varoquaux2014statistical}, and maximize the log-likelihood
of the data to infer the 3D genome organization.

Almost all existing methods make the simplifying assumption that the genome is haploid, when in fact most organisms of interest including humans are diploid, i.e. there are two copies of each chromosome known as \emph{homologous chromosomes}. For example, human cells contain two copies of 23 chromosomes each.
The challenge is that the contact frequency data from chromosome conformation capture experiments is generally \emph{unphased}, meaning that the copies of each chromosome cannot be distinguished.
%and hence each measured contact frequency is the sum of four frequencies.
%it cannot distinguish between homologous chromosomes and 
%thus the contact frequency matrix between all coordinates in the diploid genome cannot be obtained experimentally. 
As a result, if the DNA is modeled as a string of beads containing two copies of each bead $i$ for 
$1\leq i\leq n$,
%$1\leq i\neq n$, 
then the measured contact frequencies result in an $n \times n$ matrix, from which we would like to infer the 3D embedding of $2n$ points. This problem cannot be solved by classical methods for 3D genome reconstruction methods such as those mentioned above.
%, making classical methods for 3D genome reconstruction not suitable for this challenge. 
With significant experimental efforts, phased data can be obtained~\cite{1000genomes2012integrated, 1000genomes2015global} and used in order to reconstruct the 3D genome organization~\cite{cauer2019diploid}. However, such data is rare and costly.

\begin{figure}[!t]
    \centering
    \includegraphics[width=0.3\textwidth]{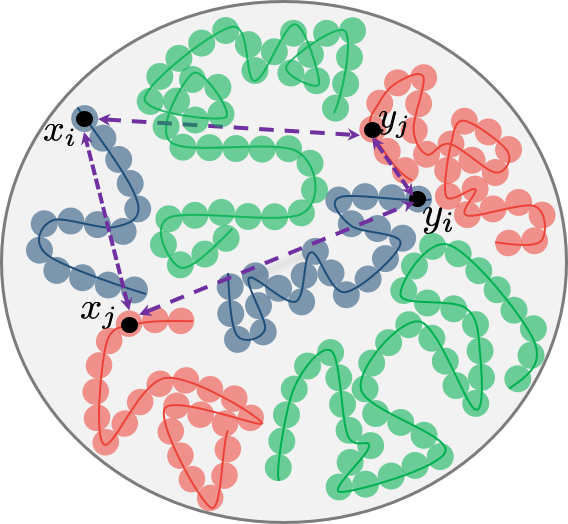}
    \caption{Schematic of the diploid genome. Nucleus with green, blue and red curves depicting three homologous pairs of chromosomes. In the unphased setting, the measured distance between loci $i$ and $j$ corresponds to the sum of the four distances (denoted in purple) between two pairs of homologous loci $x_i, y_i$ and $x_j, y_j$. 
    %The distance between loci $i$ and $j$, derived from Hi-C matrices, is a sum of four distances (denoted in purple) between pairs of coordinates.}
    }
    \label{fig:distances}
\end{figure}

In this paper, we provide a computational method for inferring the 3D diploid organization of the genome without relying on phased data.
In particular, we consider a distance-based approach and use Euclidean distance geometry to obtain the 3D diploid structure of the genome.
%based on unphased data only. 
The precise mathematical problem considered in this paper is as follows and illustrated in \Cref{fig:distances}. DNA is modeled as a string of beads, that contains two copies of each bead $i$ for $1\leq i\leq n$.
%$1\leq i\neq n$.
%$i$ such that $1\leq i\neq j\leq n$. 
We would like to infer the location of the two copies of each bead, which we denote by $x_i \in\mathbb{R}^3$  and $y_i \in\mathbb{R}^3$. Since for unphased data, the two copies of each bead cannot be distinguished, the problem is to identify the 3D configuration ($2n \times 3$ matrix), i.e.~$x_1, \dots x_n, y_1, \dots y_n\in\mathbb{R}^3$ (up to translation and rotation), from the composite distance measurements $D_{ij}$, $1\leq i\neq j\leq n$ ($n \times n$ matrix), corresponding to the sum of the distances between either copy of bead $i$ and $j$, i.e.,
$$
D_{ij} = \|x_i - x_j\|^2 + \|x_i - y_j\|^2 + \|y_i - x_j\|^2 + \|y_i - y_j\|^2.
$$

In the haploid or phased setting, this problem boils down to the standard Euclidean distance geometry problem. This problem has a long history: in the classical setting with no missing values, this problem can be solved via the classical multidimensional scaling (cMDS) algorithm that is based on spectral decomposition followed by dimensionality reduction; see~\cite{cox2000multidimensional} for an overview. Other approaches for the Euclidean embedding and completion problems, including in the presence of missing values, are non-convex formulations~\cite{Fang2012,Mishra2011} as well as semidefinite relaxations~\cite{Alfakih1999,fazel2003log,Cayton,Lu12332,weinberger2007graph,zhang2016distance}.

A naive approach in the unphased diploid setting is to assume that the four distances that make up our measured composite distance $D_{ij}$ are equal and solve the corresponding Euclidean embedding problem. However, it is evident from single-cell imaging studies that the four distances in $D_{ij}$ can be wildly different~\cite{bolzer2005three, nir2018walking}. Hence this approach cannot provide realistic embeddings. While, a simple dimension argument ($6n$ variables versus $\binom{n}{2}$ constraints) suggests that the 3D genome configuration is uniquely identifiable, one of the main results of our paper is that the 3D diploid genome configuration is not identifiable from unphased data. In fact, we show that there are infinitely many configurations that satisfy the constraints imposed by $D_{ij}$, even in the noiseless setting (\cref{section:non_identifiability}, \cref{theorem:non_identifiability}).

We therefore consider additional biologically relevant and experimentally measurable constraints and study identifiability of the 3D diploid structure under these constraints. First, we take into account distances between neighboring beads, i.e.~$\|x_i - x_{i+1}\|^2$ and $\|y_i - y_{i+1}\|^2$ on each chromosome. While we show that this yields unique identifiability for configurations in 2D, there are still infinitely many configurations in 3D, which is of primary interest for genome modeling (\cref{sec_3}, \cref{prop:uniqueness_2d_distances_between_neighboring_loci}, \cref{prop:uniqueness_3d_distances_between_neighboring_loci}). To obtain identifiability in 3D, we consider adding constraints based on contact frequencies between three or more loci simultaneously.
%additionally knowing contact frequencies between three or more loci simultaneously.}
The measurement of such higher-order contact frequencies has recently been enabled by experimental assays such as SPRITE~\cite{Quinodoz219683}, C-walks~\cite{Olivares} and GAM~\cite{beagrie2017complex}. We prove that this information can be used to uniquely identify the 3D genome organization from unphased data in the noiseless setting (\cref{sec_5}, \cref{theorem:tensor_uniqueness}).
%We prove that this information can be used to uniquely identify the 3D genome organization from unphased data without using any information regarding the distance to the center of the cell nucleus (Section~\ref{sec_5}, Theorem~\ref{theorem:tensor_uniqueness}).

%In this paper, we establish identifiability of 3D diploid structure in the noiseless setting. 

Finally, we provide an SDP formulation for obtaining the 3D diploid configuration from noisy measurements (\cref{sec_algorithms}) and show based on simulated data that our algorithm has good performance and that it is able to recover known genome organization patterns when applied to real contact frequency data collected from human lymphoblastoid cells  (\cref{sec_data}).
%We provide algorithms for obtaining the 3D configuration in the noiseless setting. We show that our algorithm has good performance in simulation and also is able to recover prior biological knowledge on real data.

%In Section~\ref{sec_algorithms}

\section{Unidentifiability from pairwise distance constraints}
\label{section:non_identifiability}

In the remainder of the paper we denote the true but unknown coordinates of the homologous loci by $x^*_i$ and $y^*_i$ and the corresponding noiseless distances by $D^*_{ij}$ while the symbols $x_i$ and $y_i$ denote the variables that we want to solve for. While from a biological perspective the relevant setting is when $x_i, y_i\in\mathbb{R}^3$, results that hold more generally will be stated in $\mathbb{R}^d$.  The main result of this section is \cref{theorem:non_identifiability}, which characterizes the set of solutions given by the constraints $D^*_{ij}$ in dimension $d\leq 3$. In particular, it establishes non-identifiability of the 3D genome structure from pairwise distance measurements in the diploid unphased setting.

\begin{theorem} \label{theorem:non_identifiability}
Let $d \leq 3$ and $n \ge 2d + 3$. 
%The set of points $(x_1,\ldots,x_n,y_1,\ldots,y_n)\in(\mathbb{R}^d)^{2n}$ satisfying
%\begin{equation} \label{equations_for_pairwise_distances}
%D^*_{ij} = \|x_i - x_j\|^2 + \|x_i - y_j\|^2 + \|y_i - x_j\|^2 + \|y_i - y_j\|^2 \;\text{ for all }\, 1 \leq i \neq j \leq n
%\end{equation} 
%is equal (up to translations and rotations in $\mathbb{R}^d$ and permutations of $x_i$ and $y_i$) to the set of points satisfying
%\begin{equation} \label{equations_for_nonidentifiable_solutions}
%x_i + y_i = x^*_i + y^*_i \;\text{ and }\; \|x_i\|^2 + \|y_i\|^2 = \|x^*_i\|^2 + \|y^*_i\|^2 \;\text{ for all } \,1\leq i \leq n.
%\end{equation}
Then $(x_1,\ldots,x_n,y_1,\ldots,y_n) \in (\mathbb{R}^d)^{2n}$ satisfies
\begin{equation} \label{equations_for_pairwise_distances}
D^*_{ij} = \|x_i - x_j\|^2 + \|x_i - y_j\|^2 + \|y_i - x_j\|^2 + \|y_i - y_j\|^2 \;\text{ for all }\, 1 \leq i \neq j \leq n
\end{equation} 
 if and only if it satisfies
\begin{equation} \label{equations_for_nonidentifiable_solutions}
x_i + y_i = x^*_i + y^*_i \;\text{ and }\; \|x_i\|^2 + \|y_i\|^2 = \|x^*_i\|^2 + \|y^*_i\|^2 \;\text{ for all } \,1\leq i \leq n
\end{equation}
up to translations and rotations in $\mathbb{R}^d$ and permutations of $x_i$ and $y_i$.
\end{theorem}

As a consequence, the measurements $D_{ij}^*$ identify the location of each pair of homologous loci $(x_i, y_i)$ up to a sphere with center  $(x^*_i + y^*_i)/2$ and radius $\|x^*_i - y^*_i\|/2$. Namely, the points $x_i, y_i$ lie opposite to each other anywhere on this sphere. 
Unless $x^*_i = y^*_i$ for all $i$, i.e., all spheres have radius $0$, this set is infinite in dimensions $d>1$ and hence the configuration is unidentifiable. 

In the remainder of this section, we will prove \cref{theorem:non_identifiability}. The two inclusions in \cref{theorem:non_identifiability} are proven in \cref{lemma:non_identifiability_from_HiC_data} and \cref{lemma:exact_solution_set_HiC_data}. In \cref{lemma:distances_between_homologue_pairs} it is shown that the distance $\|x_i-y_i\|$ within each homologous pair is fixed given the pairwise distances~$D^*_{ij}$. This result is used to prove \cref{lemma:exact_solution_set_HiC_data}.

\begin{lemma} \label{lemma:non_identifiability_from_HiC_data}
Let $(x_1,\ldots,x_n,y_1,\ldots,y_n) \in (\mathbb{R}^d)^{2n}$ satisfy
\begin{equation}\label{equation3}
x_i + y_i = x^*_i + y^*_i \text{ and } \|x_i\|^2 + \|y_i\|^2 = \|x^*_i\|^2 + \|y^*_i\|^2 \;\text{ for all } \,1\leq i \leq n.
\end{equation}
Then
$$
\|x_i - x_j\|^2 + \|x_i - y_j\|^2 + \|y_i - x_j\|^2 + \|y_i - y_j\|^2 = D^*_{ij} \;\text{ for all }\, 1 \leq i \neq j \leq n.
$$
\end{lemma}

\begin{proof}
Observe that for each pair $x_i, y_i$ satisfying the equations \cref{equation3}, it holds that
\begin{eqnarray*}
D^*_{ij} & = & 2 \cdot (\|x^*_i\|^2 + \|y^*_i\|^2) + 2 \cdot (\|x^*_j\|^2 + \|y^*_j\|^2)  - 2(x^*_i + y^*_i) \cdot (x^*_j + y^*_j) \\
& = & 2 \cdot (\|x_i\|^2 + \|y_i\|^2) + 2 \cdot (\|x_j\|^2 + \|y_j\|^2)  - 2(x_i + y_i) \cdot (x_j + y_j) \\
& = & \|x_i - x_j\|^2 + \|x_i - y_j\|^2 + \|y_i - x_j\|^2 + \|y_i - y_j\|^2.
\end{eqnarray*}
This completes the proof.
\end{proof}

Next we will show that the distance between homologous pairs is uniquely determined by the $D^*_{ij}$

\begin{lemma} \label{lemma:distances_between_homologue_pairs}
Let $d \leq 3$ and $n \ge 2d + 3$. Then for each $1\leq i \leq n$ the quantity $\|x_i - y_i\|$ is identifiable from the constraints imposed by the $D^*_{ij}$, i.e., for any solution $(x_1,\ldots,x_n,y_1,\ldots,y_n) \in (\mathbb{R}^d)^{2n}$ to the equations defined by the $D^*_{ij}$ in \cref{equations_for_pairwise_distances},
the quantity $\|x_i - y_i\|$ is constant.
\end{lemma}

The constraint $d \leq 3$ is due to our proof technique. The condition $n \ge 2d + 3$ is  necessary for unique identifiability of the distance between homologous pairs of loci.

\begin{proof}
Without loss of generality we assume that $i=1$ and show that $\|x_1 - y_1\|$ is equal to some constant. First, we perform a shift on the solution so that $x_1 = -y_1 = v$. Since shifts preserve distances, they in particular preserve the equality constraints \cref{equations_for_pairwise_distances}. Hence,
\[
D^*_{1j} = \|v - x_j\|^2 +  \|v - y_j\|^2 +  \|-v - x_j\|^2 +  \|-v - y_j\|^2.
\]
Expanding this out into dot products and simplifying yields
\[
D^*_{1j} = 4\|v\|^2 + 2(\|x_j\|^2 + \|y_j\|^2).
\]
Let $j \neq k$ be both not equal to $1$. Then substituting the above leads to
\[
D^*_{1j} + D^*_{1k} - D^*_{jk} = 8\|v\|^2 + 2(x_j + y_j) \cdot (x_k + y_k).
\]
Defining $T_{jk} := D^*_{1j} + D^*_{1k} - D^*_{jk}$ and $s_j := \sqrt{2}(x_j + y_j)$, this is equivalent to
\[
T_{jk} - 8\|v\|^2 = s_j \cdot s_k.
\]

Let $T'$ be the $(d+1) \times (d+1)$ submatrix of $T$ satisfying $T'_{ij} = T_{i+1,j + d + 2}$, i.e. the rows of $T'$ correspond to the rows $2, 3, \ldots, d+2$ of $T$ and the columns of $T'$ correspond to the columns $d+3, d+4, \ldots, 2d+3$ of $T$. We now show that for generic configurations $\det(T') \neq 0$. Since $\det(T')$ can be written as a polynomial in the coordinates $x_i$ and $y_i$, then $\det(T') \neq 0$ for generic configurations as long as it does not identically vanish. Hence it suffices to present one configuration where $\det(T')$ is nonzero. For $d \le 3$ we can check this using random configurations.

Since $T'$ has full rank, then the matrix determinant lemma implies that 
\begin{equation} \label{equation_matrix_determinant_lemma}
\det(T' - 8J\|v\|^2) = (1 - 8\|v\|^2\mathds{1}^T(T')^{-1}\mathds{1}) \det (T'),
\end{equation}
  where $\mathds{1}$ denotes the all ones vector. Note that the scalar $\mathds{1}^TT'^{-1}\mathds{1}$ is fixed and $(\det T') \neq 0$. Furthermore, since $T' - 8J\|v\|^2$ is formed from the dot products between $d$-dimensional vectors, it has rank at most $d$ and therefore $\det(T' - 8J\|v\|^2) = 0$ due to $T' - 8J\|v\|^2$ being a $(d+1) \times (d+1)$ matrix. Hence, $(1 - 8\|v\|^2\mathds{1}^T(T')^{-1}\mathds{1}) \det (T')=0$, which is a linear equation in terms of $\|v\|^2$. As a consequence, it has a unique solution for $\|v\|^2$ and thus the distance between the homologous pair $x_1, y_1$ is fixed as long as $n \ge 2d + 3$.
\end{proof}

We next characterize all solutions to the constraints imposed by the  $D^*_{ij}$.

\begin{lemma} \label{lemma:exact_solution_set_HiC_data}
Let $d \leq 3$ and $n \ge 2d + 3$. Let $(x_1,\ldots,x_n,y_1,\ldots,y_n) \in (\mathbb{R}^d)^{2n}$ be a solution to
$$
\|x_i - x_j\|^2 + \|x_i - y_j\|^2 + \|y_i - x_j\|^2 + \|y_i - y_j\|^2 = D^*_{ij} \;\text{ for all }\, 1 \leq i \neq j \leq n.
$$
Then
$$
x_i + y_i = x^*_i + y^*_i \text{ and } \|x_i\|^2 + \|y_i\|^2 = \|x^*_i\|^2 + \|y^*_i\|^2 \;\text{ for all } \,1\leq i \leq n
$$
up to translations and rotations in $\mathbb{R}^d$ and permutations of $x_i$ and $y_i$.
\end{lemma}

\begin{proof}
Without loss of generality we perform a translation on the solution  such that $x_1 = -y_1 = v$ for some vector $v$. By \cref{lemma:distances_between_homologue_pairs} the quantity $\|x_k - y_k\|$ is constant for each $1\leq k\leq n$ and thus also $\|v\|$ is constant. Since for any $j \neq 1$ it holds that $D^*_{1j} = 4\|v\|^2 + 2(\|x_j\|^2 + \|y_j\|^2)$, also $\|x_j\|^2 + \|y_j\|^2$ is constant and hence $\|x_i\|^2 + \|y_i\|^2 = \|x^*_i\|^2 + \|y^*_i\|^2$ for all $1\leq i \leq n$.

Similarly to the proof of \cref{lemma:distances_between_homologue_pairs}, if we define $T_{jk} = D^*_{1j} + D^*_{1k} - D^*_{jk}$ and $s_j = \sqrt{2}(x_j + y_j)$, we find that
\[
T_{jk} - 8 \|v\|^2 = s_j \cdot s_k.
\]
Because we have access to the diagonal constraints now, this relationship holds for all $j,k$ and not just $j \neq k$. Thus $T - 8J \|v\|^2$ is a symmetric $(n - 1) \times (n - 1)$ matrix admitting a rank $d$ factorization. Let $S$ be the matrix formed with the vectors $s_j$. We then have $T - 8J \|v\|^2 = SS^T$. There is a result on rank factorizations of symmetric matrices that any other factorization $T - 8J \|v\|^2 = S'S'^T$ satisfies $S = S'Q$ for some orthogonal matrix $Q$~\cite[Proposition 3.2]{krislock2010semidefinite}. Thus for any other solution $s'_j$, we have $s_j = s'_jQ$, implying all solutions are simply orthogonal transformations of each other (rotations, reflections, etc.)

In summary, we have shown that once we have fixed $x_1 + y_1 = 0$ via translation, then  the quantities $x_j + y_j$ are unique up to orthogonal transformations and the quantities $\|x_j\|^2 + \|y_j\|^2$ are unique. 
\end{proof}

\section{Distance constraints between neighboring loci}
\label{sec_3}
\begin{figure}[!t]
    \centering
    \includegraphics[width=0.3\textwidth]{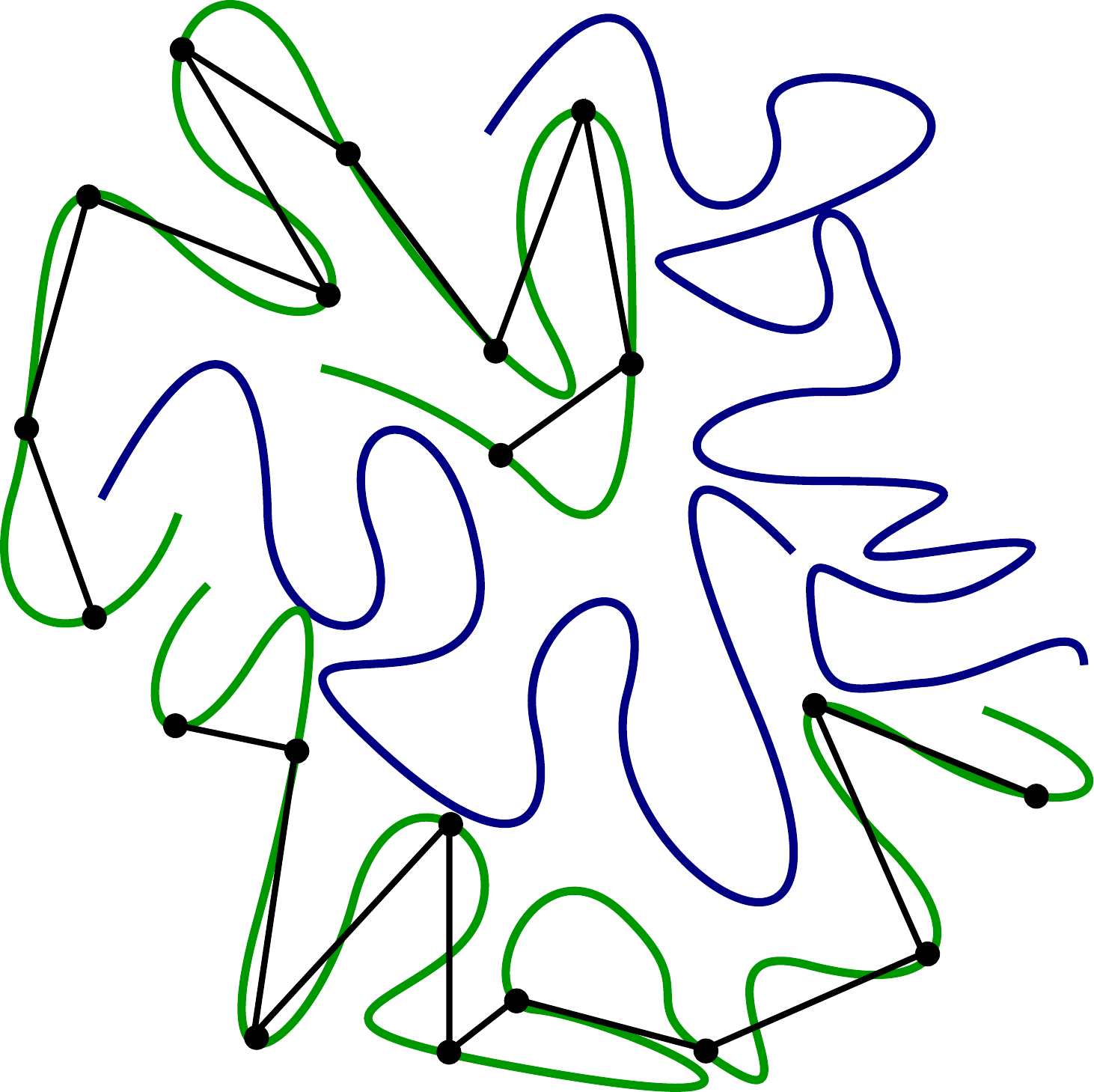}
    \caption{Distance constraints between neighboring beads. Green and blue curves depict two homologous pairs of chromosomes. For the green curves distances between neighboring genomic regions are shown by black lines.}
    \label{fig:distances_between_neighboring_loci}
\end{figure}

In \cref{section:non_identifiability}, we showed that the 3D genome configuration is not identifiable from pairwise distance constraints available from typical (unphased) contact frequency maps. In order to gain identifiability, we next consider adding other biological constraints to the problem formulation that are generally available or can be measured. In particular, since DNA can be viewed as a string of connected beads, we use the distance between adjacent beads as an additional constraint. The distance between neighboring beads can be derived empirically for example from imaging studies~\cite{muller2010stable, jungmann2014DNAPAINT}; see also our experimental results in \cref{sec_data}. The additional mathematical constraints are:
\begin{equation*} 
\|x_i - x_{i+1}\| = \|x^*_i - x^*_{i+1}\| \text{ and } \|y_i - y_{i+1}\| = \|y^*_i - y^*_{i+1}\| \text{ for } 1 \leq i \leq n-1,
\end{equation*}
where $x^*_1, x^*_2, \ldots, x^*_n$ and $y^*_1, y^*_2, \ldots, y^*_n$ correspond to consecutive beads on homologous chromosomes; see \Cref{fig:distances_between_neighboring_loci}.

%corresponding to each chromosome and we know the distances $\|x^*_i - x^*_{i+1}\|$ and $\|y^*_i - y^*_{i+1}\|$ between neighboring genomic regions, 
In this section we show the following results: under the additional distance constraints between neighboring loci, we prove that identifiability can be obtained in the 2D setting (\cref{prop:uniqueness_2d_distances_between_neighboring_loci}). However, in the 3D setting we prove that there are still infinitely many 3D configurations even with these additional distance constraints (\cref{prop:uniqueness_3d_distances_between_neighboring_loci}).

%but in Proposition~\ref{prop:uniqueness_3d_distances_between_neighboring_loci} we prove that in the 3D setting there are still infinitely many 3D configurations.

% In this section, we assume that additional information is available. Let $x^*_1, x^*_2, \ldots, x^*_n$ and $y^*_1, y^*_2, \ldots, y^*_n$ form chains of genomic regions. We know the distances $\|x^*_i - x^*_{i+1}\|$ and $\|y^*_i - y^*_{i+1}\|$ between neighboring genomic regions, see Figure~\ref{fig:distances_between_neighboring_loci}. We consider the polynomial system:
% \begin{equation} \label{equation_pairwise_distances}
% \begin{gathered}
% x_i + y_i = x^*_i + y^*_i \text{ and } \|x_i\|^2 + \|y_i\|^2 = \|x^*_i\|^2 + \|y^*_i\|^2 \text{ for } 1 \leq i \leq n,\\
% \|x_i - x_{i+1}\| = \|x^*_i - x^*_{i+1}\| \text{ and } \|y_i - y_{i+1}\| = \|y^*_i - y^*_{i+1}\| \text{ for } 1 \leq i \leq n-1.
% \end{gathered}
% \end{equation}

For the proofs of \Cref{prop:uniqueness_2d_distances_between_neighboring_loci} and \Cref{prop:uniqueness_3d_distances_between_neighboring_loci} we recall from \cref{theorem:non_identifiability} that $(x_i, y_i)$ and $(x^*_i, y^*_i)$ are diametrically opposite points on the same sphere. Denote the $i$-th sphere by $S_i$ and let it have center $c_i$ and radius $r_i$.  Then $\|c_i - x_i\| = r_i$ and $2c_i - x_i = y_i$. 

\begin{proposition}\label{prop:uniqueness_2d_distances_between_neighboring_loci}
For $n \geq 3$ and generic $(x^*_1, \ldots, x^*_n,y^*_1, \ldots, y^*_n) \in (\mathbb{R}^2)^{2n}$, there is a unique point $(x_1,\ldots,x_n,y_1,\ldots,y_n) \in (\mathbb{R}^2)^{2n}$ satisfying the equations
\begin{equation} \label{equation_pairwise_distances}
\begin{gathered}
x_i + y_i = x^*_i + y^*_i \text{ and } \|x_i\|^2 + \|y_i\|^2 = \|x^*_i\|^2 + \|y^*_i\|^2 \text{ for } 1 \leq i \leq n,\\
\|x_i - x_{i+1}\| = \|x^*_i - x^*_{i+1}\| \text{ and } \|y_i - y_{i+1}\| = \|y^*_i - y^*_{i+1}\| \text{ for } 1 \leq i \leq n-1.
\end{gathered}
\end{equation}
\end{proposition}

\begin{proof}
We have $y_1 = 2c_1 - x_1$ and $y_2 = 2c_2 - x_2$. Plugging this into $\|y_1 - y_2\| = \|y^*_1 - y^*_2\|$ gives
\begin{eqnarray*}
\|y^*_1 - y^*_2\| & = & \|(2c_1 - x_1) - (2c_2 - x_2)\|^2 \\
& = & \|(2c_1 - 2c_2) - (x_1 - x_2)\|^2 \\
& = &\|2c_1 - 2c_2\|^2 + \|x_1 - x_2\|^2 - 2 (2c_1 - 2c_2) \cdot (x_1 - x_2).
\end{eqnarray*}
The quantities $\|2c_1 - 2c_2\|^2$ and $\|x_1 - x_2\|^2$ are fixed. This implies that the quantity $(2c_1 - 2c_2) \cdot (x_1 - x_2)$ is fixed. Since we know $\|x_1 - x_2\|$ and $c_1 \neq c_2$ holds by genericness, then there are two possible angles for $x_1 - x_2$ (this is where we use the 2D constraint) and thus that there are two possible solutions  for $x_1 - x_2$.

Because $x_1, x_2$ are constrained to lie on circles, the solutions for $x_1$ are the intersection points of the first circle and the second circle translated by $x_1-x_2$ and the solutions for $x_2$ are the intersection points of the second circle and the first circle translated by $x_2-x_1$. Hence each solution for $x_1 - x_2$ leads to at most two possible solutions for $(x_1, x_2)$. In turn this implies there are at most four solutions for $x_2$. 

\begin{figure}[!b]
	\centering
    \includegraphics[scale=0.2]{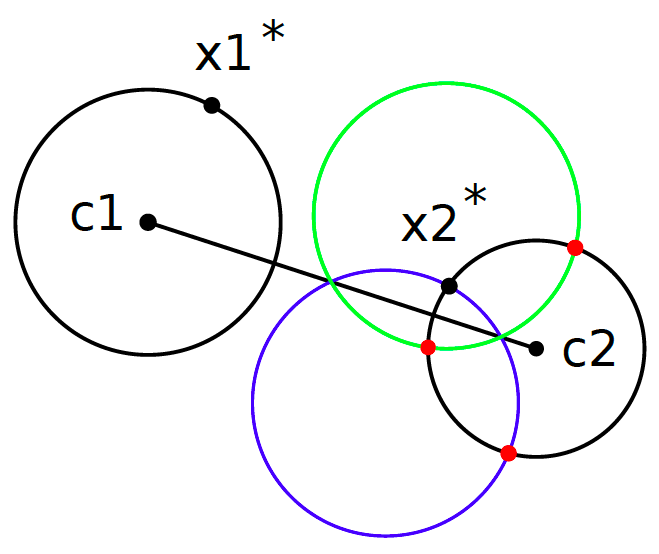}
    \caption{Identifiability in the 2D setting with neighboring distance constraints. Two solutions for $x_2$ are obtained by translating the circle centered at $c_1$ by $x^*_2-x^*_1$ (this new circle is colored blue) and intersecting it with the circle centered at $c_2$. The other two solutions are obtained by reflecting the blue circle over the line through $c_1$ and $c_2$ (this new circle is colored green) and intersecting it with the circle centered at $c_2$. The true solution for $x_2$ is colored black and the three alternative solutions for $x_2$ are colored red.}
    \label{figure_solutions_neighboring_circles}
\end{figure}

We now investigate the four solutions. The first two solutions are obtained by translating the circle centered at $c_1$ by $x^*_2 - x^*_1$ and intersecting it with the circle centered at $c_2$, see \Cref{figure_solutions_neighboring_circles}. One of the two solutions is $x^*_2$. The other two solutions are reflections of these two solutions over the line from $c_1$ to $c_2$.

Let $x^*_1,x^*_2,c_1,c_2$ be fixed. They determine four possible solutions for $x_2$. We will show that these four solutions are different from the four solutions we get from considering $x^*_2,x^*_3,c_2,c_3$ for generic $x^*_3,c_3$ (apart from $x^*_2$).

If either of the reflected solutions over the line from $c_2$ to $c_3$ coincides with one of the four original solutions, then we can perturb $c_3$ away from the line from $c_2$ to $c_3$ to change these solutions. If the solution that is the intersection point of the circle centered at $c_2$ and the translation by $x^*_2-x^*_3$ of the circle centered at $c_3$ (different from $x^*_2$) coincides with one of the four original solutions, then we can perturb $x^*_3$. This changes $x^*_2-x^*_3$ and hence the second intersection point of the circle centered at $c_2$ and the translation by $x^*_2-x^*_3$ of the circle centered at $c_3$.

A similar argument can be used to show that $x_3,\ldots,x_{n-1}$ have unique solutions. Given a unique solution for $x_2$, there are two solutions for $x_1$ if and only if $x^*_2$ lies on the line from $c_1$ to $c_2$. This is however not a generic configuration. A similar argument applies for $x_n$.
\end{proof}

Despite having uniqueness in 2D, we do not have uniqueness in 3D as shown in the following proposition.

\begin{proposition}\label{prop:uniqueness_3d_distances_between_neighboring_loci}
Let $n \in \mathbb{N}$. For generic $(x^*_1,  \ldots, x^*_n,y^*_1,  \ldots, y^*_n) \in (\mathbb{R}^3)^{2n}$, there are infinitely many points $(x_1,\ldots,x_n,y_1,\ldots,y_n) \in (\mathbb{R}^3)^{2n}$ satisfying equations \cref{equation_pairwise_distances}.
\end{proposition}

\begin{proof}
If $n=1$, then $x_1^*,y_1^*$ can be chosen randomly with the constraint that $x_1^* \neq y_1^*$. Then $x_1$ and $y_1$ can be any points on the sphere $S_1$ defined by $x_1^*,y_1^*$. Now assume that $n \geq 2$. Fix any $x_1^*,y_1^*$ such that $x_1^* \neq y_1^*$. Choose two circles $C_1$ and $C'_1$ on the sphere $S_1$ defined by $x_1^*,y_1^*$ that intersect at two points one of which is $x_1^*$. The circle $C_1$ is the intersection of $S_1$ and another sphere $T_1$. Let $x_2^*$ be the center of the sphere $T_1$. Let $C''_1$ be the circle on $S_1$ that consists of points antipodal to $C'_1$. Then $C''_1$ is also an intersection of $S_1$ and another sphere $T''_2$. Let $y_2^*$ be the center of the sphere $T''_2$. We use the same procedure to construct $x_3^*$ and $y_3^*$ from $x_2^*$ and $y_2^*$, $x_4^*$ and $y_4^*$ from $x_3^*$ and $y_3^*$ etc.

The only condition on $x_1^*$ and $y_1^*$ is $x_1^* \neq y_1^*$, hence $(x_1^*,y_1^*)$ is a generic point in $\mathbb{R}^3 \times \mathbb{R}^3$. The condition that $C_1$ is a circle on the sphere $S_1$ containing $x_1^*$ is equivalent to $x_2^*$ being any point in $\mathbb{R}^3$ outside the line through $x_1^*$ and $y_1^*$. Similarly, the condition that $C'_1$ is a circle on the sphere $S_1$ containing $x_1^*$ is equivalent to $y_2^*$ being any point in $\mathbb{R}^3$ outside the line through $x_1^*$ and $y_1^*$. The condition that $C_1$ and $C'_1$ intersect at two different points of $S_1$ is equivalent to the normal vector of the tangent plane of $S_1$ at $x_1^*$ and the normal vectors of the planes defined by $C_1$ and $C'_1$ being linearly independent. Hence $(x_1^*,x_2^*,y_1^*,y_2^*)$ is a generic point in $(\mathbb{R}^3)^4$. Similar arguments can be used to show that $(x^*_1, x^*_2, \ldots, x^*_n,y^*_1, y^*_2, \ldots, y^*_n)$ is a generic point in $(\mathbb{R}^3)^{2n}$.

Now consider points $x_n$ and $y_n$ in an $\varepsilon$-neighborhood of $x_n^*$ and 
$y_n^*$. Consider the spheres that are centered at $x_n$ and $y_n$ and have radii $ \|x^*_{n-1} - x^*_n\| \text{ and }  \|y^*_{n-1} - y^*_n\|$. The intersections of these spheres with $S_{n-1}$ give circles $\tilde{C}_{n-1}$ and $\tilde{C}''_{n-1}$ that are perturbations of circles $C_{n-1}$ and $C''_{n-1}$. In particular, the intersection of the circle $\tilde{C}_{n-1}$ and the circle $\tilde{C}'_{n-1}$ that consists of points antipodal to $\tilde{C}''_{n-1}$ consists of two points for $\varepsilon$ small enough. Choosing $x_{n-1}$ to be the intersection point corresponding to $x_{n-1}^*$ and $y_{n-1}$ its antipodal gives points $x_{n-1},y_{n-1}$ satisfying $\|x_{n-1} - x_n\|= \|x^*_{n-1} - x^*_n\| \text{ and } \|y_{n-1} - y_n\|= \|y^*_{n-1} - y^*_n\|$.

Assuming that $\varepsilon$ is small enough, then $x_{n-1}$ and $y_{n-1}$ are in small neighborhoods of $x_{n-1}^*$ and $y_{n-1}^*$, and we can continue the same procedure to find $x_{n-2}$ and $y_{n-2}$ from $x_{n-1}$ and $y_{n-1}$, $x_{n-3}$ and $y_{n-3}$ from $x_{n-2}$ and $y_{n-2}$ etc. In particular, we can find $x_1,\ldots,x_{n-1},y_1,\ldots,y_{n-1}$  satisfying equations \cref{equation_pairwise_distances} for every $x_n$ and $y_n$ in an $\varepsilon$-neighborhood of $x_n^*$ and 
$y_n^*$.
\end{proof}

The previous proposition suggests that there are two degrees of freedom for choosing $x_1,\ldots,x_n,y_1,\ldots,y_n$ on each homologous pair and thus that finite identifiability requires two additional algebraically independent constraints per homologous pair. Similarly this suggests that unique identifiability requires three additional algebraically independent constraints per homologous pair, where each endpoint of a chromosome needs to be included in at least one of the additional constraints.

%one expects that three additional algebraically independent constraints per homologous pair such that each endpoint of a chromosome or its homolog is included in at least one of the additional constraints gives unique identifiability.

\section{Identifiability from higher-order contact constraints}
\label{sec_5}

In \cref{sec_3}, we showed that considering distances between neighboring beads only yields identifiability in 2D but not in 3D. In the following, we consider adding further constraints
that are becoming widely available from experimental data, namely higher-order contact frequencies between three or more loci as measured by experimental assays such as SPRITE~\cite{Quinodoz219683}, C-walks~\cite{Olivares} and GAM~\cite{beagrie2017complex}. We express these constraints mathematically by letting $F \in \mathbb{R}^{m \times m \times \cdots \times m}$ be a contact frequency tensor, where $F_{x_{i_1}, x_{i_2}, \ldots ,x_{i_k}}$ measures the contact frequency between loci $i_1,i_2,\ldots,i_k$ with coordinates $x_{i_1},x_{i_2},\ldots,x_{i_k}$. In the unphased setting, we can only measure a combination of contact frequencies over the homologous loci $\{x_{i_1},y_{i_1}\} \times \{x_{i_2},y_{i_2}\} \times \ldots \times \{x_{i_k},y_{i_k}\}$, which we denote by $F_{i_1 i_2 \ldots i_k}$. In addition, as for 2-way interactions, we turn contact frequencies into ``distances'' by defining $D_{i_1 i_2 \ldots i_k}:=1/F_{i_1 i_2 \ldots i_k}$.

\begin{figure*}[!t]
    \centering
  \subfloat[]{\label{fig:tensor_constraints_a}\includegraphics[width=0.45\textwidth]{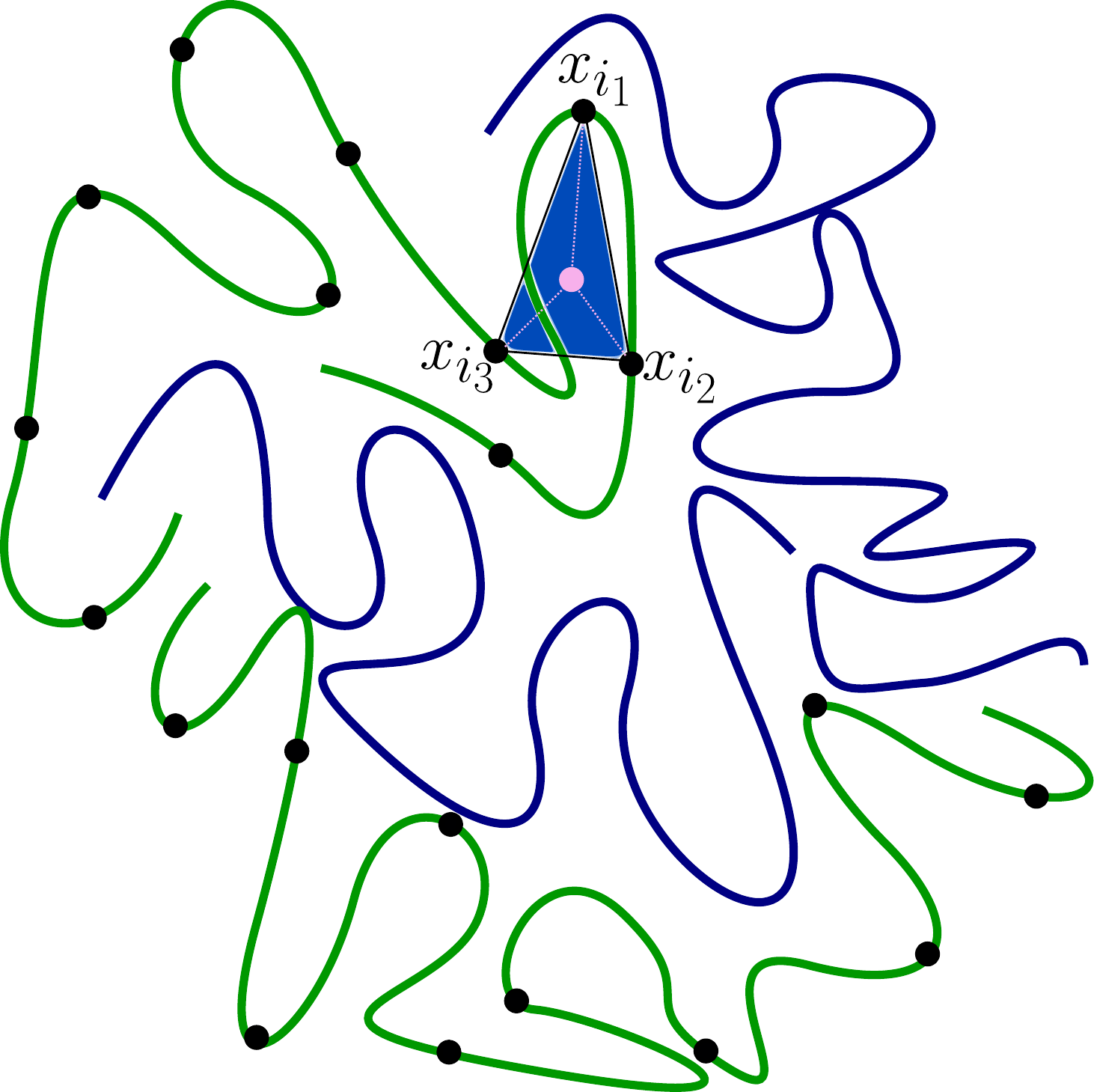}}
  \subfloat[]{\label{fig:tensor_constraints_b}\includegraphics[width=0.53\textwidth]{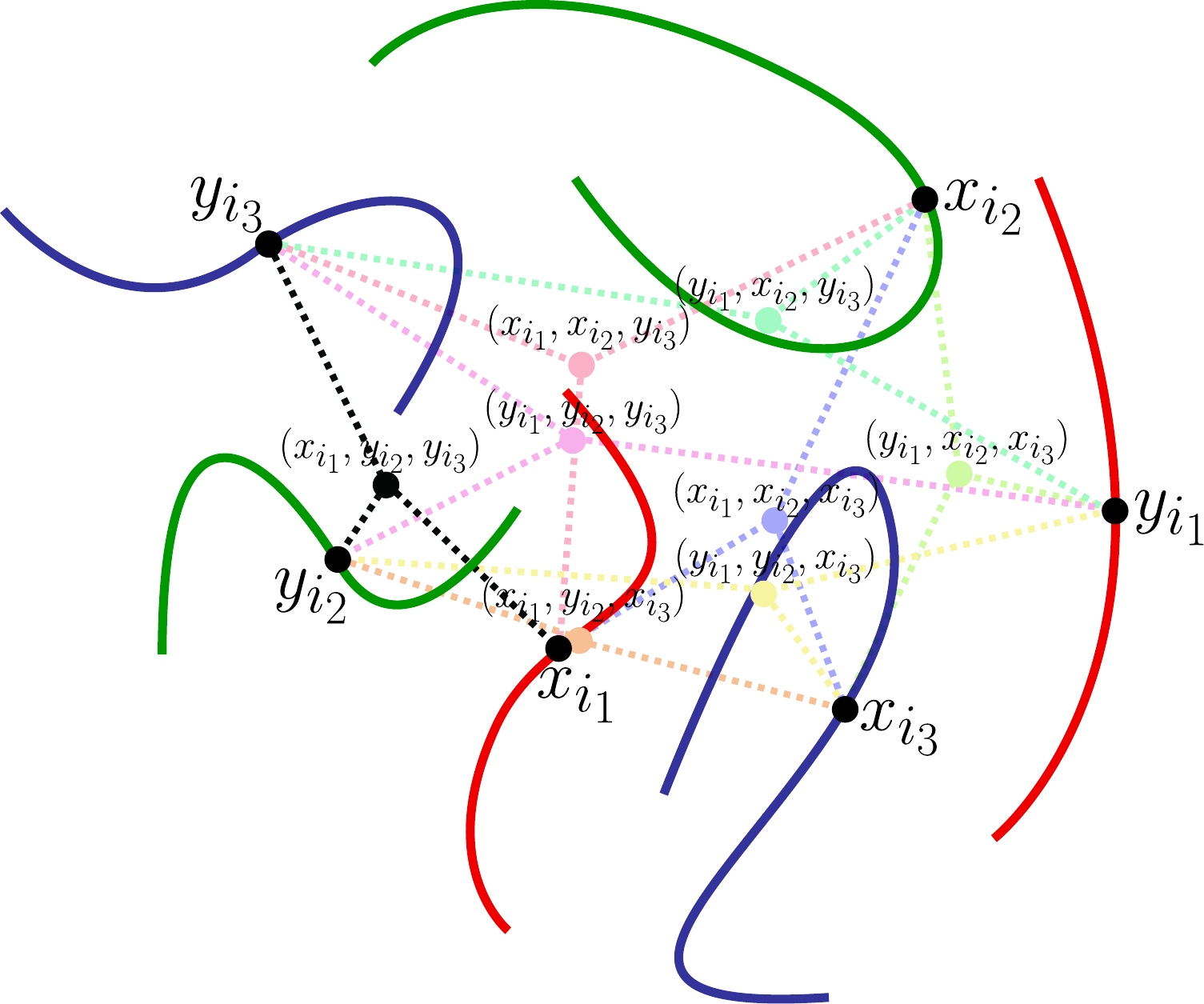}}
    \caption{Higher-order distance constraints. (a) Three loci $x_{i_1},x_{i_2}, x_{i_3}$, located on the same chromosome, are depicted. In the phased setting, the higher-order distance $D_{x_{i_1} x_{i_2} x_{i_3}}$ is defined as the sum of the distances (pink dashed lines) of the three loci $x_{i_1},x_{i_2}, x_{i_3}$ to their centroid (pink point). Green and blue curves depict two different chromosomes. (b) Illustrates the definition of $D_{i_1 i_2 i_3}$ in the unphased setting. Green, blue and red curves depict neighborhoods around three homologous loci $(x_{i_1}, y_{i_1}), (x_{i_2}, y_{i_2})$ and $(x_{i_3}, y_{i_3})$. From these homologous loci 8 possible higher-order distances can be defined (colored dashed lines) based on the 8 centroids depicted in the figure. The higher-order distance  $D_{i_1 i_2 i_3}$ is defined as the minimum of these 8 distances (achieved here by the three black dashed line segments).}
    \label{fig:tensor_constraints}
\end{figure*}

%\textcolor{green}{
In the following, we provide our interpretation of distances in the higher-order setting. %describe how to derive constraints on the 3D location of the genomic loci from such measured higher-order distances. 
For simplicity, we first describe the higher-order distances for three loci in the phased setting. Since $F_{x_{i_1} x_{i_2} x_{i_3}}$ counts how often the three loci come together, we interpret  $D_{x_{i_1} x_{i_2} x_{i_3}}$ as the sum of the distances of the three loci $x_{i_1},x_{i_2}, x_{i_3}$ to their centroid (\Cref{fig:tensor_constraints_a}).
% paragraph: phased setting, we're looking at triples, we now need to define a distance among three points, contact frequencies are points coming thogether, we define D123 is the sum of distances to the centroid
We next provide a generalization to the unphased setting. For three homologous loci $(x_{i_1}, y_{i_1}), (x_{i_2}, y_{i_2})$ and $(x_{i_3}, y_{i_3})$, their contact frequency can be formed by 8 possible triples, namely $(x_{i_1}, x_{i_2}, x_{i_3})$, $(x_{i_1}, x_{i_2}, y_{i_3})$, $(x_{i_1}, y_{i_2}, x_{i_3})$, $(y_{i_1}, x_{i_2}, x_{i_3})$, $(x_{i_1}, y_{i_2}, y_{i_3})$, $(y_{i_1}, y_{i_2}, x_{i_3})$, $(y_{i_1}, x_{i_2}, y_{i_3})$, and $(y_{i_1}, y_{i_2}, y_{i_3})$. We will assume that one of the triples constitutes the majority of the observed contact frequency count and hence we define $D_{i_1 i_2 i_3}$ as the minimum over all 8 higher order distances. This is illustrated in \Cref{fig:tensor_constraints_b}.
Generalizing from three to $k$ loci, our higher-order distance definition then becomes
%}
$$
D_{i_1 i_2 \ldots i_k} = \min_{z_{i_j} \in \{x_{i_j},y_{i_j}\}}\left(\sum_{j=1}^k \|z_{i_j}-(z_{i_1}+\ldots+z_{i_k})/k\|^2\right).
$$
%\textcolor{blue}{
%Figure~\ref{fig:tensor_constraints} graphically illustrates the definition of $D_{i_1 i_2 \ldots i_k}$ and tensor constraints. 

In the following, we prove our main result; namely we show that the distance constraints of order 3 (3-way distances) together with the previously considered pairwise distance constraints and distance constraints among consecutive beads results in unique identifiability of the 3D genome configuration (\cref{theorem:tensor_uniqueness}). In fact, only very few order 3 distance constraints are required for unique identifiability. As we show in \cref{theorem:tensor_uniqueness} it is sufficient that the first and last bead of each chromosome be contained in an order 3 distance constraint. This is a reasonable constraint given that methods such as SPRITE, C-walks and GAM measure higher-order interactions over the whole genome. These insights are of interest experimentally since they suggest that the methods can restrict the measurement of such higher order constraints to first and last beads of each chromosome, known as telomeres.

%The assumption that the beginnings and ends of chromosomes are contained in at least one triple is a reasonable constraint on the data given that methods such as SPRITE, C-walks and GAM measure higher-order interactions over the whole genome and are not restricted to particular loci. Moreover, we assume that genomic regions in a tensor constraint coincide.

%we show that if the beginnings and ends of chromosomes are contained in at least one triple tensor constraint when considered in conjunction with pairwise distance constraints from Section~\ref{section:non_identifiability} and distances between neighboring beads from Section~\ref{sec_3} result in unique identifiability of the 3D genome configuration (Theorem~\ref{theorem:tensor_uniqueness}). 

\begin{theorem}\label{theorem:tensor_uniqueness}
Let $m$ be the number of chromosome pairs, let $n_1,n_2,\ldots,n_m$ be the number of domains on chromosomes $1,2,\ldots,m$ and define $n=n_1+n_2+\ldots+n_m$. Let $I \subseteq [n] \times [n] \times [n]$ be such that each of $1,n_1,n_1+1,n_1+n_2,\ldots,n_1+n_2+\ldots+n_{m-1}+1,n$ (labels of domains at the beginning and at the end of each chromosome) is contained in at least one triple in $I$. Let $x^*_1,\ldots,x^*_n,y^*_1,\ldots,y^*_n \in \mathbb{R}^3$ be fixed~such~that

\begin{equation*}
\begin{gathered}
\min_{z^*_i \in \{x^*_i,y^*_i\} \text{ for } i=k_1,k_2,k_3}\left(\sum_{j \in \{k_1,k_2,k_3\}}\|z^*_{j}-(z^*_{k_1}+z^*_{k_2}+z^*_{k_3})/3\|^2\right) = 0 \\
\text{ for } (k_1,k_2,k_3) \in I.
\end{gathered}
\end{equation*}
%}
 Consider the polynomial system:
\begin{equation}\label{equations:tensor_uniqueness_constraints}
\begin{gathered}
x_i + y_i = x^*_i + y^*_i \text{ and } \|x_i\|^2 + \|y_i\|^2 = \|x^*_i\|^2 + \|y^*_i\|^2 \text{ for } 1 \leq i \leq n,\\
\|x_i - x_{i+1}\| = \|x^*_i - x^*_{i+1}\| \text{ and } \|y_i - y_{i+1}\| = \|y^*_i - y^*_{i+1}\| \\
\text{ for } i \in [n]\backslash \{n_1,n_1+n_2,\ldots,n\},\\
\min_{z_i \in \{x_i,y_i\} \text{ for } i=k_1,k_2,k_3}\left(\sum_{j \in \{k_1,k_2,k_3\}}\|z_{j}-(z_{k_1}+z_{k_2}+z_{k_3})/3\|^2\right) = 0 \\
\text{ for } (k_1,k_2,k_3) \in I.
\end{gathered}
\end{equation}
%}
Then for generic $x^*_1,\ldots,x^*_n,y^*_1,\ldots,y^*_n$, this system has a unique solution in $(\mathbb{R}^3)^{2n}$.
\end{theorem}

To prove \cref{theorem:tensor_uniqueness}, we will need two lemmas. \cref{LemmaOnePointTwoSpheres} states that for a fixed solution $(x_1^*,y_1^*)$ on a sphere $S_1$ and given distances between solutions on $S_1$ and $S_2$, there are finitely many solutions $(x_2,y_2)$ on the sphere $S_2$. \cref{corollary:one_point_many_spheres} is an extension of \cref{LemmaOnePointTwoSpheres}. It states that if one has finitely many solutions on a sphere $S_i$, then given distances between neighboring beads, there are finitely many solutions on any sphere connected to $S_i$.  

\begin{lemma} \label{LemmaOnePointTwoSpheres}
\label{lemma:one_point_two_spheres}
Let $x^*_1,x^*_2,y^*_1,y^*_2 \in \mathbb{R}^3$ be fixed. Consider the polynomial system:
\begin{equation} \label{equation:one_point}
\begin{gathered}
x_2 + y_2 = x^*_2 + y^*_2, \|x_2\|^2 + \|y_2\|^2 = \|x^*_2\|^2 + \|y^*_2\|^2,\\
\|x^*_1-x_2\|=\|x^*_1-x^*_2\| \text{ and } \|y^*_1-y_2\|=\|y^*_1-y^*_2\|.
\end{gathered}
\end{equation}
For generic $x^*_1,x^*_2,y^*_1,y^*_2$, this system has finitely many solutions in $(\mathbb{R}^3)^{2n}$.
\end{lemma}

\begin{proof}
The first two equations of \cref{equation:one_point} say that $x_2,y_2$ and $x_2^*,y_2^*$ are pairs of antipodal points on the same sphere. We denote this sphere by $S_2$. The third equation says that $x_2$ is the same distance from $x_1^*$ as $x_2^*$ is from $x_1^*$. Hence $x_2$ must lie on the circle $C_{x_2}$ that is the intersection of $S_2$ and the sphere centered at $x_1^*$ and with radius $\|x^*_1-x^*_2\|$. The last equation says that $y_2$ must lie on the circle $C_{y_2}$ that is the intersection of $S_2$ and the sphere centered at $y_1^*$ with radius $\|y^*_1-y^*_2\|$. We consider the circle $C'_{x_2}$ that consists of antipodal points to the circle $C_{y_2}$ on the sphere $S_2$. The intersection of the circles $C_{x_2}$ and $C'_{x_2}$ gives the solutions for $x_2$. Unless the two circles are equal, they intersect at at most two points. Since $y_2$ is antipodal to $x_2$, then for each $x_2$ there is a unique $y_2$. The circles coincide if and only if $x_1^*,y_1^*$ and the center of $S_2$ are collinear. 
\end{proof}

\begin{lemma}
\label{corollary:one_point_many_spheres}
Let $x^*_1,\ldots,x^*_n,y^*_1,\ldots,y^*_n \in \mathbb{R}^3$ be fixed. Consider the polynomial system:
\begin{equation*} \label{equation:one_point_many_spheres}
\begin{gathered}
x_i + y_i = x^*_i + y^*_i \text{ and } \|x_i\|^2 + \|y_i\|^2 = \|x^*_i\|^2 + \|y^*_i\|^2 \text{ for } 2 \leq i \leq n,\\
\|x^*_1-x_2\|=\|x^*_1-x^*_2\|, \|y^*_1-y_2\|=\|y^*_1-y^*_2\|,\\
\|x_i-x_{i+1}\|=\|x^*_i-x^*_{i+1}\| \text{ and } \|y_i-y_{i+1}\|=\|y^*_i-y^*_{i+1}\| \text{ for } 2 \leq i \leq n-1.
\end{gathered}
\end{equation*}
For generic $x^*_1,\ldots,x^*_n,y^*_1,\ldots,y^*_n$, this system has finitely many solutions in $(\mathbb{R}^3)^{2n-2}$.
\end{lemma}

\begin{proof}
By \cref{lemma:one_point_two_spheres}, there are finitely many antipodal pairs $(x_{2},y_2)\in \mathbb{R}^3 \times \mathbb{R}^3$ on $S_{2}$ such that $\|x^*_1 - x_{2}\|=\|x^*_1 - x^*_{2}\|$ and $\|y^*_1 - y_{2}\|=\|y^*_1 - y^*_{2}\|$. Similarly, for each of these antipodal pairs $(x_{2},y_2)\in \mathbb{R}^3 \times \mathbb{R}^3$ on $S_{2}$, there are finitely many antipodal pairs $(x_{3},y_3)\in \mathbb{R}^3 \times \mathbb{R}^3$ on $S_{3}$ satisfying $\|x_{2} - x_{3}\|=\|x^*_{2} - x^*_{3}\|$ and $\|y_{2} - y_{3}\|=\|y^*_{2} - y^*_{3}\|$ etc. 
\end{proof}

\begin{proof}[Proof of \cref{theorem:tensor_uniqueness}]
We recall that the first line of the polynomial system \cref{equations:tensor_uniqueness_constraints} gives that $x_i,y_i$ are antipodal points on a sphere $S_i$. Consider a triple $(k_1,k_2,k_3) \in I$ that contains $1$ and the equation on the last line of the polynomial system \cref{equations:tensor_uniqueness_constraints} corresponding to this triple. This equation gives that $z_{k_1},z_{k_2},z_{k_3}$, where $z_i \in \{x_i,y_i\}$, coincide. Hence $z_{k_1},z_{k_2},z_{k_3}$ lie on the intersection of $S_{k_1},S_{k_2},S_{k_3}$. Generically, if the intersection of three spheres is non-empty in $\mathbb{R}^3$, then it consists of two points $P$ and $P'$. This gives four possible solutions for $x_1,y_1$: the points $P,P'$ and their antipodals on $S_1$. By \cref{corollary:one_point_many_spheres}, there are finitely many solutions for $x_2,\ldots,x_{n_1},y_2,\ldots,y_{n_1}$ given these fixed solutions $x_1,y_1$ on $S_1$. In the next two paragraphs we will show that generically these finitely many solutions do not contain antipodal points on any of the spheres $S_2,\ldots,S_{n_1}$. 

If there are two antipodal solutions on $S_i$, then we may assume that they come either from the same solution on $S_1$ or antipodal solutions on $S_1$, because we can perturb $S_{k_1},S_{k_2},S_{k_3}$ slightly to change the other pair of solutions. First we will show that generically a solution for $x_i$ on $S_i$ does not give a pair of antipodal solutions for $x_{i+1}$ on $S_{i+1}$. If this was the case, then both the solution for $x_i$ and its antipodal would have to lie on the plane that is perpendicular to the line through the antipodal pair of solutions for $x_{i+1}$ on $S_{i+1}$. This plane contains the centers of $S_i$ and $S_{i+1}$. Hence for a solution for $x_i$, there is only one antipodal pair on solutions on $S_{i+1}$. Thus for a generic distance between the solutions on $S_{i}$ and $S_{i+1}$, a solution on $S_i$ does not give an antipodal pair of solutions on $S_{n+1}$.

Secondly, suppose that two different solutions on $S_{i}$ give a pair of antipodal solutions on $S_{i+1}$. We will show that when we perturb the distance between solutions on $S_{i}$ and $S_{i+1}$, then we do not get an antipodal pair anymore. Let $x_i$ and $x'_i$ be two different solutions on $S_i$ that give solutions $x_{i+1}$ and $2c_{i+1}-x_{i+1}$ on $S_{i+1}$. Hence $\|2c_{i+1}-x_{i+1}-x'_i\|^2=\|x_{i+1}-x_i \|^2$. We want to show that generically  
$$
\|2c_{i+1}-(x_{i+1}+\epsilon)-x'_i\|^2 \neq \|x_{i+1} +\epsilon -x_i \|^2  ,
$$
where $x_{i+1} + \epsilon$ is the perturbed solution. 
Indeed, using the identity $\|x_{i+1}-x_i \|^2=\|2c_{i+1}-x_{i+1}-x'_i\|^2$ gives
$$
\|2c_{i+1}-(x_{i+1}+\epsilon)-x'_i\|^2 - \|x_{i+1} +\epsilon -x_i \|^2 = 2\epsilon (x_i + x'_i -2c_{i+1}).
$$
This quantity is equal to zero if and only if $\epsilon=0$ or $c_{i+1}$ is the middle point of the line segment from $x_i$ to $x'_i$. This is generically not the case. 

Using a triple $(k'_1,k'_2,k'_3) \in I$ containing $n_1$ and the equation for this triple, we get four possible solutions for $x_{n_1},y_{n_1}$. Generically, only one of them coincides with the finitely many solutions on $S_{n_1}$ that we get from the solutions on $S_1$, because perturbing the spheres slightly (with keeping the coinciding points fixed) perturbs the second intersection point of the three spheres and we know that generically the finitely many points do not contain antipodal points. 

The unique solution on $S_n$ comes from one solution on each of the spheres $S_1,\ldots,S_{n_1-1}$: If this was not the case then two different solutions on $S_i$ give the same solution on $S_{i+1}$. By the proof of \cref{prop:uniqueness_2d_distances_between_neighboring_loci}, the dot product $(c_i-c_{i+1}) \cdot (x_i-x_{i+1})$ is fixed. Hence for a fixed $x_{i+1}$, all possible solutions for $x_i$ lie on a hyperplane and this hyperplane is perpendicular to $c_i-c_{i+1}$. Therefore, if two solutions on $S_i$ give the same solution on $S_{i+1}$, then they lie on a hyperplane perpendicular to $c_i-c_{i+1}$. By slightly perturbing the sphere $S_{i+1}$, this is not the case anymore, and hence generically a solution on $S_{i+1}$ comes from a unique solution on $S_i$.
\end{proof}

\section{Algorithms and implementation}
\label{sec_algorithms}

%\subsection{Implementation: SDP formulation} \label{sec:implementation}

So far, we derived a theoretical framework to establish when we have unique and finite identifiability of the 3D configuration in the noiseless setting. However, a unique solution does not necessarily mean that we can find it efficiently, as in many cases finding the solution may be NP-hard. In addition, we have so far not yet considered the noisy setting. 
%In this section we construct optimization problems which can be solved efficiently to yield the solution. 
In this section, we show how to construct an optimization formulation to determine the 3D configuration efficiently.

We frame the 3D reconstruction problem as a Euclidean embedding problem, where the coordinates~$x_1, \dots x_n, y_1, \dots y_n\in\mathbb{R}^3$ are inferred from distances. Similar to ChromSDE~\cite{chromsde}, we formulate all distances in terms of entries in the Gram matrix $G$, which tracks the dot products between the $2n$ genomic regions. Namely, letting the column/row $i$ of $G$ correspond to $x_i$  and the column/row $n + i$ correspond to its homologous locus $y_i$, then
the distances are given by $\|x_i - x_j\|^2 = G_{i,i} + G_{j,j}-2G_{i,j}$, $\|x_i - y_j\|^2 = G_{i,i} + G_{n+j,n+j}-2G_{i,n+j}$ and $\|y_i - y_j\|^2 = G_{n+i,n+i} + G_{n+j,n+j}-2G_{n+i,n+j}$. 
It is natural to work with the Gram matrix $G$, 
since it is rotation invariant. By imposing the constraint $\sum_{i,j} G_{i,j} = 0$ we can also fix the translational axis. Also the additional distance constraints that we introduced in the previous sections (\cref{{theorem:non_identifiability}}, \cref{prop:uniqueness_2d_distances_between_neighboring_loci}, \cref{theorem:tensor_uniqueness})  can be represented as linear constraints in terms of entries in $G$ as follows:

%The distance constraints in Theorem~\ref{theorem:tensor_uniqueness} and distances between homologous pairs which can be computed from the pairwise distance constraints using Lemma~\ref{lemma:distances_between_homologue_pairs} can be represented as linear constraints in $G$:

\begin{itemize}
    
    \item Pairwise distances:
    $$
    g_{ij}(G) := G_{i,i} + G_{j,j} + G_{n+i, n+i} + G_{n+j, n+j} -  G_{i,j} - G_{n+i, j} - G_{i, n+j} - G_{n+i, n+j}
    $$
    \item Distances between homologous pairs: 
    $$
    g_{ii}(G) := G_{i,i} + G_{n+i, n+i} - 2G_{i,n+i}
    $$
    \item Distances between neighboring beads:
    $$
    g_{i+}(G) := G_{i,i} + G_{i+1, i+1} - 2G_{i,i+1}
    $$
    \item Distances of order 3 (can be generalized to higher orders):
    $$
    g_{ijk}(G) := \min_l(g_{ijkl}:l=1,\ldots,8)
    $$
    where
    \begin{align*}
    g_{ijk1}(G) := &G_{i,i} + G_{j,j} + G_{k,k} - G_{i,j} - G_{i,k} - G_{j,k},\\
    g_{ijk2}(G):=&G_{i,i} + G_{j,j} + G_{n+k,n+k} - G_{i,j} - G_{i,n+k} - G_{j,n+k},\\
    g_{ijk3}(G) :=&G_{i,i} + G_{n+j,n+j} + G_{k,k} - G_{i,n+j} - G_{i,k} - G_{n+j,k},\\
    g_{ijk4}(G) :=&G_{i,i} + G_{n+j,n+j} + G_{n+k,n+k} - G_{i,n+j} - G_{i,n+k} - G_{n+j,n+k},\\
    g_{ijk5}(G) :=&G_{n+i,n+i} + G_{j,j} + G_{k,k} - G_{n+i,j} - G_{n+i,k} - G_{j,k},\\
    g_{ijk6}(G) :=&G_{n+i,n+i} + G_{j,j} + G_{n+k,n+k} - G_{n+i,j} - G_{n+i,n+k} - G_{j,n+k},\\
    g_{ijk7}(G) :=&G_{n+i,n+i} + G_{n+j,n+j} + G_{k,k} - G_{n+i,n+j} - G_{n+i,k} - G_{n+j,k},\\
    % g_{ijk8}(G) :=&G_{n+i,n+i} + G_{n+j,n+j} + G_{n+k,n+k} - G_{n+i,n+j} - G_{n+i,n+k} - G_{n+j,n+k}
    g_{ijk8}(G) :=&G_{n+i,n+i} + G_{n+j,n+j} + G_{n+k,n+k}\\
    &- G_{n+i,n+j} - G_{n+i,n+k} - G_{n+j,n+k}.
    \end{align*}
    
\end{itemize}

Our objective is to determine a rank 3 solution of $G$, satisfying the above constraints. However, this optimization problem is non-convex due to the rank constraint, and we instead consider the standard relaxation: we minimize the trace of the Gram matrix as an approximation to matrix rank~\cite{fazel2001rank}. The resulting optimization problem then becomes the following semidefinite program (SDP):

\begin{equation} \label{SDP_noiseless}
\begin{aligned}
& \underset{G}{\text{minimize}}
& & \tr(G) \\
& \text{subject to}
& & g_{ii}(G) = D^*_{ii}, \; 1 \leq i \leq n,\\
&&& g_{ij}(G) = D^*_{ij}, \; 1 \le i < j \le n,\\
&&& g_{i+}(G) = D^{*}_{i+}, \; i \in \Omega_1, \\
&&& g_{ijk}(G) = D^*_{ijk}, \; (i,j,k) \in \Omega_2, \\
&&& \sum_{1 \leq i,j \leq 2n} G_{i,j} = 0, \\
&&& G \succeq 0.
\end{aligned}
\end{equation}
Here, $D^*_{ii}$ denote the distances between homologous pairs computed from the pairwise distances using \cref{lemma:distances_between_homologue_pairs}, $D^*_{ij}$ denote the pairwise distances, $D^{*}_{i+}$ denote the distances between neighboring beads, and $D^*_{ijk}$ denote the distances between three loci (while one could also consider 4 or higher order distance constraints, in our implementation we only used 3-way distance constraints since higher-order contacts are extremely sparse). The index set $\Omega_1=[2n]\backslash \{n_1,n_1+n_2,\ldots,n,n+n_1,n+n_1+n_2,\ldots,2n\}$ corresponds to all beads that are not the last bead on a chromosome. The index set $\Omega_2 \subseteq [n]^3$ corresponds to all triples of beads with non-zero contact frequencies.
%for which we have observed tensor constraints. 
%The objective function $\tr(G)$ is concave.

In the noisy setting, which is relevant for 
%To deal with the fact that 
biological data, we replace the equality constraints by penalties in the loss function. Namely, using  $D^*$ for the noiseless and $D$ for the noisy distances, we replace the equality constraints of the form $g(G) = D^*$ by adding $(g(G) - D)^2$ to the objective function. 
%True biological data is noisy and strict constraints such as the ones we used before become ineffective. To combat this, we modify the SDP formulation by allowing strict constraints to be slightly off at some penalty. We will use $D^*$ for noiseless and $D$ for noisy distances. Equality constraints of the form $g(G) = D^*$ are translated by adding $(g(G) - D)^2$ to the objective function. 
For the higher-order distance constraints of the form $D^*_{ijk} = \min(g_{ijk1}(G),  \ldots, g_{ijk8}(G))$ for $(i,j,k) \in \Omega_2$ we use slack variables and a convex relaxation using an atomic norm that combines the $\ell_2$- and $\ell_1$-norms. More precisely, we propose the use of the following transformation in the noisy setting,
\[
D_{ijk} + \lambda_{ijkl} = g_{ijkl}(G) + s_{ijkl} \text{ for } l = 1, 2, \ldots, 8,
\]
where $\lambda_{ijkl}, s_{ijkl} \ge 0$ for all $i,j,k,l$ act as slack variables. In general, for each triple $(i,j,k)$ we want one of the $\lambda_{ijkl}$ to be close to $0$ and the sum over all $s_{ijkl}$ to be small. Naively this can be done by placing $\sum s_{ijkl} +\sum \lambda_{ijkl}$ into the objective function. 
However, this would not enforce 
%But the issue with this idea is that we are not enforcing 
for each $(i,j,k)$ at least one $\lambda_{ijkl}$ to be close to $0$. Instead we propose to use
\[
\sum_{(i,j,k) \in \Omega_2, 1 \leq l \leq 8} s_{ijkl} + \sqrt{\sum_{(i,j,k) \in \Omega_2} \left ( \sum_{1 \leq l \leq 8} \lambda_{ijkl} \right )^2}.
\]
The $\ell_2$-norm will push down the $\sum_l \lambda_{ijkl}$ for each $(i,j,k)$, while the $\ell_1$ norm will drive at least one of the $\lambda_{ijkl}$ to zero, which is precisely the desired behavior.
The quantity $\sqrt{\sum_{i,j,k} \left ( \sum_l \lambda_{ijkl} \right )^2}$ is an atomic norm as defined in~\cite{chandrasekaran2012convex} with the set of atoms
%This way, because we have an $\ell^2$-norm in general the solution will try to minimize $\sum_l \lambda_{ijkl}$ for each $(i,j,k)$. This is an $\ell^1$ norm, so the solution will try to drive at least one of the $\lambda_{ijkl}$ zero. This is precisely the desired behavior. The quantity $\sqrt{\sum_{i,j,k} \left ( \sum_l \lambda_{ijkl} \right )^2}$ is an atomic norm as defined in~\cite{chandrasekaran2012convex} with the set of atoms
% $$
% \mathcal{A} = \{(\lambda_{ijkl}):\sum_{i,j,k} \left ( \sum_l \lambda_{ijkl} \right )^2=1 \text{ and } \\
% \sum_{i,j,k} \lambda_{ijkl_{ijk}}^2 =1 \text{ for } l_{ijk}=1,\ldots,8,(i,j,k) \in \Omega_2\}.
% $$
\begin{equation*}
\begin{gathered}
\mathcal{A} = \{(\lambda_{ijkl}):\sum_{i,j,k} \left ( \sum_l \lambda_{ijkl} \right )^2=1 \text{ and } \\
\qquad\qquad\qquad\qquad\qquad\sum_{i,j,k} \lambda_{ijkl_{ijk}}^2 =1 \text{ for } l_{ijk}=1,\ldots,8,(i,j,k) \in \Omega_2\}.
\end{gathered}
\end{equation*}

Then the optimization problem in the noisy setting becomes:
%After these transformations, for tensor constraints our optimization problem will resemble:
\begin{equation} \label{SDP_noisy}
\begin{aligned}
& \underset{G,s,\lambda}{\text{minimize}}
& & \rho \tr(G) + \sum_{1 \leq i \leq n} ( g_{ii}(G) - D_{ii})^2 + \sum_{1 \leq i < j \leq n} (g_{ij}(G) - D_{ij})^2  
  \\
 &&&+ \sum_{i \in \Omega_1} (g_{i+}(G) - D_{i+})^2 + \sum_{(i,j,k) \in \Omega_2, 1 \leq l \leq 8} s_{ijkl} \\
 &&&+  \sqrt{\sum_{(i,j,k) \in \Omega_2} \left ( \sum_{1 \leq l \leq 8} \lambda_{ijkl} \right )^2}\\
& \text{subject to}
& & D_{ijk} + \lambda_{ijkl} = g_{ijkl}(G) + s_{ijkl}, \; (i,j,k) \in \Omega_2, 1 \leq l \leq 8,\\
&&&  s_{ijkl} \geq 0, \; (i,j,k) \in \Omega_2, 1 \leq l \leq 8, \\
&&& \lambda_{ijkl} \geq 0, \; (i,j,k) \in \Omega_2, 1 \leq l \leq 8, \\
&&&  \sum_{1 \leq i,j \leq 2n} G_{i,j} = 0, \\
&&&  G \succeq 0.
\end{aligned}
\end{equation}

We use a tuning parameter $\rho$ for the trace in the objective function, which can be used to balance obtaining a low-rank solution versus satisfying the constraints.
%We attach a tuning constant $\rho$ to the trace minimization part of the objective in order to balance alignment with the constraints and approximately minimizing rank. 
The tuning parameter $\rho$ can be chosen using cross-validation or by selecting it so that the resulting solution has small $(d+1)^{th}$ eigenvalue.
%$\rho$, which results in a solution with
%small $(d+1)^{th}$ eigenvalue. 
%As shown in \cref{appendix:simulations_rho} and~\cref{appendix:real_data_rho}, 
As shown in \textcolor{green!50!black}{section SM3} and~\textcolor{green!50!black}{section SM7},
we observe on synthetic and real data that the solution is robust to the choice of $\rho$.

 The theoretical results from \cref{lemma:distances_between_homologue_pairs} allow us to compute the distances between homologous pairs from the pairwise distances $D_{ij}$. 
 We recall that we need to compute $\|v\|^2$ such that
\begin{equation*} 
\det(T' - 8J\|v\|^2) = 0,
\end{equation*}
where $T'$ is an invertible matrix constructed from the pairwise distance matrix by selecting a set of $2d + 2$ indices. 
One step of computing $\|v\|$ involves inverting $T'$.  Even if the error in the measurements is small, noise can propagate and severely impact  this computation. 
 In order to obtain a robust estimate of homolog-homolog distances, for each locus $i$, we sample 100 $T'$ matrices and obtain 100 solutions to the equation for $\| v\|^2$. We then take the median of the solutions to be the homolog-homolog distance for locus $i$ and use these homolog-homolog distances for the evaluation of our algorithms on synthetic and real data in the following section.
% . We identify the homolog-homolog distances from the pairwise distance matrix for simulations and in real data.

To solve the two convex optimization problems presented in this section for the noiseless and noisy setting, we make use of the solver MOSEK implemented in CVX within MATLAB. This results in the Gram matrix. In order to reconstruct the coordinates of the genomic regions from the Gram matrix, we use an eigenvector decomposition as also done in \cite{chromsde}, namely: letting $\gamma_1,\ldots,\gamma_d$ be the top $d$ eigenvalues and $\nu_1,\ldots,\nu_d$ the corresponding eigenvectors of $G$, then
%, we mimic the approach in \cite{chromsde} to reconstruct the coordinates of the genomic regions as follows. Let $\gamma_1,\ldots,\gamma_d$ be the top $d$ eigenvalues and $\nu_1,\ldots,\nu_d$ be the top $d$ eigenvectors of $G$.
$$
x_i = (\sqrt{\gamma_1} \cdot \nu_{1,i}, \ldots,\sqrt{\gamma_d} \cdot \nu_{d,i}) \text{ and } y_i = (\sqrt{\gamma_1} \cdot \nu_{1,n+i}, \ldots,\sqrt{\gamma_d} \cdot \nu_{d,n+i}) \text{ for } i=1,\ldots,n.
$$
Since we are interested in recovering the genome configuration in 3D, we use $d=3$, thereby obtaining the desired 3D diploid configuration. We provide the code for our algorithm at \url{https://github.com/uhlerlab/diploid-3D-reconstruction}.

\section{Evaluation on synthetic and real data}\label{sec_data}

\subsection{Synthetic data}
We start by testing our method on simulated data. For this we construct three different types of 3D structures: (a) a Brownian motion model using a standard normal distribution to generate successive points; (b) points sampled uniformly along a spiral with random translations sampled uniformly within $(0,0.5)$ range and orientations sampled uniformly within $(-\frac{\pi}{4}, \frac{\pi}{4}$); (c) points sampled uniformly in a unit sphere. 

\begin{figure*}[htbp]
    \centering
  \subfloat[]{\includegraphics[width=0.3\textwidth]{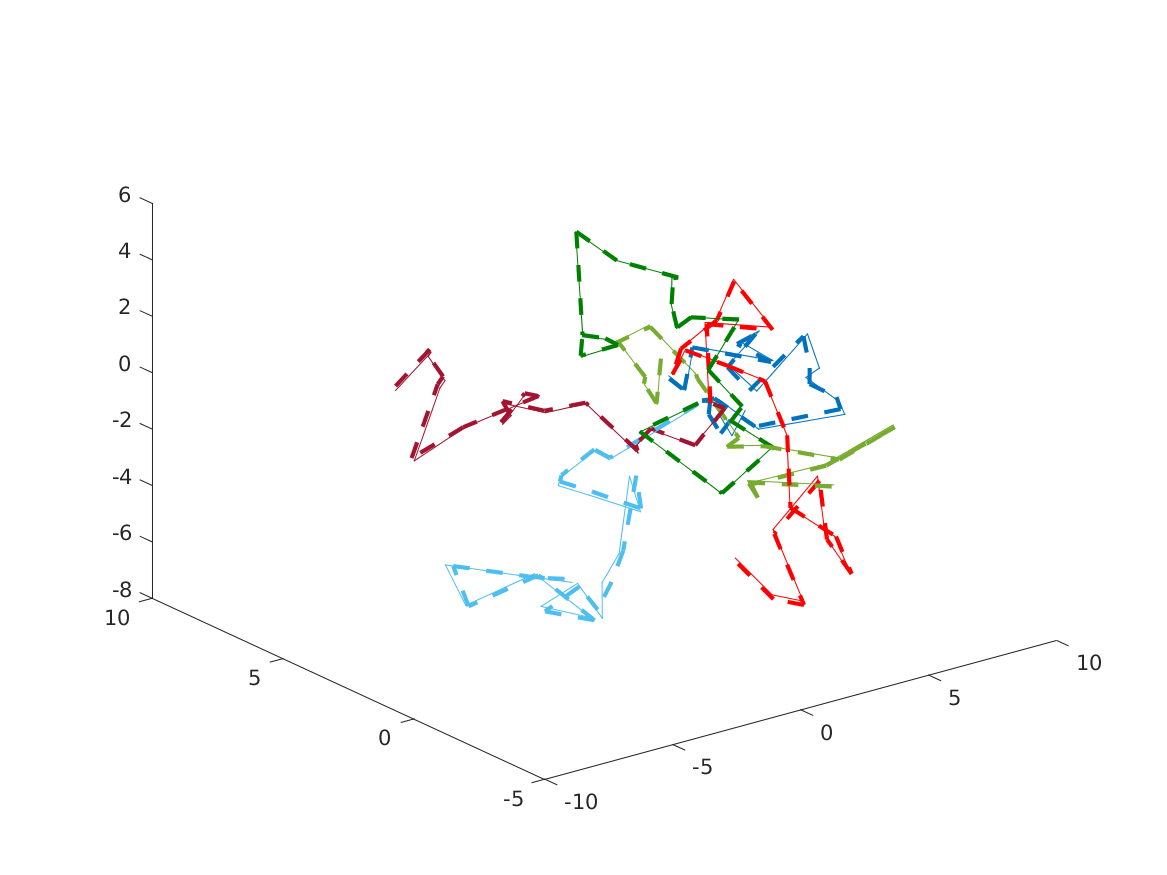}}
  ~
   \subfloat[]{\includegraphics[width=0.3\textwidth]{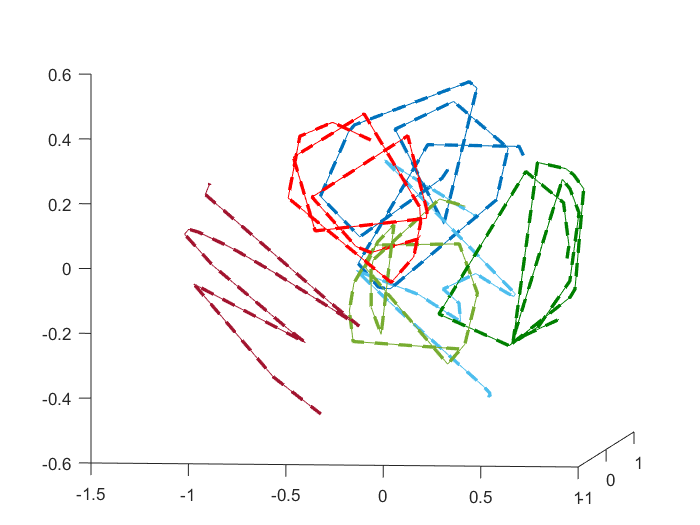}}
   ~
  \subfloat[]{\includegraphics[width=0.3\textwidth]{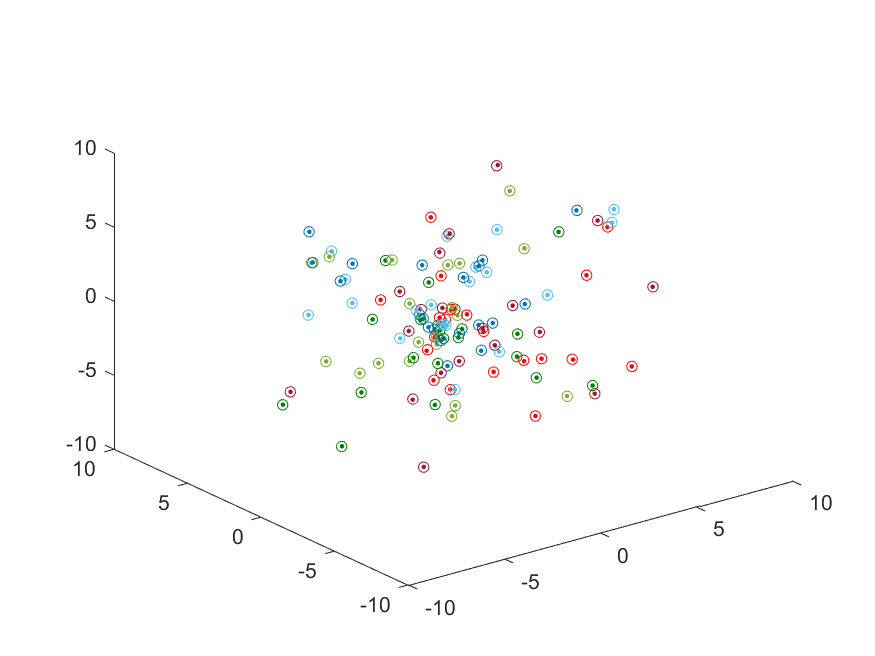}}
    \caption{
    Examples of true and reconstructed points on simulated data.
    (a) Brownian motion model. (b) Spirals. (c) Random points in a sphere. We generate six chromosomes with  in total of 120 domains, corresponding to three homologous pairs with 20 domains per chromosome  in the noiseless setting.
    Solid lines / points correspond to true 3D coordinates and dashed lines / unfilled points to reconstructions via our method. Each color represents a different chromosome. 
    }
    \label{fig:reconstruction_simulation}
\end{figure*}

\begin{figure*}[htbp]
    \centering
  \subfloat[]{\includegraphics[width=0.3\textwidth]{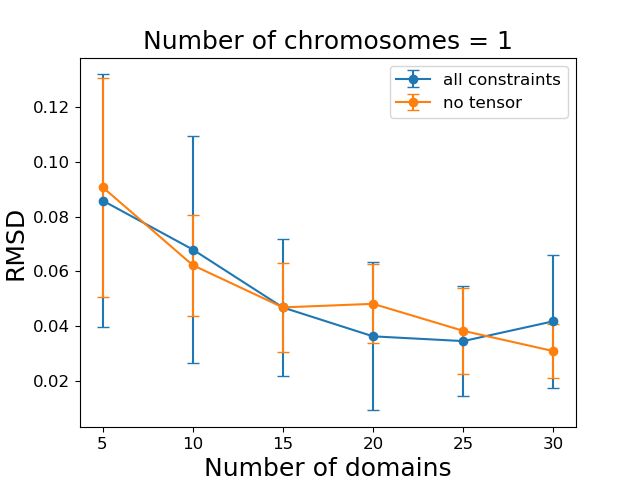}}
  ~
  \subfloat[]{\includegraphics[width=0.3\textwidth]{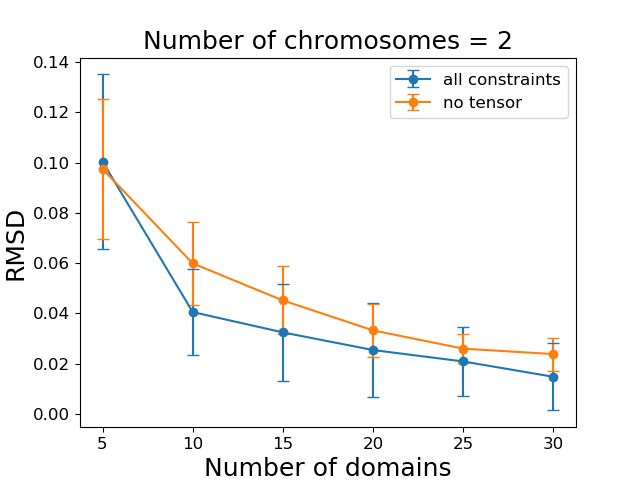}}
  ~
  \subfloat[]{\includegraphics[width=0.3\textwidth]{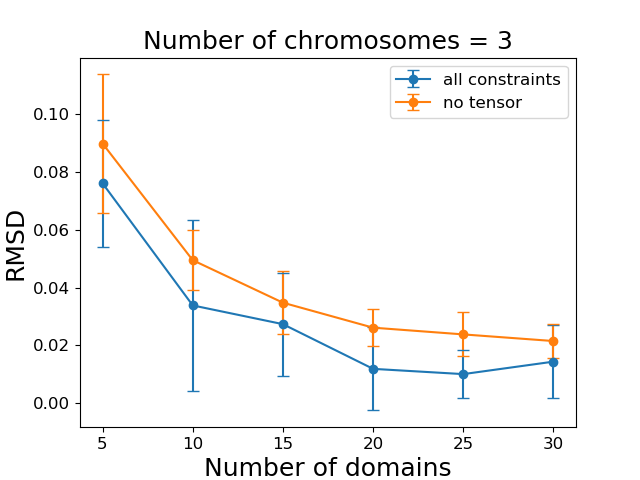}}
    \caption{Performance of our method in the noiseless setting. Root-mean-square deviation (RMSD) between true and reconstructed structure computed with and without higher-order distance constraints. Simulated data was generated using a Brownian motion model with (a) one (b) two and (c) three chromosomes. Mean and standard deviation over $20$ trials are shown.}
    \label{fig:rmsd_comparison_notensor}
\end{figure*}

\vspace{0.2cm}
\textbf{Performance of our method in the noiseless setting.}  For the 1D setting we deduced in \cref{section:non_identifiability} that the pairwise distance constraints by themselves are sufficient to identify the underlying 3D configuration. For the 2D setting we proved in \cref{sec_3} that knowing additionally the distances between neighboring beads leads to uniqueness. We here perform simulations in 3D since this is the biologically relevant setting.
These results are depicted in \Cref{fig:reconstruction_simulation} with additional examples in \textcolor{green!50!black}{Figure SM1}.
%\Cref{fig:noiseless_recon_addn}.
%, \cref{appendix:simulations_recon}.
The input to our algorithm are the pairwise distances (which are summed over homologs), all 3-way distances, the distances between homologous loci, and the distances between neighboring beads. In the noiseless setting considered here we solve the SDP formulation in \Cref{SDP_noiseless}. \Cref{fig:reconstruction_simulation} and \textcolor{green!50!black}{Figure SM1}
%in \cref{appendix:simulations_recon} 
show that the true and reconstructed structures highly overlap, thereby indicating that our optimization formulation is able to recover the 3D structure of the full diploid genome in the noiseless setting.  
When the 3-way distance constraints are removed, the reconstructions are less aligned with the true structures. This is shown in \Cref{fig:rmsd_comparison_notensor}, where we measure the root-mean-square deviation (RMSD) between true and reconstructed 3D coordinates over $20$ trials. In line with our theoretical results, these experimental results in the noiseless setting indicate the importance of higher-order contact frequencies for recovering the 3D diploid configuration, especially when the number of chromosomes is larger.

\vspace{0.2cm}
\textbf{Performance of our method in the noisy setting.} 
Next, we consider noisy distance observations
$D_{ij} = D_{ij}^*(1 + \delta)$ and noisy 3-way distance observations $D_{i_1 i_2 \ldots i_k} = D_{i_1 i_2 \ldots i_k}^*(1 + \delta)$ by sampling $\delta$ uniformly within $(-\epsilon, \epsilon)$ as in~\cite{chromsde}, where $\epsilon$ is a given noise level. %; similar simulations to.
For our simulations we sample a maximum of $1000$ 3-way distance constraints.
%use $1000$ 3-way distance constraints, however, if the number of all possible triplet constraints is smaller than $1000$, then all 3-way distance constraints are used. 
As shown in \textcolor{green!50!black}{Figure SM2},
%\Cref{fig:impact_numc},
%in \cref{appendix:simulations_numc}, 
we observe that the number of constraints does not have a major effect on the reconstruction accuracy. While for all simulations shown in this section, we set the tuning parameter $\rho=0.000001$, \textcolor{green!50!black}{Figure SM3}
%\Cref{fig:impact_rho} 
%in \cref{appendix:simulations_rho} 
shows that the performance is not significantly different when using different choices of $\rho$.

\begin{figure*}[htbp]
    \centering
  \subfloat[]{\includegraphics[width=0.3\textwidth]{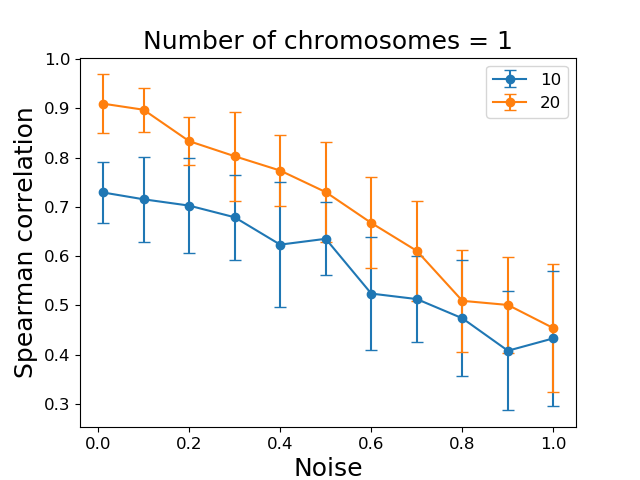}}
  ~
  \subfloat[]{\includegraphics[width=0.3\textwidth]{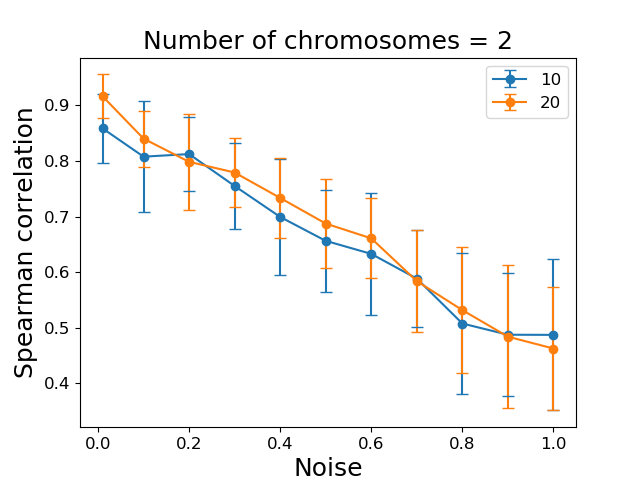}}
  ~
  \subfloat[]{\includegraphics[width=0.3\textwidth]{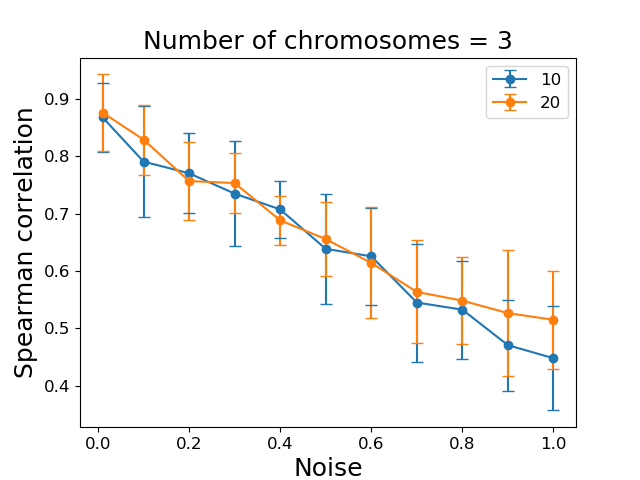}}
    \caption{Performance of our method in the noisy setting. Spearman correlation under different noise levels for (a) one, (b) two and (c) three chromosomes. Simulated data was generated using a Brownian motion model where each chromosome has $10$ or $20$ domains.  Mean and standard deviation over $20$ trials are shown.}
    \label{fig:spearman_corr_noisy_data}
\end{figure*}

 In \Cref{fig:spearman_corr_noisy_data} we numerically assess the accuracy of our predicted structure for the Brownian motion model for different number of chromosomes (one, two,  or three) and different number of domains per chromosome ($10$ or $20$) by computing the Spearman correlation between reconstructed and true pairwise distances, similar to~\cite{chromsde}. As expected, \Cref{fig:spearman_corr_noisy_data} shows that when the noise level increases, then the Spearman correlation between the original and reconstructed configuration decreases. For the simulations with one chromosome, the Spearman correlation is higher for $20$ domains than $10$.

\subsection{Application to 3D diploid genome reconstruction}

 We apply our algorithm to the problem of reconstructing the diploid genome from contact frequency data derived from experiments. We obtain pairwise and 3-way contact frequencies collected via SPRITE in human lymphoblastoid cells from~\cite{Quinodoz219683}. Since we aim to reconstruct the whole diploid genome, which consists of approximately 6 billion base pairs, for computational reasons we bin the contact frequencies in the SPRITE dataset into 10 Mega-base pair (Mb) regions. While some previous studies considered higher resolutions, the majority of the studies~\cite{cauer2019diploid, hu2013bayesian, rousseau2011three, varoquaux2014statistical, chromsde} did not attempt to reconstruct the whole diploid genome and focused only on reconstructing one chromosome, thus enabling them to consider higher resolutions.

 After filtering out regions with a small number of total contacts, we obtain 514 unphased points on the chromosomes. We convert the pairwise contact frequencies to pairwise distances using the previously observed relationship $D_{ij}=F_{ij}^{-1/2}$~\cite{rousseau2011three} and use \cref{lemma:distances_between_homologue_pairs} to obtain the distances between homolog pairs from this data. As in our simulations in the noisy setting, we randomly sample 1000 3-way distance constraints from all nonzero 3-way contact frequencies (for the transformation from 3-way contact frequencies to 3-way distances, see \cref{sec_5}). Finally, we obtain the distances between neighboring 10Mb beads by empirically evaluating the 3D reconstructions under different input distances; see \textcolor{green!50!black}{section SM4} and \textcolor{green!50!black}{Figure SM4}, \textcolor{green!50!black}{Figure SM5}.

\begin{figure*}[!t]
    \centering
  \subfloat[]{\label{fig:real_data_a}\includegraphics[width=0.33\textwidth]{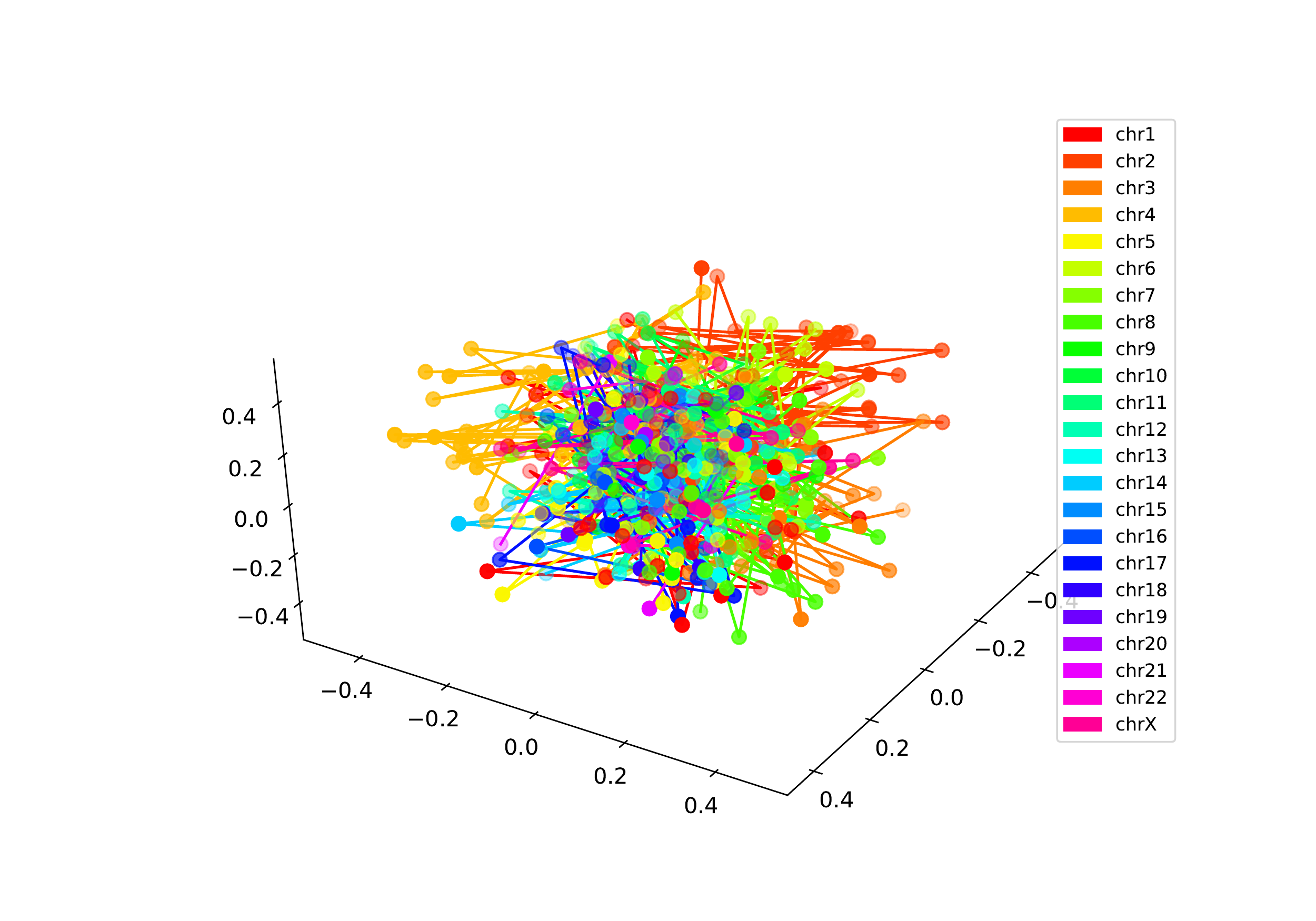}}
  \subfloat[]{\label{fig:real_data_b}\includegraphics[width=0.33\textwidth]{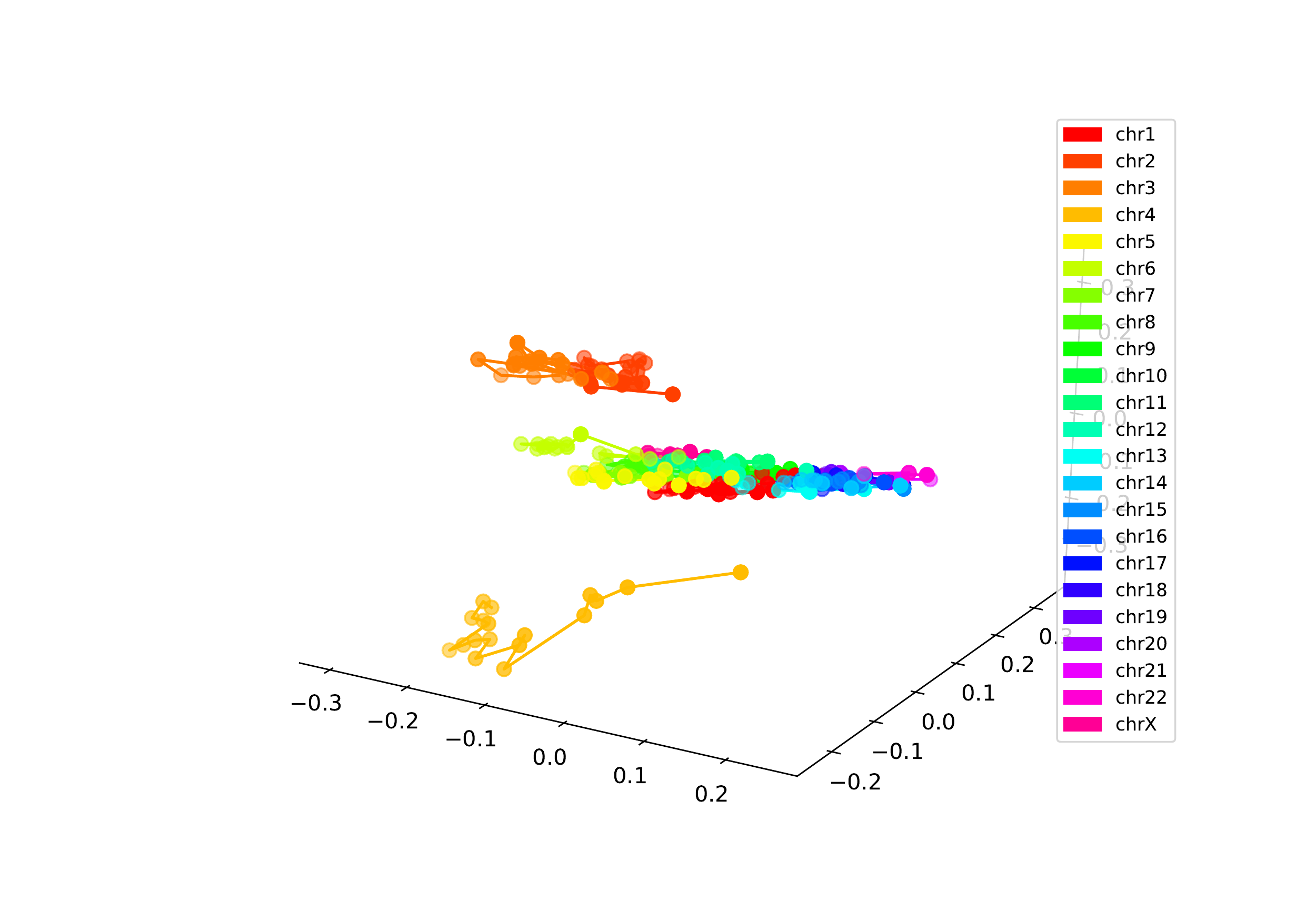}}
  \subfloat[]{\label{fig:real_data_c}\includegraphics[width=0.33\textwidth]{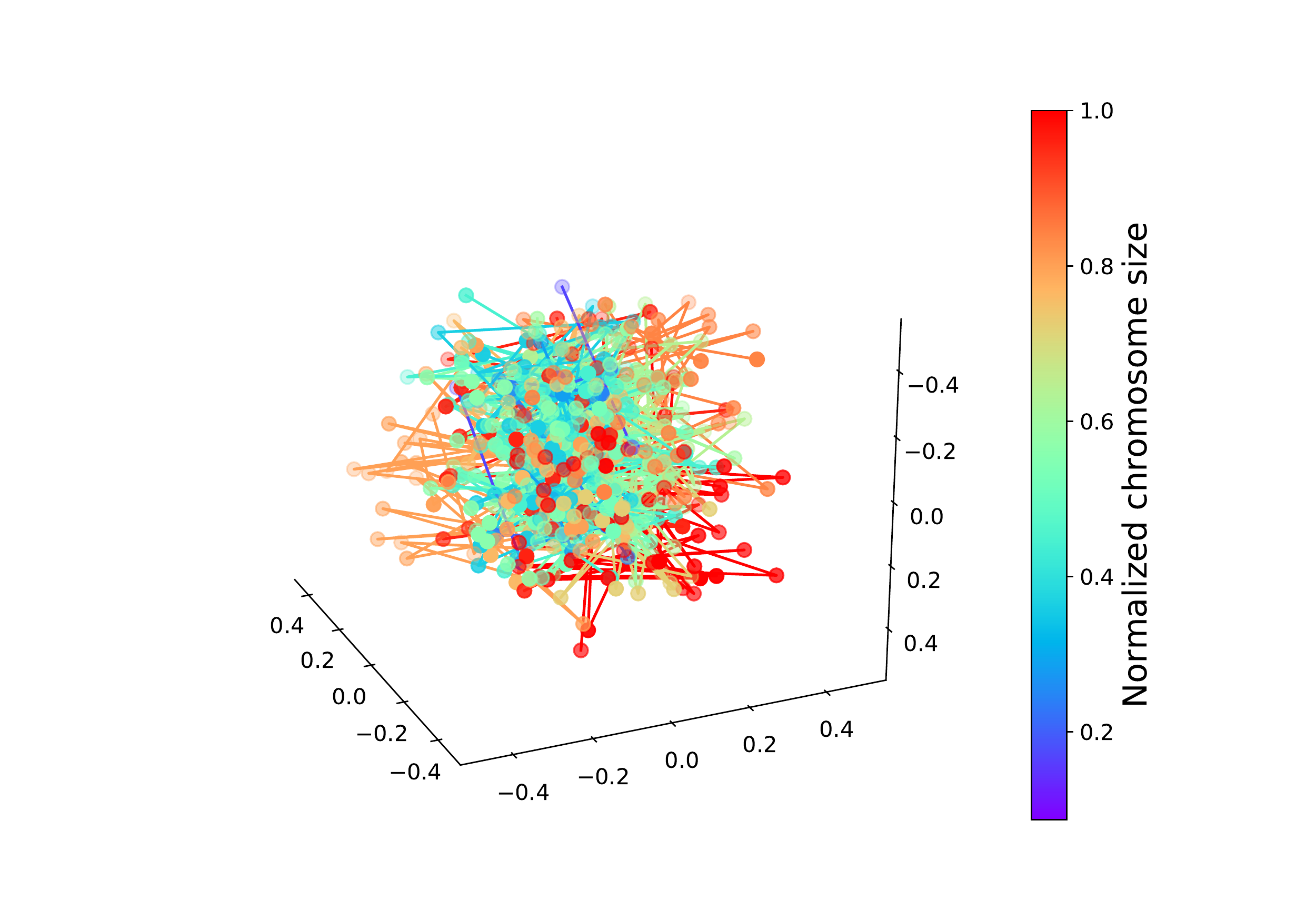}}
  \newline
  \subfloat[]{\label{fig:real_data_d}\includegraphics[width=0.38\textwidth]{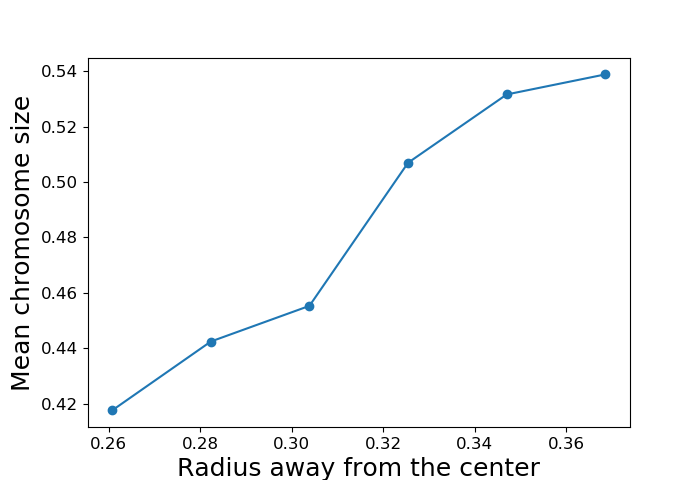}}
  ~
  \subfloat[]{\label{fig:real_data_e}\includegraphics[width=0.35\textwidth]{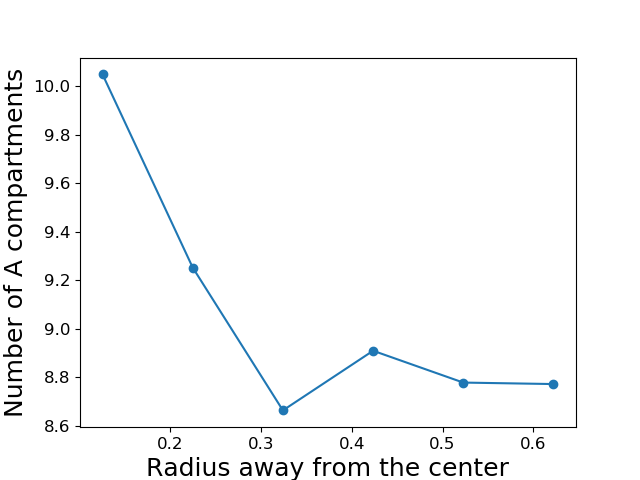}}
    \caption{3D Diploid Genome Reconstruction. Estimated 3D positions of all chromosomes and their corresponding homologs at 10Mb resolution. 3D positions obtained using (a) our method and (b) using ChromSDE with chromosomes colored according to chromosome number. (c) Whole diploid organization obtained via our method, colored by chromosome size. (d) Mean chromosome size as the distance from the center increases. (e) The number of A compartments as the distance from the center increases.}
    \label{fig:real_data}
\end{figure*}

Using the pairwise constraints, homolog-homolog constraints, neighboring bead constraints, and 3-way distance constraints, we solve the SDP problem in \Cref{SDP_noisy} for the noisy setting and analyze the corresponding 3D coordinates.
Our diploid reconstruction is shown in \Cref{fig:real_data_a}. We compare this diploid genome reconstruction to the 3D structure obtained via ChromSDE, shown in \Cref{fig:real_data_b} obtained under the assumption that the observed contact frequencies and the corresponding distances are a sum of four equal quantities, i.e.,
$\|x_i - x_j\|^2, \|x_i - y_j\|^2, \|y_i - x_j\|^2$, and $\|y_i - y_j\|^2$ are equal. In
%\cref{appendix:real_data_chromsde}, 
\textcolor{green!50!black}{Figure SM6},
%\Cref{fig:chromsde}, 
we show that the reconstruction obtained using ChromSDE with equal distances does not recapitulate known biology as described in the following paragraphs.

Experimental (imaging) studies have shown that chromosomes are organized by size within the nucleus, with small chromosomes in the interior and larger chromosomes on the periphery~\cite{bolzer2005three}. We colored each chromosome according to its size and computed the mean chromosome size versus distance away from the center. The results of the 3D configuration obtained using our method are shown in \Cref{fig:real_data_c,fig:real_data_d} and recapitulate prior studies: smaller chromosomes are preferentially located in the center, whereas larger chromosomes are preferentially on the periphery; 
see also \textcolor{green!50!black}{section SM6}, \textcolor{green!50!black}{Figure SM7}.
%see also \cref{appendix:real_data_analysis}, \Cref{fig:chromsize_realdata}. 
This is especially apparent for chromosomes 2 and 4, which are some of the largest chromosomes, and in our reconstruction they are located on the periphery as expected.

Experimental studies on the spatial organization of the genome have also shown that the center of the nucleus is enriched in active compartments (known as A compartments), while the periphery contains inactive compartments (known as B compartments)~\cite{stevens20173d}. From previously published data on the location of A and B compartments along the genome in human lymphoblastoid cells~\cite{rao20143d}, we counted the
number of A
compartments per 10Mb bin. Then dividing our 3D reconstruction into concentric circles of increasing radius away from the center, we found the mean number of A compartments in each concentric circle. \Cref{fig:real_data_e} shows that  with increasing distance away from the center, the number of A compartments decreases. Thus, our reconstruction recovers the experimentally observed trend for A compartments to be preferentially located near the nucleus center. As shown in 
%\Cref{fig:impact_rho_real,fig:real_data_rho}
\textcolor{green!50!black}{Figures SM8 and SM9}
 in \textcolor{green!50!black}{section SM7}, we note that our results are robust to the choice of the tuning parameter $\rho$ resulting in biologically plausible configurations independent of the choice of $\rho$. % $\rho$ has very little effect on the observed 3D reconstructions.

Currently, many studies such as~\cite{rieber2017miniMDS} simply ignore the fact that the genome is diploid and infer the 3D genome organization as if the data was collected from a haploid organism, assuming that the homologous loci have the same 3D structure. However, we show in \textcolor{green!50!black}{section SM8}, \textcolor{green!50!black}{Figure SM10} that the haploid distance matrices, computed by including only one copy of each of the homologous loci,
 are different between the two copies with a mean Spearman correlation of only 0.08. This shows that modeling the diploid aspect of the genome provides valuable information regarding the 3D structure of each of the homologs, which may be substantially different.

\section{Discussion}
\label{sec_7}

%{\color{blue}
In this paper, we proved that for a diploid organism the 3D genome structure is not identifiable from pairwise distance measurements alone. This implies that applying any algorithm for the reconstruction of the 3D genome structure from typical chromosome conformation capture data for a diploid organism can result in any of the infinitely many configurations with the same pairwise contact frequencies. We showed that unique idenfiability is obtained using distance constraints between neighboring genomic loci as well as 3-way distance constraints in addition to the pairwise distance constraints that can be obtained from typical contact frequency data.
%chromosome capture data.  
Distances between neighboring genomic loci can be obtained empirically, e.g. from imaging studies, while 3-way distance constraints can be obtained from the most recently developed sequencing-based methods for obtaining contact frequencies such as SPRITE~\cite{Quinodoz219683}, C-walks~\cite{Olivares} and GAM~\cite{beagrie2017complex}. %knowing pairwise distances, distances between neighboring genomic loci and certain distances between three loci simulatenously (every endpoint of a chromosome has to be included in one such distance). 
We also presented SDP formulations for determining the 3D genome reconstruction both in the noiseless and the noisy setting. Finally, we applied our algorithm to contact frequency data from human lymphoblastoid cells collected using SPRITE and showed that our results recapitulate known biological trends; in particular, in the 3D configuration identified using our method, the  small chromosomes are preferentially situated in the interior of the cell nucleus, while the larger chromosomes are preferentially situated at the periphery of the cell nucleus. In addition,  in the 3D configuration identified using our method the number of A domains is higher in the interior versus the periphery, which is in line with  experimental results. Our work shows the importance of higher-order contact frequencies that can be measured using SPRITE~\cite{Quinodoz219683}, C-walks~\cite{Olivares} and GAM~\cite{beagrie2017complex} for obtaining the 3D organization of the genome in diploid organisms.
%This work provides algorithms for using data from Hi-C-like assays such as SPRITE and suggests that such assays provide additional value for understanding the 3D organization of the genome. 
This is particularly relevant for the reconstruction of cancer genomes, where copy number variations are frequent and hence the genome may contain even more than two copies of each locus.

 We conjecture that identifiability of the 3D genome structure can also be achieved by replacing the higher-order contact constraints  by distance constraints to the center of the cell nucleus. Such constraints are also biologically relevant, since these distances can be measured via imaging experiments, or inferred by measuring whether a particular locus is in a lamin-associated domain or a telomere, both of which tend to lie at the boundary of the cell nucleus~\cite{crabbe2012human, guelen2008LADs, van2017LADs_review}.  %In Appendix~\ref{appendix:proofs}, we suggest an approach for proving that pairwise distances, distances between neighboring genomic loci and distances to the center of the cell nucleus for at least two genomic loci per chromosome give finite identifiability.
%First, we studied knowing distances of loci to the center of the nucleus, which can be measured via FISH experiments, or determined by measuring whether a particular locus is in a lamin-associated domain or a telomere, both of which tend to lie at the boundary of the cell nucleus~\cite{crabbe2012human, guelen2008LADs, van2017LADs_review}.
%A further problem left open is whether knowing pairwise distances, distances between neighboring genomic loci and distances to the center of the cell nucleus for at least three genomic loci per chromosome including the endpoints gives unique identifiability of the 3D genome structure. 
%A future research direction is to study identifiability of 3D genome reconstruction in the presence of SNP data.
Another future research direction is the development of specialized solvers to enable reconstruction of the genome at higher resolution. In this study we used a 10Mbp resolution due to the computational constraints imposed by SDP solvers. 
%While in this study, we inferred the whole diploid 3D organization at 10Mbp resolution due to computational constraints imposed by the SDP solvers, in the future, development of specialized solvers for this problem can enable reconstruction of the genome at higher resolution.
Finally, the theoretical results in this paper build on the assumption that distances are inverses of square roots of pairwise and higher-order contact frequencies. An  interesting future research direction is to develop a method for estimating the map between higher-order contact frequencies and distances, and then prove identifiability as well as build reconstruction algorithms for these different maps. % functions mapping contact frequencies to distances.
% Another future research direction is the study of identifiability of the 3D genome structure from single cell data.}

\section*{Acknowldegements}
We thank Mohab Safey El Din for helpful discussions.

\bibliographystyle{siamplain}
\bibliography{references}

\newcommand{\noopsort}[1]{} \newcommand{\printfirst}[2]{#1}
  \newcommand{\singleletter}[1]{#1} \newcommand{\switchargs}[2]{#2#1}
\begin{thebibliography}{10}

\bibitem{Alfakih1999}
{\sc A.~Y. Alfakih, A.~Khandani, and H.~Wolkowicz}, {\em Solving {E}uclidean
  distance matrix completion problems via semidefinite programming},
  Computational Optimization and Applications, 12 (1999), pp.~13--30.

\bibitem{beagrie2017complex}
{\sc R.~A. Beagrie, A.~Scialdone, M.~Schueler, D.~C.~A. Kraemer, M.~Chotalia,
  S.~Q. Xie, M.~Barbieri, I.~de~Santiago, L.-M. Lavitas, M.~R. Branco, et~al.},
  {\em Complex multi-enhancer contacts captured by genome architecture
  mapping}, Nature, 543 (2017), p.~519.

\bibitem{bolzer2005three}
{\sc A.~Bolzer, G.~Kreth, I.~Solovei, D.~Koehler, K.~Saracoglu, C.~Fauth,
  S.~M{\"u}ller, R.~Eils, C.~Cremer, M.~R. Speicher, et~al.}, {\em
  Three-dimensional maps of all chromosomes in human male fibroblast nuclei and
  prometaphase rosettes}, PLoS Biology, 3 (2005), p.~e157.

\bibitem{cauer2019diploid}
{\sc A.~G. Cauer, G.~Yardimci, J.-P. Vert, N.~Varoquaux, and W.~S. Noble}, {\em
  Inferring diploid {3D} chromatin structures from {Hi-C} data},
  bioRxiv:644294,  (2019).

\bibitem{Cayton}
{\sc L.~Cayton and S.~Dasgupta}, {\em Robust {E}uclidean embedding}, in
  Proceedings of the 23rd International Conference on Machine Learning, ICML
  '06, New York, NY, USA, 2006, ACM, pp.~169--176.

\bibitem{chandrasekaran2012convex}
{\sc V.~Chandrasekaran, B.~Recht, P.~A. Parrilo, and A.~S. Willsky}, {\em The
  convex geometry of linear inverse problems}, Foundations of Computational
  Mathematics, 12 (2012), pp.~805--849.

\bibitem{1000genomes2012integrated}
{\sc .~G.~P. Consortium et~al.}, {\em An integrated map of genetic variation
  from 1,092 human genomes}, Nature, 491 (2012), pp.~56--65.

\bibitem{1000genomes2015global}
{\sc .~G.~P. Consortium et~al.}, {\em A global reference for human genetic
  variation}, Nature, 526 (2015), pp.~68--74.

\bibitem{cox2000multidimensional}
{\sc T.~F. Cox and M.~A.~A. Cox}, {\em Multidimensional scaling}, Chapman and
  Hall / CRC, 2000.

\bibitem{crabbe2012human}
{\sc L.~Crabbe, A.~J. Cesare, J.~M. Kasuboski, J.~A.~J. Fitzpatrick, and
  J.~Karlseder}, {\em Human telomeres are tethered to the nuclear envelope
  during postmitotic nuclear assembly}, Cell Reports, 2 (2012), pp.~1521--1529.

\bibitem{gene}
{\sc J.~Dekker}, {\em Gene regulation in the third dimension}, Science, 319
  (2008), pp.~1793--1794.

\bibitem{Dekker1306}
{\sc J.~Dekker, K.~Rippe, M.~Dekker, and N.~Kleckner}, {\em Capturing
  chromosome conformation}, Science, 295 (2002), pp.~1306--1311.

\bibitem{dipierro2016transferablemodel}
{\sc M.~Di~Pierro, B.~Zhang, E.~L. Aiden, P.~G. Wolynes, and J.~N. Onuchic},
  {\em Transferable model for chromosome architecture}, Proceedings of the
  National Academy of Sciences, 113 (2016), pp.~12168--12173.

\bibitem{duan2010three}
{\sc Z.~Duan, M.~Andronescu, K.~Schutz, S.~McIlwain, Y.~J. Kim, C.~Lee,
  J.~Shendure, S.~Fields, C.~A. Blau, and W.~S. Noble}, {\em A
  three-dimensional model of the yeast genome}, Nature, 465 (2010),
  pp.~363--367, \url{http://dx.doi.org/10.1038/nature08973}.

\bibitem{Fang2012}
{\sc H.~Fang and D.~P. O'Leary}, {\em Euclidean distance matrix completion
  problems}, Optimization Methods and Software, 27 (2012), pp.~695--717.

\bibitem{fazel2003log}
{\sc M.~Fazel, H.~Hindi, and S.~P. Boyd}, {\em Log-det heuristic for matrix
  rank minimization with applications to {H}ankel and {E}uclidean distance
  matrices}, in Proceedings of the 2003 American Control Conference, vol.~3,
  IEEE, 2003, pp.~2156--2162.

\bibitem{fazel2001rank}
{\sc M.~Fazel, H.~Hindi, S.~P. Boyd, et~al.}, {\em A rank minimization
  heuristic with application to minimum order system approximation}, in
  Proceedings of the American Control Conference, vol.~6, Citeseer, 2001,
  pp.~4734--4739.

\bibitem{guelen2008LADs}
{\sc L.~Guelen, L.~Pagie, E.~Brasset, W.~Meuleman, M.~B. Faza, W.~Talhout,
  B.~H. Eussen, A.~de~Klein, L.~Wessels, W.~de~Laat, et~al.}, {\em Domain
  organization of human chromosomes revealed by mapping of nuclear lamina
  interactions}, Nature, 453 (2008), pp.~948--951.

\bibitem{hu2013bayesian}
{\sc M.~Hu, K.~Deng, Z.~Qin, J.~Dixon, S.~Selvaraj, J.~Fang, B.~Ren, and J.~S.
  Liu}, {\em Bayesian inference of spatial organizations of chromosomes}, PLoS
  Computational Biology, 9 (2013), p.~e1002893.

\bibitem{Hughes}
{\sc J.~R. Hughes, N.~Roberts, S.~McGowan, D.~Hay, E.~Giannoulatou, M.~Lynch,
  M.~De~Gobbi, S.~Taylor, R.~Gibbons, and D.~R. Higgs}, {\em Analysis of
  hundreds of cis-regulatory landscapes at high resolution in a single,
  high-throughput experiment}, Nature Genetics, 46 (2014), pp.~205--212.

\bibitem{jungmann2014DNAPAINT}
{\sc R.~Jungmann, M.~S. Avenda{\~n}o, J.~B. Woehrstein, M.~Dai, W.~M. Shih, and
  P.~Yin}, {\em Multiplexed 3d cellular super-resolution imaging with dna-paint
  and exchange-paint}, Nature Methods, 11 (2014), p.~313.

\bibitem{krislock2010semidefinite}
{\sc N.~Krislock}, {\em Semidefinite facial reduction for low-rank {E}uclidean
  distance matrix completion}, PhD thesis, University of Waterloo, 2010,
  \url{http://hdl.handle.net/10012/5093}.

\bibitem{lesne20143d}
{\sc A.~Lesne, J.~Riposo, P.~Roger, A.~Cournac, and J.~Mozziconacci}, {\em 3{D}
  genome reconstruction from chromosomal contacts}, Nature Methods, 11 (2014),
  pp.~1141--1143.

\bibitem{genedist}
{\sc E.~Lieberman-Aiden, N.~L.~v. Berkum, L.~Williams, M.~Imakaev, T.~Ragoczy,
  A.~Telling, I.~Amit, B.~R. Lajoie, P.~J. Sabo, M.~O. Dorschner, R.~Sandstrom,
  B.~Bernstein, M.~A. Bender, M.~Groudine, A.~Gnirke, J.~Stamatoyannopoulos,
  L.~A. Mirny, E.~S. Lander, and J.~Dekker}, {\em {Comprehensive mapping of
  long-range interactions reveals folding principles of the human genome}},
  Science, 326 (2009).

\bibitem{lieberman2009comprehensive}
{\sc E.~Lieberman-Aiden, N.~L. Van~Berkum, L.~Williams, M.~Imakaev, T.~Ragoczy,
  A.~Telling, I.~Amit, B.~R. Lajoie, P.~J. Sabo, M.~O. Dorschner, R.~Sandstrom,
  B.~Bernstein, M.~A. Bender, M.~Groudine, A.~Gnirke, J.~Stamatoyannopoulos,
  L.~A. Mirny, E.~S. Lander, and J.~Dekker}, {\em Comprehensive mapping of
  long-range interactions reveals folding principles of the human genome},
  science, 326 (2009), pp.~289--293.

\bibitem{Lu12332}
{\sc F.~Lu, S.~Kele{\c s}, S.~J. Wright, and G.~Wahba}, {\em Framework for
  kernel regularization with application to protein clustering}, Proceedings of
  the National Academy of Sciences, 102 (2005), pp.~12332--12337.

\bibitem{mirny2011fractal}
{\sc L.~A. Mirny}, {\em The fractal globule as a model of chromatin
  architecture in the cell}, Chromosome research, 19 (2011), pp.~37--51.

\bibitem{Mishra2011}
{\sc B.~Mishra, G.~Meyer, and R.~Sepulchre}, {\em Low-rank optimization for
  distance matrix completion}, in 2011 50th IEEE Conference on Decision and
  Control and European Control Conference, 2011, pp.~4455--4460.

\bibitem{muller2010stable}
{\sc I.~M{\"u}ller, S.~Boyle, R.~H. Singer, W.~A. Bickmore, and J.~R. Chubb},
  {\em Stable morphology, but dynamic internal reorganisation, of interphase
  human chromosomes in living cells}, PloS One, 5 (2010), p.~e11560.

\bibitem{nir2018walking}
{\sc G.~Nir, I.~Farabella, C.~P. Estrada, C.~G. Ebeling, B.~J. Beliveau, H.~M.
  Sasaki, S.~H. Lee, S.~C. Nguyen, R.~B. McCole, S.~Chattoraj, et~al.}, {\em
  Walking along chromosomes with super-resolution imaging, contact maps, and
  integrative modeling}, PLoS Genetics, 14 (2018), p.~e1007872.

\bibitem{Olivares}
{\sc P.~Olivares-Chauvet, Z.~Mukamel, A.~Lifshitz, O.~Schwartzman, N.~O.
  Elkayam, Y.~Lubling, G.~Deikus, R.~P. Sebra, and A.~Tanay}, {\em Capturing
  pairwise and multi-way chromosomal conformations using chromosomal walks},
  Nature, 540 (2016), pp.~296--300.

\bibitem{Qi2019MD}
{\sc Y.~Qi and B.~Zhang}, {\em Predicting three-dimensional genome organization
  with chromatin states}, PLoS computational biology, 15 (2019), p.~e1007024.

\bibitem{Quinodoz219683}
{\sc S.~A. Quinodoz, N.~Ollikainen, B.~Tabak, A.~Palla, J.~M. Schmidt,
  E.~Detmar, M.~M. Lai, A.~A. Shishkin, P.~Bhat, Y.~Takei, et~al.}, {\em
  Higher-order inter-chromosomal hubs shape 3{D} genome organization in the
  nucleus}, Cell, 174 (2018), pp.~744--757.

\bibitem{rao20143d}
{\sc S.~S.~P. Rao, M.~H. Huntley, N.~C. Durand, E.~K. Stamenova, I.~D. Bochkov,
  J.~T. Robinson, A.~L. Sanborn, I.~Machol, A.~D. Omer, E.~S. Lander, et~al.},
  {\em A 3{D} map of the human genome at kilobase resolution reveals principles
  of chromatin looping}, Cell, 159 (2014), pp.~1665--1680.

\bibitem{rieber2017miniMDS}
{\sc L.~Rieber and S.~Mahony}, {\em {miniMDS}: 3{D} structural inference from
  high-resolution {Hi-C} data}, Bioinformatics, 33 (2017), pp.~i261--i266.

\bibitem{rousseau2011three}
{\sc M.~Rousseau, J.~Fraser, M.~A. Ferraiuolo, J.~Dostie, and M.~Blanchette},
  {\em Three-dimensional modeling of chromatin structure from interaction
  frequency data using {M}arkov chain {M}onte {C}arlo sampling}, BMC
  Bioinformatics, 12 (2011), p.~414.

\bibitem{Simonis}
{\sc M.~Simonis, P.~Klous, E.~Splinter, Y.~Moshkin, R.~Willemsen, E.~de~Wit,
  B.~van Steensel, and W.~de~Laat}, {\em Nuclear organization of active and
  inactive chromatin domains uncovered by chromosome conformation
  capture--on-chip (4{C})}, Nature Genetics, 38 (2006), pp.~1348--1354.

\bibitem{stevens20173d}
{\sc T.~J. Stevens, D.~Lando, S.~Basu, L.~P. Atkinson, Y.~Cao, S.~F. Lee,
  M.~Leeb, K.~J. Wohlfahrt, W.~Boucher, A.~O’Shaughnessy-Kirwan, et~al.},
  {\em {3D} structures of individual mammalian genomes studied by single-cell
  {Hi-C}}, Nature, 544 (2017), p.~59.

\bibitem{Uhler_Trends}
{\sc C.~Uhler and G.~V. Shivashankar}, {\em Chromosome intermingling:
  mechanical hotspots for genome regulation}, Trends in Cell Biology, 27
  (2017), pp.~810--819.

\bibitem{Uhler_Nat_Rev}
{\sc C.~Uhler and G.~V. Shivashankar}, {\em The regulation of genome
  organization and gene expression by nuclear mechanotransduction}, Nature
  Reviews Molecular Cell Biology, 18 (2017), pp.~717--727.

\bibitem{van2017LADs_review}
{\sc B.~Van~Steensel and A.~S. Belmont}, {\em Lamina-associated domains: links
  with chromosome architecture, heterochromatin, and gene repression}, Cell,
  169 (2017), pp.~780--791.

\bibitem{varoquaux2014statistical}
{\sc N.~Varoquaux, F.~Ay, W.~S. Noble, and J.-P. Vert}, {\em A statistical
  approach for inferring the 3{D} structure of the genome}, Bioinformatics, 30
  (2014), pp.~i26--i33.

\bibitem{wang2018crispr}
{\sc H.~Wang, X.~Xu, C.~M. Nguyen, Y.~Liu, Y.~Gao, X.~Lin, T.~Daley, N.~H.
  Kipniss, M.~La~Russa, and L.~S. Qi}, {\em {CRISPR}-mediated programmable {3D}
  genome positioning and nuclear organization}, Cell, 175 (2018),
  pp.~1405--1417.

\bibitem{weinberger2007graph}
{\sc K.~Q. Weinberger, F.~Sha, Q.~Zhu, and L.~K. Saul}, {\em Graph {L}aplacian
  regularization for large-scale semidefinite programming}, in Advances in
  Neural Information Processing Systems, 2007, pp.~1489--1496.

\bibitem{zhang2016distance}
{\sc L.~Zhang, G.~Wahba, and M.~Yuan}, {\em Distance shrinkage and {E}uclidean
  embedding via regularized kernel estimation}, Journal of the Royal
  Statistical Society: Series B, 78 (2016), pp.~849--867.

\bibitem{chromsde}
{\sc Z.~Zhang, G.~Li, K.-C. Toh, and W.-K. Sung}, {\em {3D Chromosome Modeling
  with Semi-Definite Programming and Hi-C Data}}, Journal of Computational
  Biology, 20 (2013).

\end{thebibliography}


\newcommand{\noopsort}[1]{} \newcommand{\printfirst}[2]{#1}
  \newcommand{\singleletter}[1]{#1} \newcommand{\switchargs}[2]{#2#1}
\begin{thebibliography}{1}

\bibitem{bolzer2005three}
{\sc A.~Bolzer, G.~Kreth, I.~Solovei, D.~Koehler, K.~Saracoglu, C.~Fauth,
  S.~M{\"u}ller, R.~Eils, C.~Cremer, M.~R. Speicher, et~al.}, {\em
  Three-dimensional maps of all chromosomes in human male fibroblast nuclei and
  prometaphase rosettes}, PLoS Biology, 3 (2005), p.~e157.

\bibitem{nir2018walking}
{\sc G.~Nir, I.~Farabella, C.~P. Estrada, C.~G. Ebeling, B.~J. Beliveau, H.~M.
  Sasaki, S.~H. Lee, S.~C. Nguyen, R.~B. McCole, S.~Chattoraj, et~al.}, {\em
  Walking along chromosomes with super-resolution imaging, contact maps, and
  integrative modeling}, PLoS Genetics, 14 (2018), p.~e1007872.

\bibitem{chromsde}
{\sc Z.~Zhang, G.~Li, K.-C. Toh, and W.-K. Sung}, {\em {3D Chromosome Modeling
  with Semi-Definite Programming and Hi-C Data}}, Journal of Computational
  Biology, 20 (2013).

\end{thebibliography}

\end{document}

\maketitle

%\section{Simulations} \label{appendix:simulations}

\section{Simulations: reconstructions in the noiseless setting}\label{appendix:simulations_recon}
\begin{figure*}[h]
    \centering
  \subfloat{\includegraphics[width=0.23\textwidth]{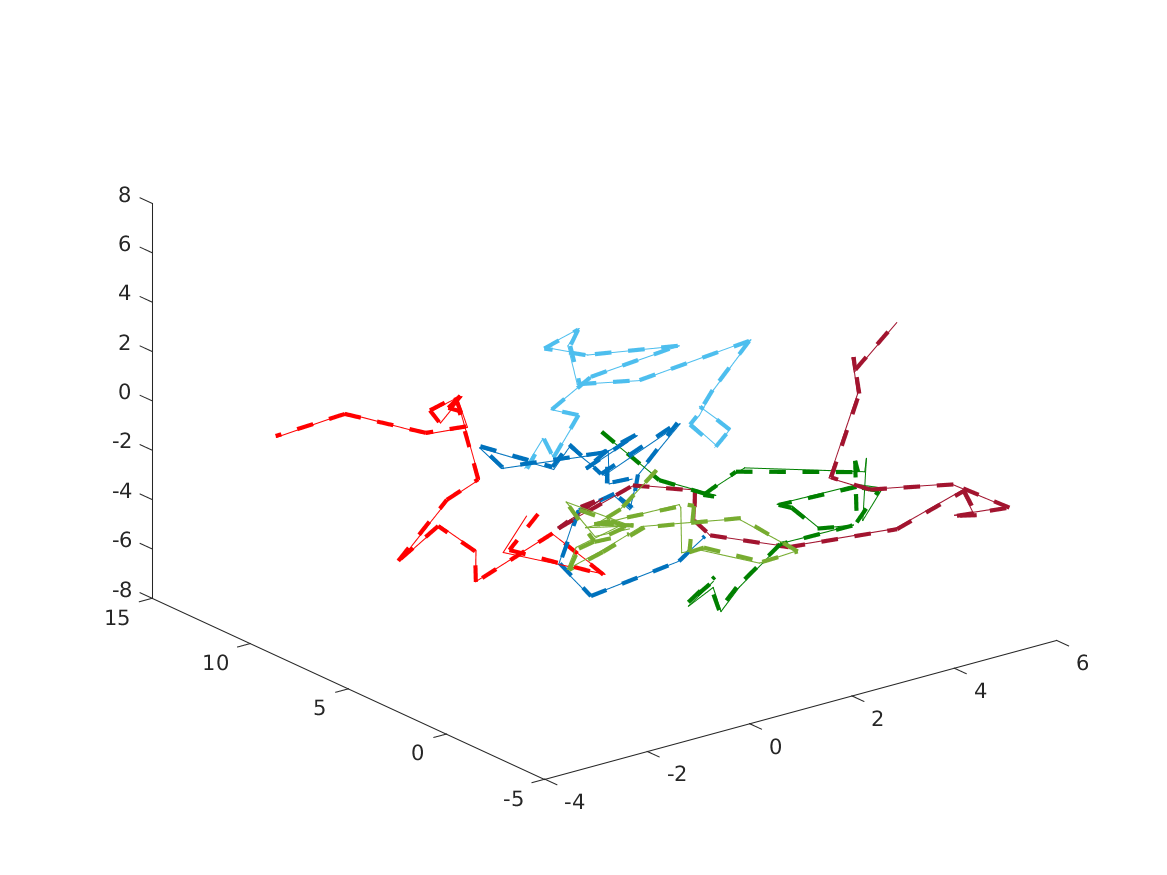}}    
  \subfloat{\includegraphics[width=0.23\textwidth]{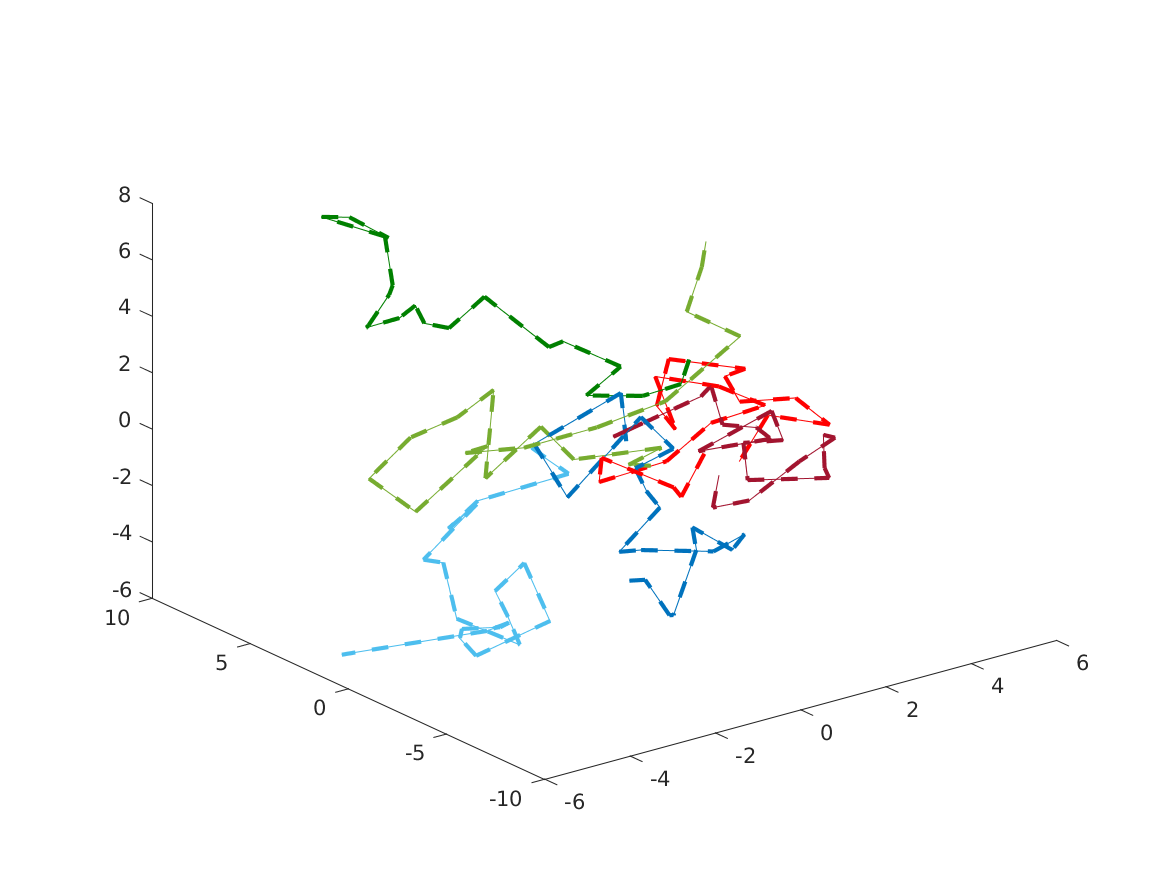}} 
  \subfloat{\includegraphics[width=0.23\textwidth]{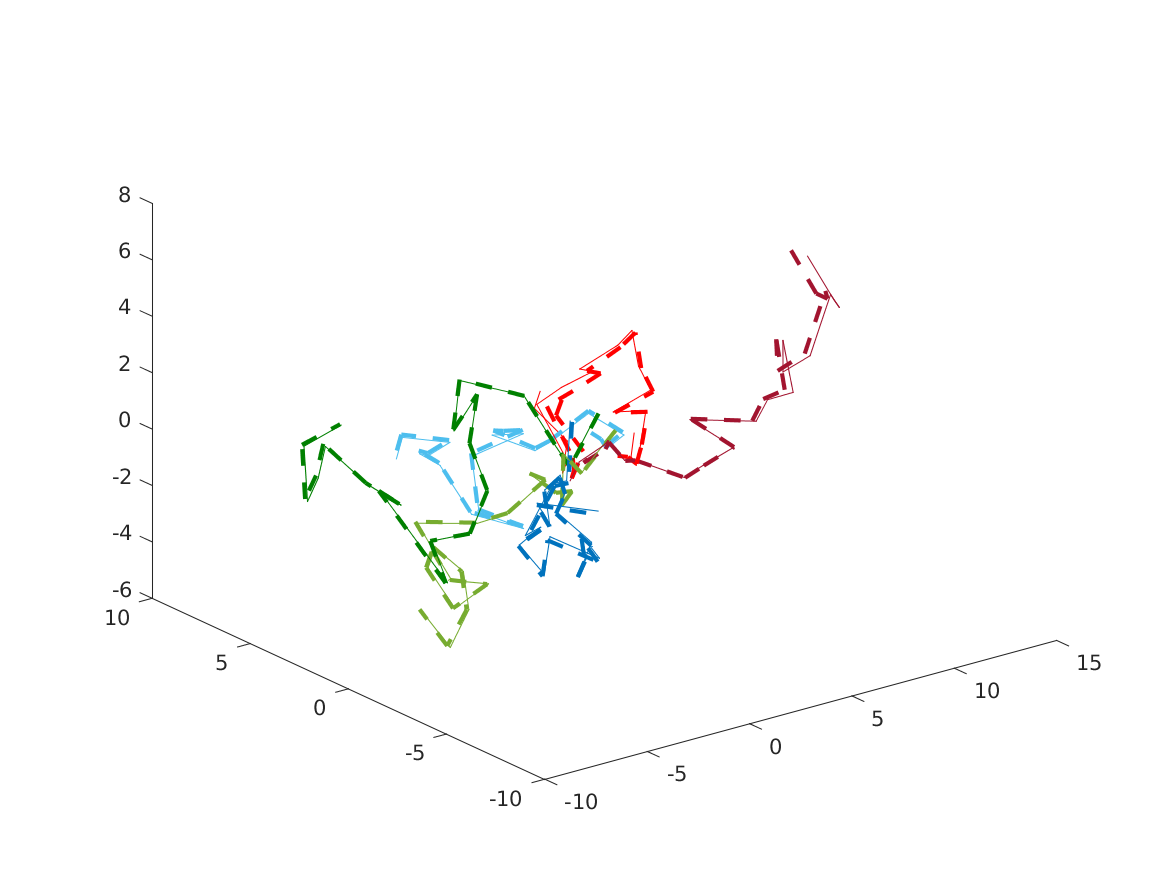}} 
  \subfloat{\includegraphics[width=0.23\textwidth]{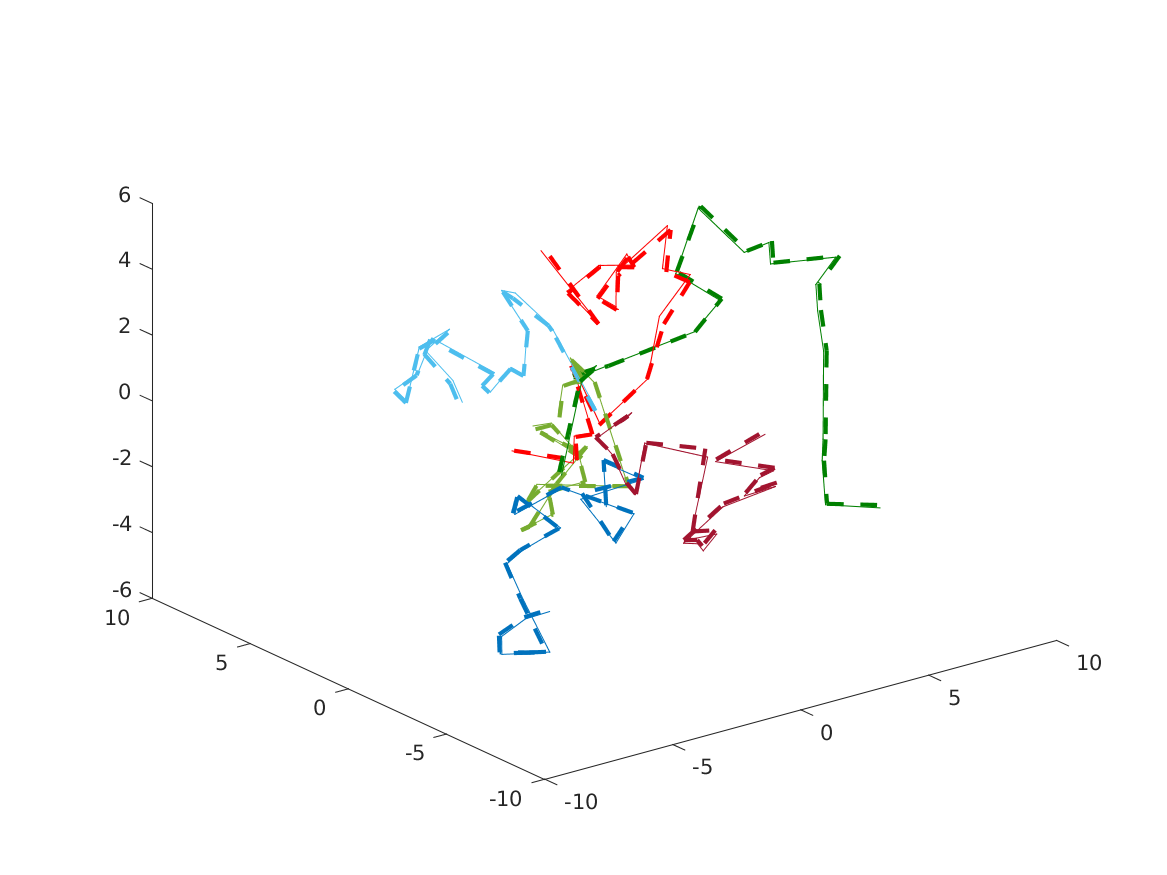}}
  \newline
  \subfloat{\includegraphics[width=0.23\textwidth]{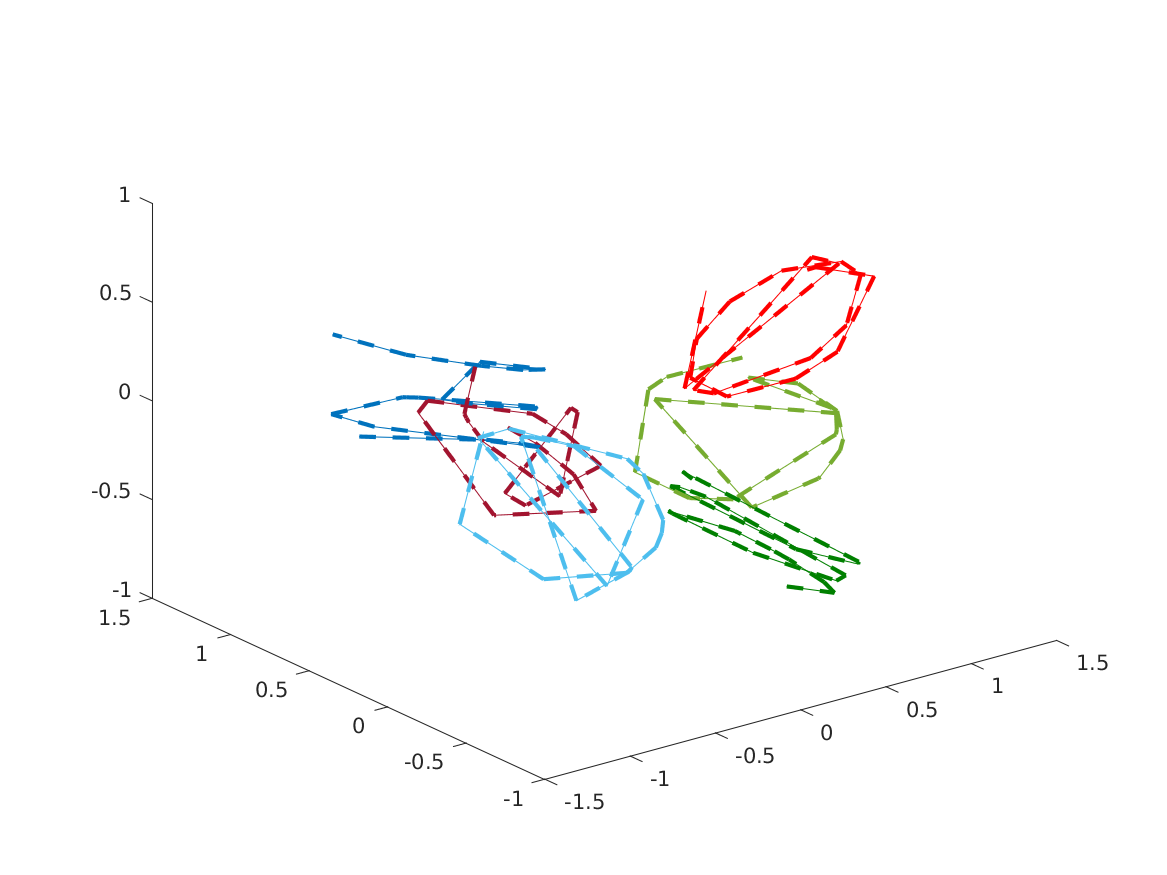}} 
  \subfloat{\includegraphics[width=0.23\textwidth]{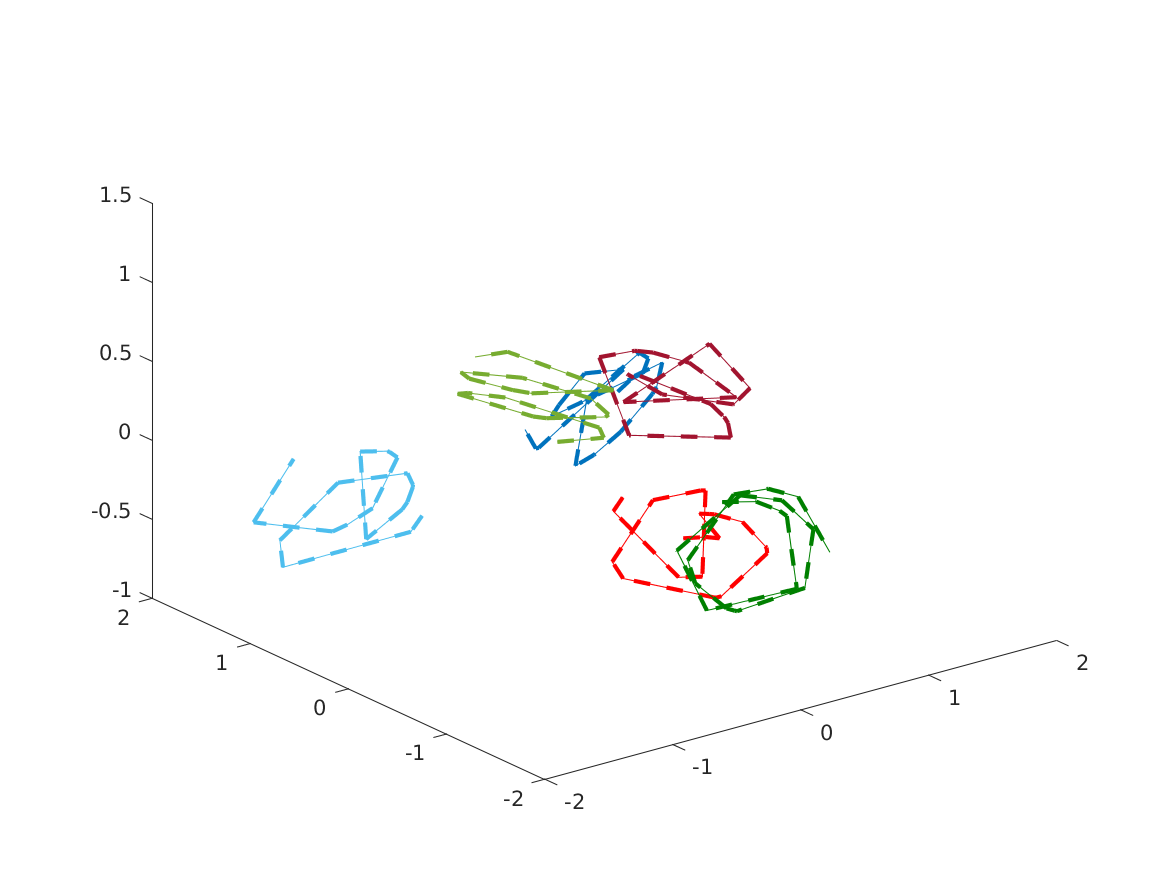}} 
  \subfloat{\includegraphics[width=0.23\textwidth]{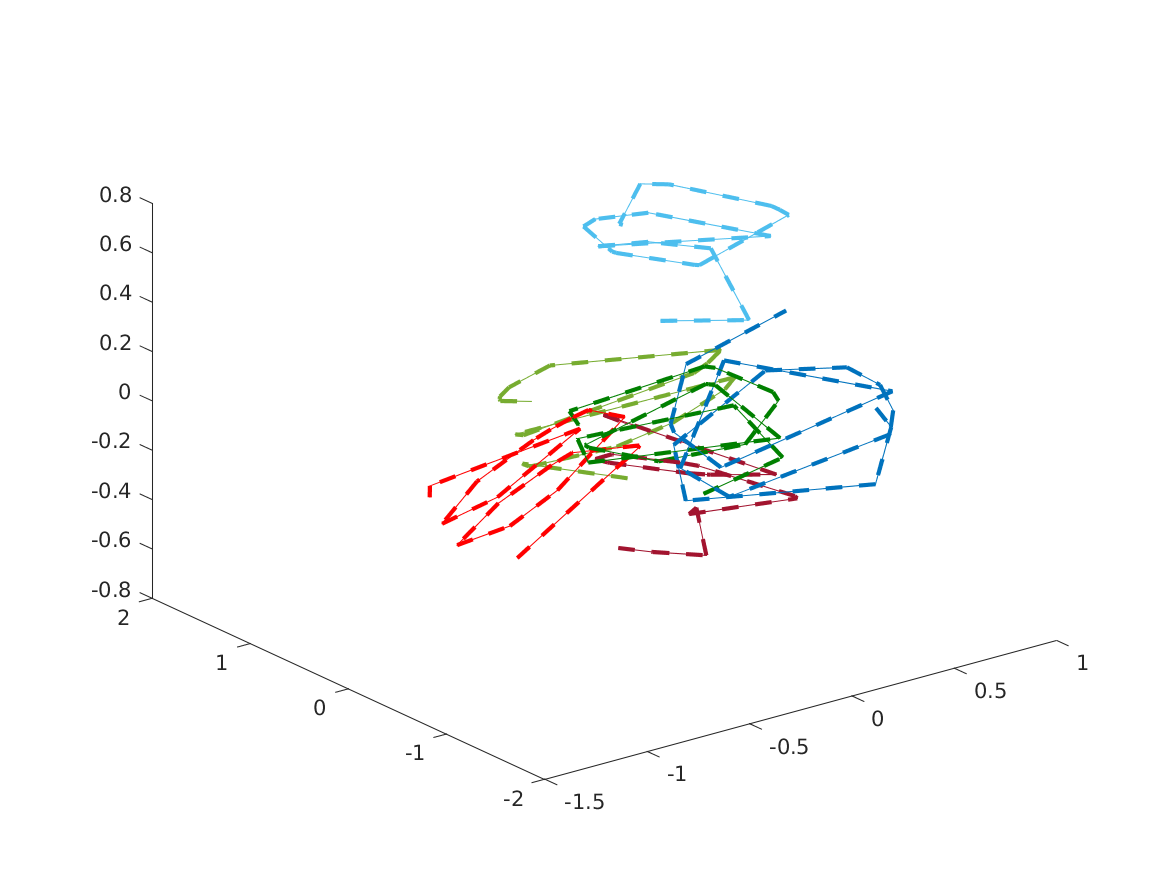}} 
  \subfloat{\includegraphics[width=0.23\textwidth]{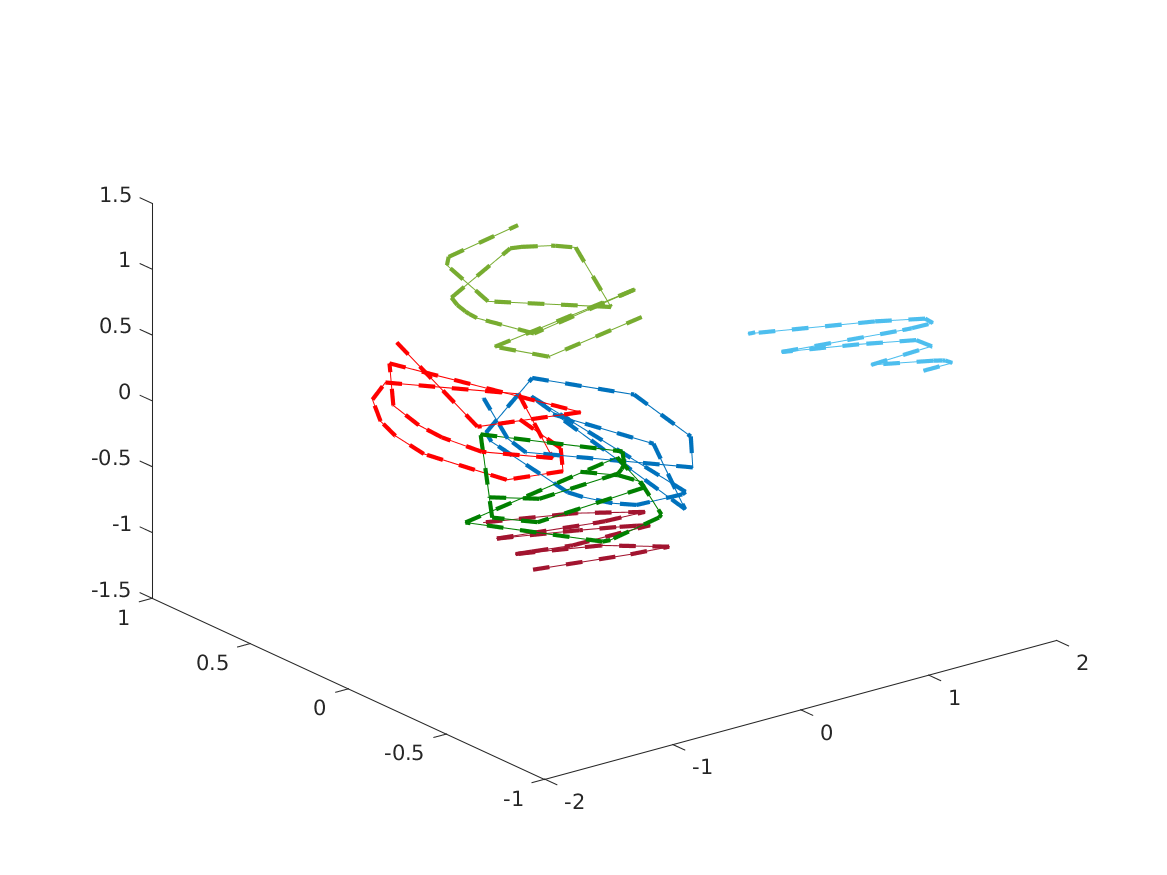}}
  \newline
  \subfloat{\includegraphics[width=0.23\textwidth]{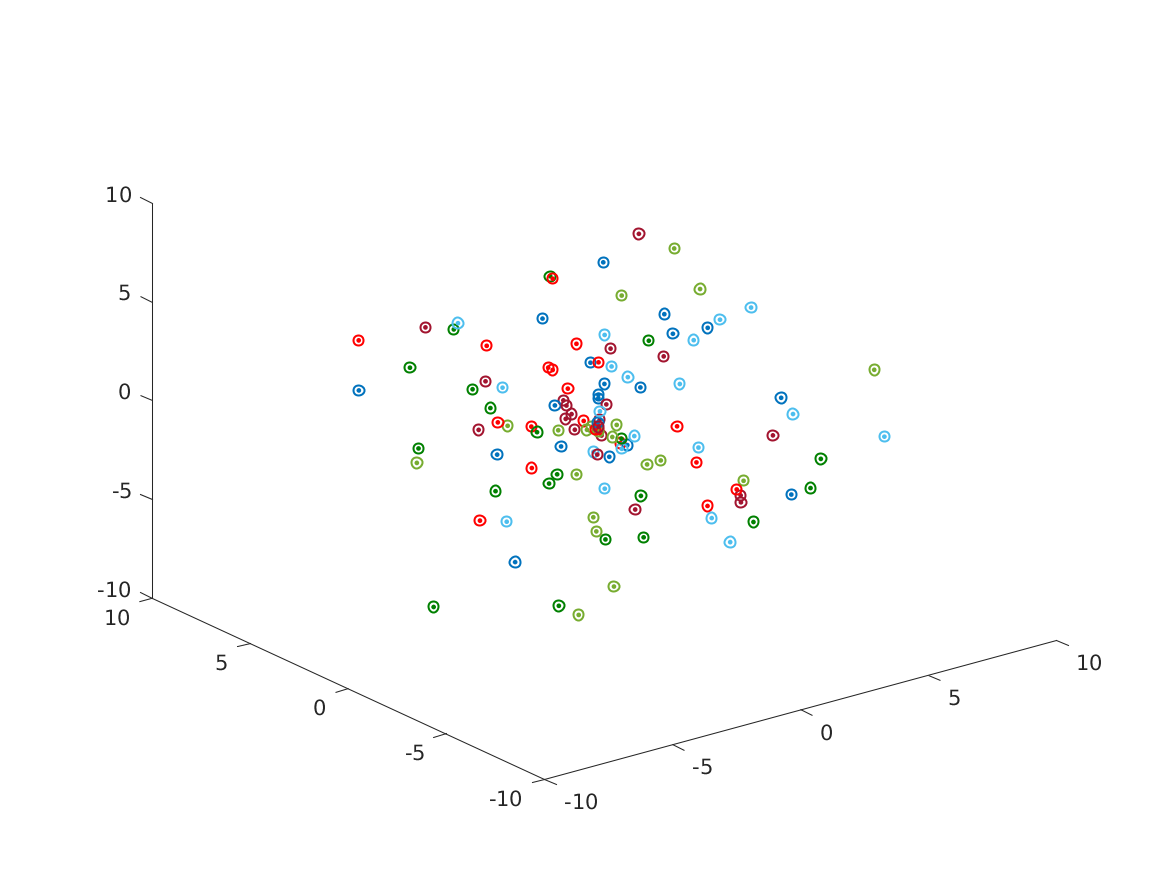}} 
  \subfloat{\includegraphics[width=0.23\textwidth]{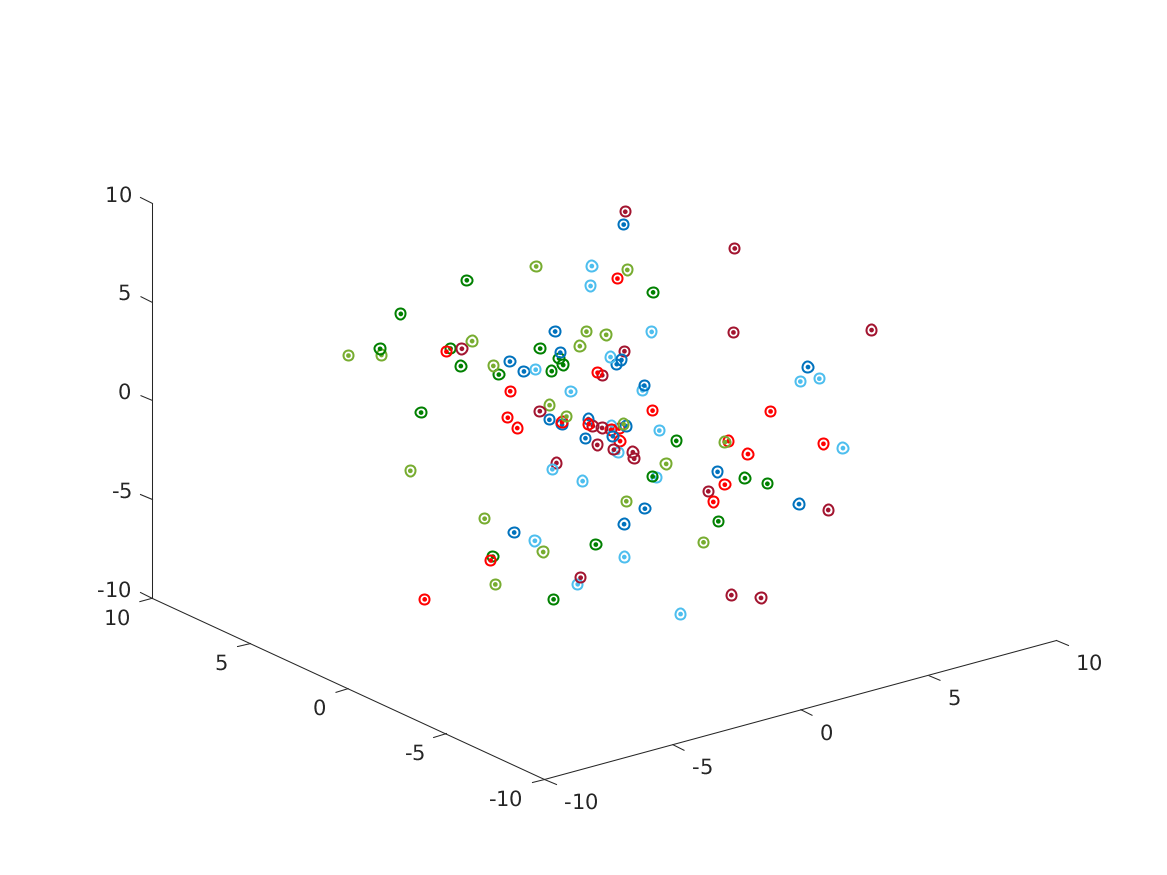}}
  \subfloat{\includegraphics[width=0.23\textwidth]{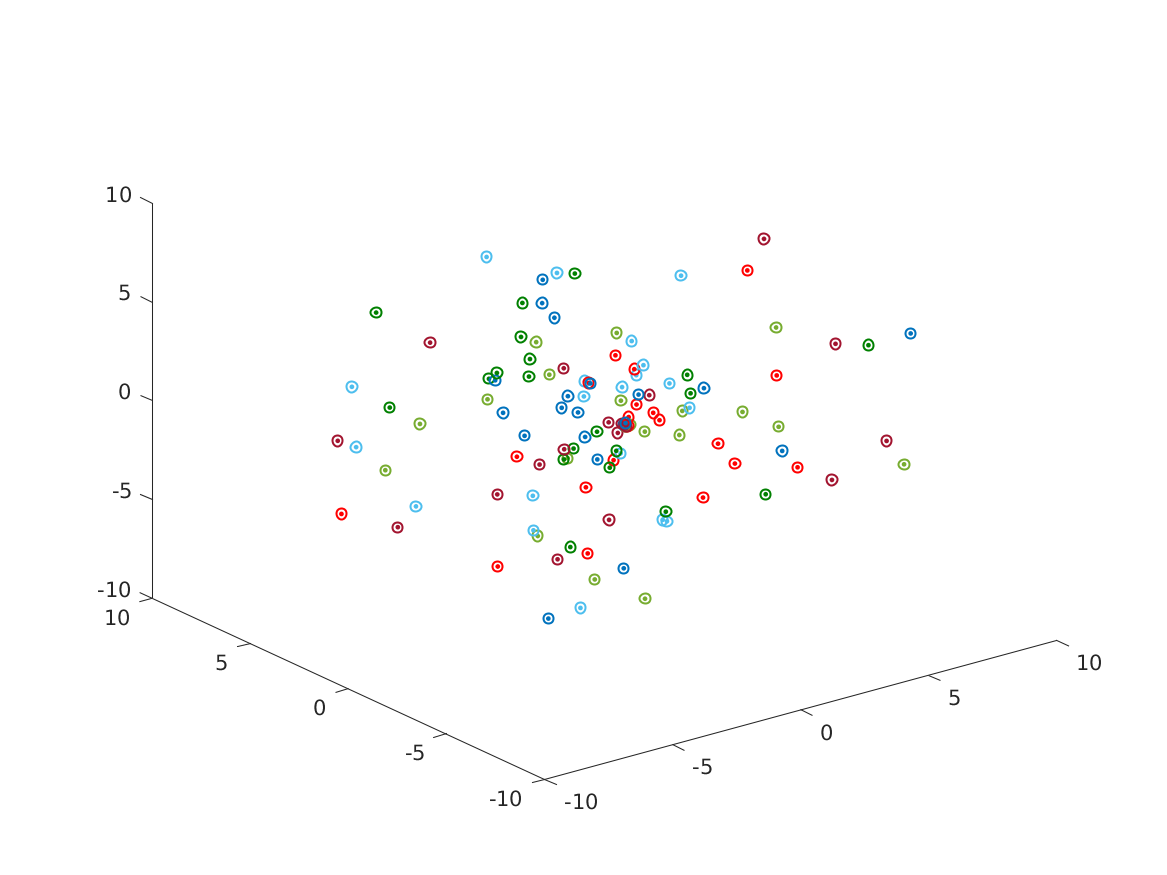}}
  \subfloat{\includegraphics[width=0.23\textwidth]{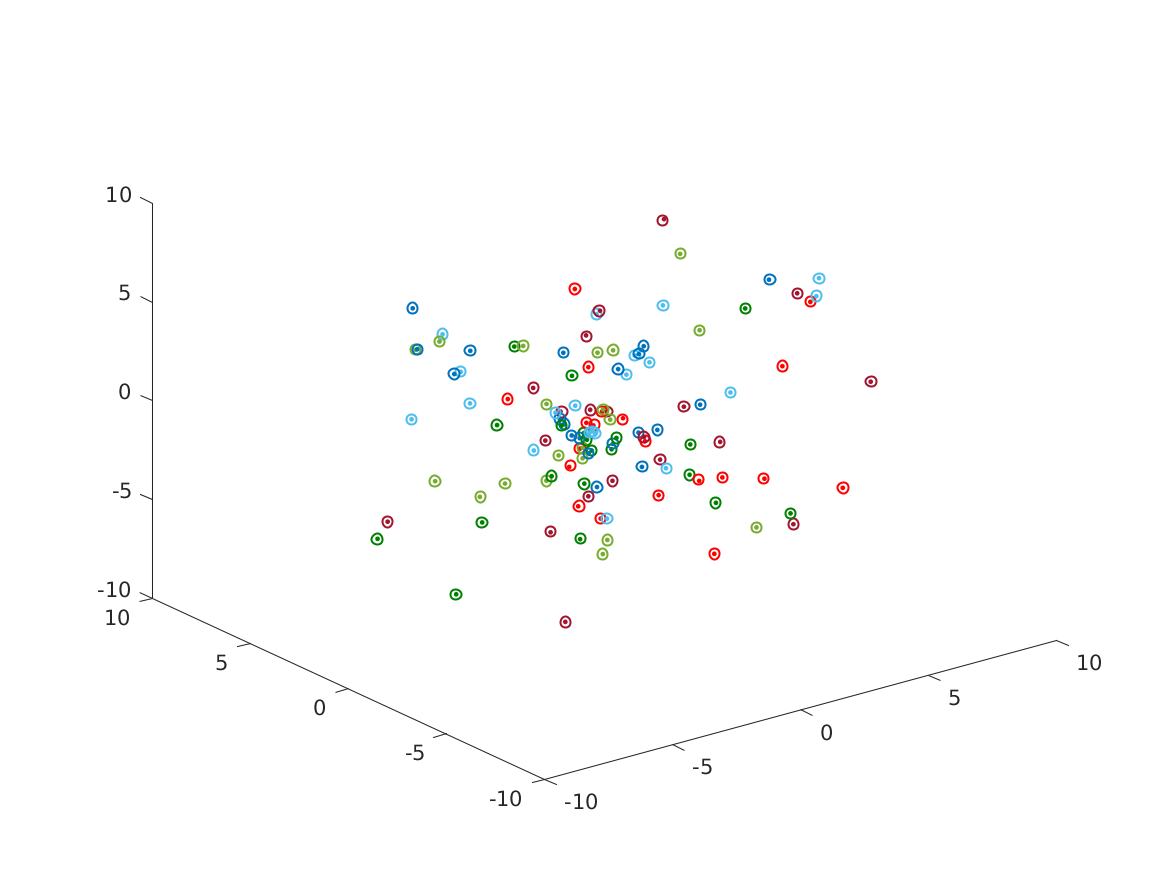}}
    \caption{Additional examples of true and reconstructed points on simulated data. True points were generated using Brownian motion model (first row), spirals (second row) and random points in a sphere (third row). We generated six chromosomes, corresponding to three homologous pairs with 20 domains per chromosome in the noiseless setting.
    Solid lines / points correspond to true 3D coordinates and dashed lines / unfilled points to reconstructions via our method. Each color represents a different chromosome. }
    \label{fig:noiseless_recon_addn}
\end{figure*}

\Cref{fig:noiseless_recon_addn} shows additional reconstructions of simulated data in the noiseless setting. The true structures are consistently recovered under different data generation models.

%\subsection{Impact of the number of tensor constraints}\label{appendix:simulations_numc}
\section{Simulations: impact of the number of 3-way distance constraints}\label{appendix:simulations_numc}
\begin{figure*}[h]
    \centering
  \subfloat[]{\includegraphics[width=0.4\textwidth]{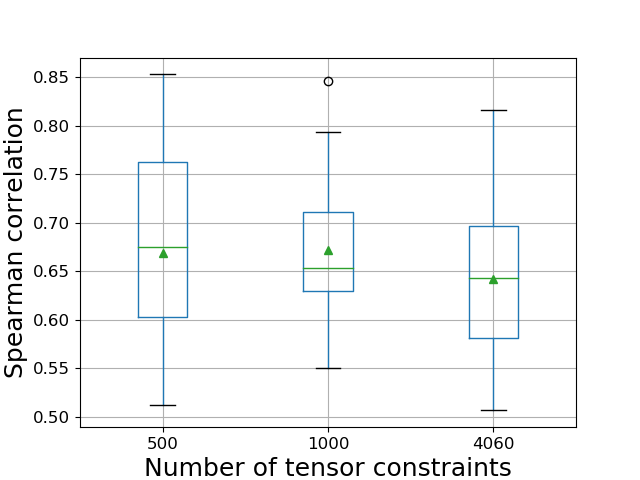}}
  ~
  \subfloat[]{\includegraphics[width=0.4\textwidth]{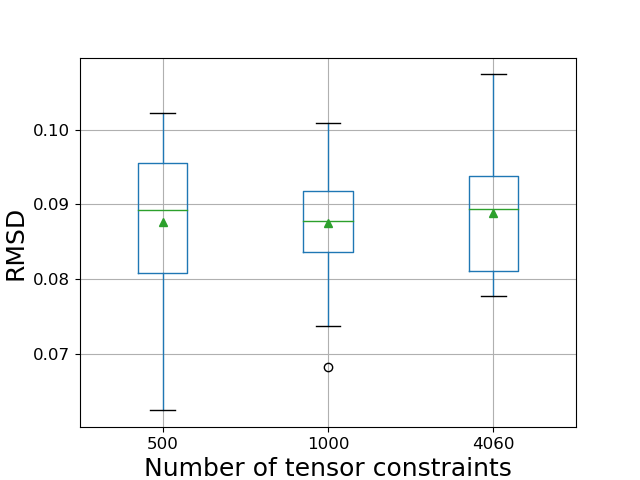}}     
    \caption{The impact of the number of 3-way distance constraints in the noisy setting. Boxplots showing Spearman correlation and root-mean-square deviation (RMSD) for different number of 3-way distance constraints over $20$ trials. Simulated data was generated using Brownian motion model with three chromosomes, where each chromosome had $10$ domains. Noise level of $0.5$ was added. We used $\rho = 0.000001$ to solve the SDP. Green triangles and lines indicate the mean and median performance respectively.}
    \label{fig:impact_numc}
\end{figure*}

\Cref{fig:impact_numc} shows the impact of the number of 3-way distance constraints on the solution in the noisy setting. We explored the impact of the number of 3-way distance constraints specifically when the number of chromosomes is higher (three chromosomes) since higher-order constraints seem to play a more critical role in that setting.
%, as shown in \Cref{fig:rmsd_comparison_notensor}. 
We evaluate the performance when $500$, $1000$ or all ($4060$ for 30 domains) 3-way distance constraints are used. \Cref{fig:impact_numc} shows that the choice of the number of 3-way distance constraints has little impact on the accuracy of reconstruction, so we used $1000$ 3-way distance constraints (or all possible triplets if that number was smaller) for the simulations and the real data analysis.

%\subsection{Impact of the tuning parameter $\rho$}\label{appendix:simulations_rho}
\section{Simulations: impact of the tuning parameter $\rho$}\label{appendix:simulations_rho}
\begin{figure*}[h]
    \centering
  \subfloat[]{\includegraphics[width=0.39\textwidth]{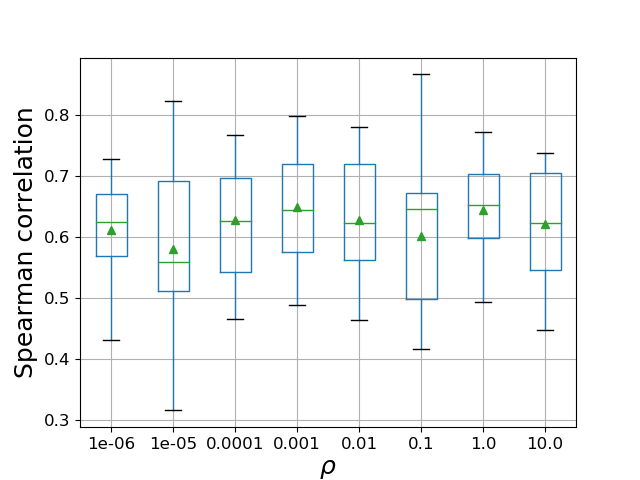}}
  ~
  \subfloat[]{\includegraphics[width=0.39\textwidth]{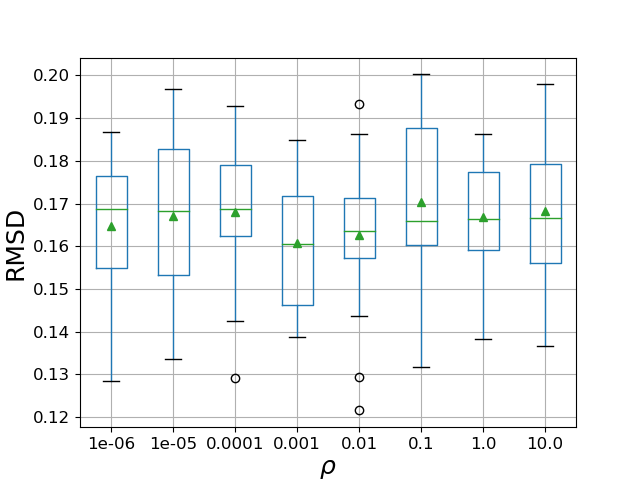}}
    \caption{The impact of $\rho$ in the noisy setting. Boxplots showing Spearman correlation and root-mean-square deviation (RMSD) for different values of $\rho$ over $20$ trials. Simulated data was generated using a Brownian motion model with one chromosome and $10$ domains per chromosome as well as noise level of $0.5$. We used the maximum number of triplet tensor constraints ($120$) to solve the SDP. Green triangles and lines indicate the mean and median performance respectively.}
    \label{fig:impact_rho}
\end{figure*}

\Cref{fig:impact_rho} explores the impact of the tuning parameter $\rho$ 
%from \Cref{SDP_noisy} 
in the noisy setting. \Cref{fig:impact_rho} shows that the choice of $\rho$ has little impact on the accuracy of the reconstruction. For the simulations in the noisy setting and for real data we chose $\rho = 0.000001$.
%, shown in \Cref{fig:spearman_corr_noisy_data} and real data analysis in \Cref{fig:real_data}, we chose $\rho = 0.000001$.

% \section{Real contact frequency data} \label{appendix:real_data}
\section{Distance between neighboring beads}\label{appendix:dist_neighboring}
\begin{figure*}[!h]
    \centering
  \subfloat[]{\label{fig:dist_neighbor_choice_a}\includegraphics[width=0.4\textwidth]{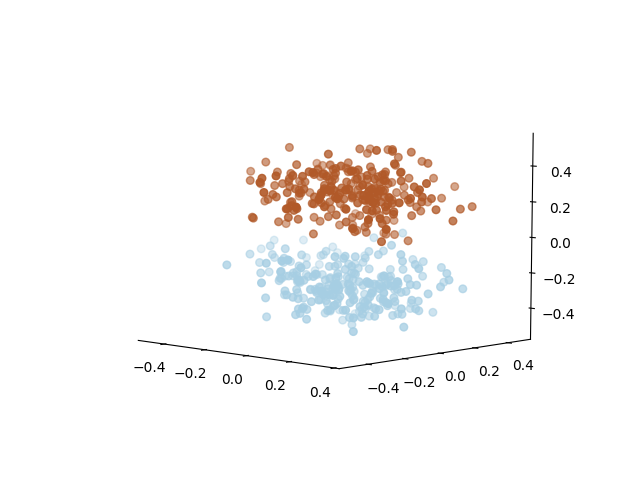}}
  ~
  \subfloat[]{\label{fig:dist_neighbor_choice_b}\includegraphics[width=0.4\textwidth]{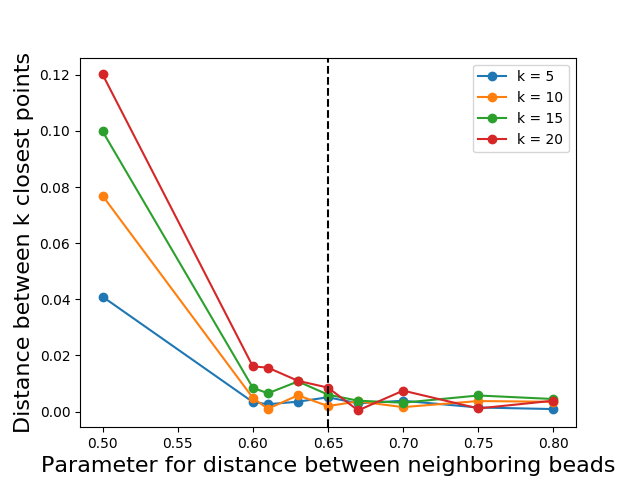}}
  \newline
  \subfloat[]{\label{fig:dist_neighbor_choice_c}\includegraphics[width=0.4\textwidth]{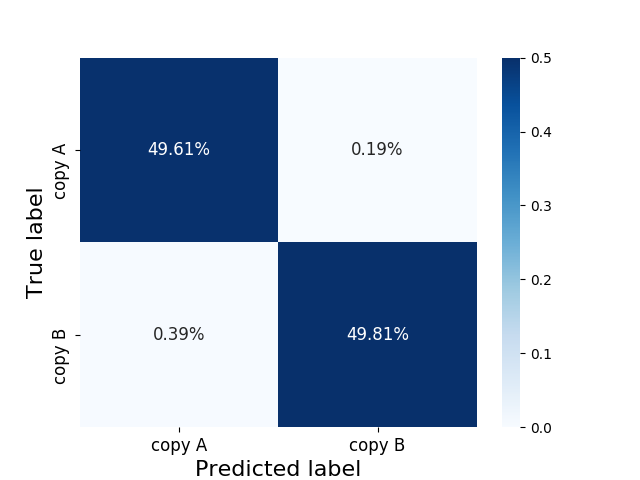}}
  ~  
  \subfloat[]{\label{fig:dist_neighbor_choice_d}\includegraphics[width=0.4\textwidth]{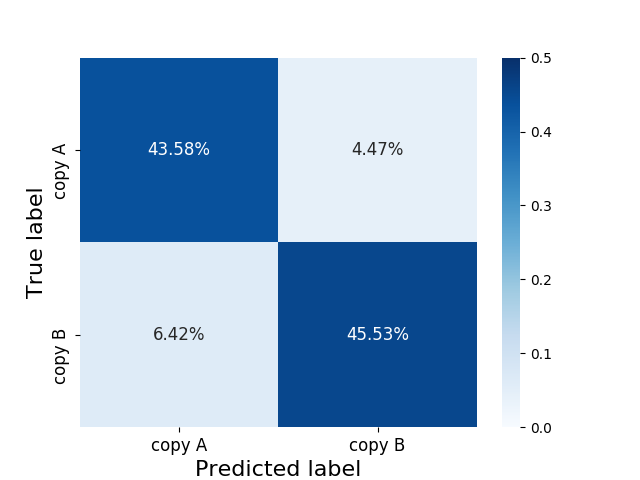}}
\caption{Empirical choice of parameter for the distance between neighboring beads. (a) The 3D genome reconstruction with parameter for the distance between neighboring beads set to 0.5. The homologous loci $x_1, \dots x_n$ (copy A) and $y_1, \dots y_n$ (copy B), colored by red and blue are completely separated. (b) The distance of centroids corresponding to $k$ closest points to the SVM hyperplane separating copy A from copy B (red and blue points) for different parameter settings. The black dashed line corresponds to the chosen parameter of 0.65. (c) Confusion matrix quantifying how often points clustered via $k$-means (predicted label) were assigned their true label (copy A or copy B) when parameter of 0.5 was used. (d) Same as (c) for the chosen parameter 0.65. Higher confusion across labels indicates that points belonging to copy A and copy B are not clearly separated, as desired.}
\label{fig:dist_neighbor_choice}
\end{figure*}

We consider different values for the distance between neighboring beads as input to our algorithm. If the distance between neighboring beads is chosen to be too small, the resulting 3D diploid reconstruction of the data results in the homologous loci $x_1, \dots x_n$ (copy A) and $y_1, \dots y_n$ (copy B) being  completely separated as shown in \Cref{fig:dist_neighbor_choice_a}. We gradually increased the values for the distance between neighboring beads and quantified the separation between $x_1, \dots x_n$ and $y_1, \dots y_n$ as follows: We obtained the hyperplane separating $x_1, \dots x_n$ and $y_1, \dots y_n$ by fitting a support-vector machine (SVM) classifier. Next, we identified the $k$ points among $x_1, \dots x_n$ as well as among $y_1, \dots y_n$ that are closest to the hyperplane and computed their centroids. \Cref{fig:dist_neighbor_choice_b} shows the sum of the distances of the two centroids to the separating hyperplane, thereby quantifying the separation of the copy A points  from the copy B points. This distance should approach 0 as the copy A and copy B points come closer together. Indeed, \Cref{fig:dist_neighbor_choice_b} shows that using a parameter of 0.65, the distance between the $k$ closest points stabilizes close to 0 and thus we used 0.65 as the distance between neighboring beads.

We provide additional quantification regarding the separation of points in copy A and copy B by clustering the 3D structure using $k$-means into two clusters and computing a confusion matrix, where the true labels are given by copy A and copy B. If the points $x_1, \dots x_n$ and $y_1, \dots y_n$ are completely separated, then $k$-means would result in near perfect accuracy of separation of all points into copy A and copy B. \Cref{fig:dist_neighbor_choice_c} shows that this is indeed the case when using a distance parameter of 0.5. For the chosen parameter of 0.65, the confusion matrix is shown in \Cref{fig:dist_neighbor_choice_d}, reinforcing the observation that indeed copy A and copy B are getting mixed. 

\begin{figure*}[h]
    \centering
  \subfloat[]{\includegraphics[width=0.4\textwidth]{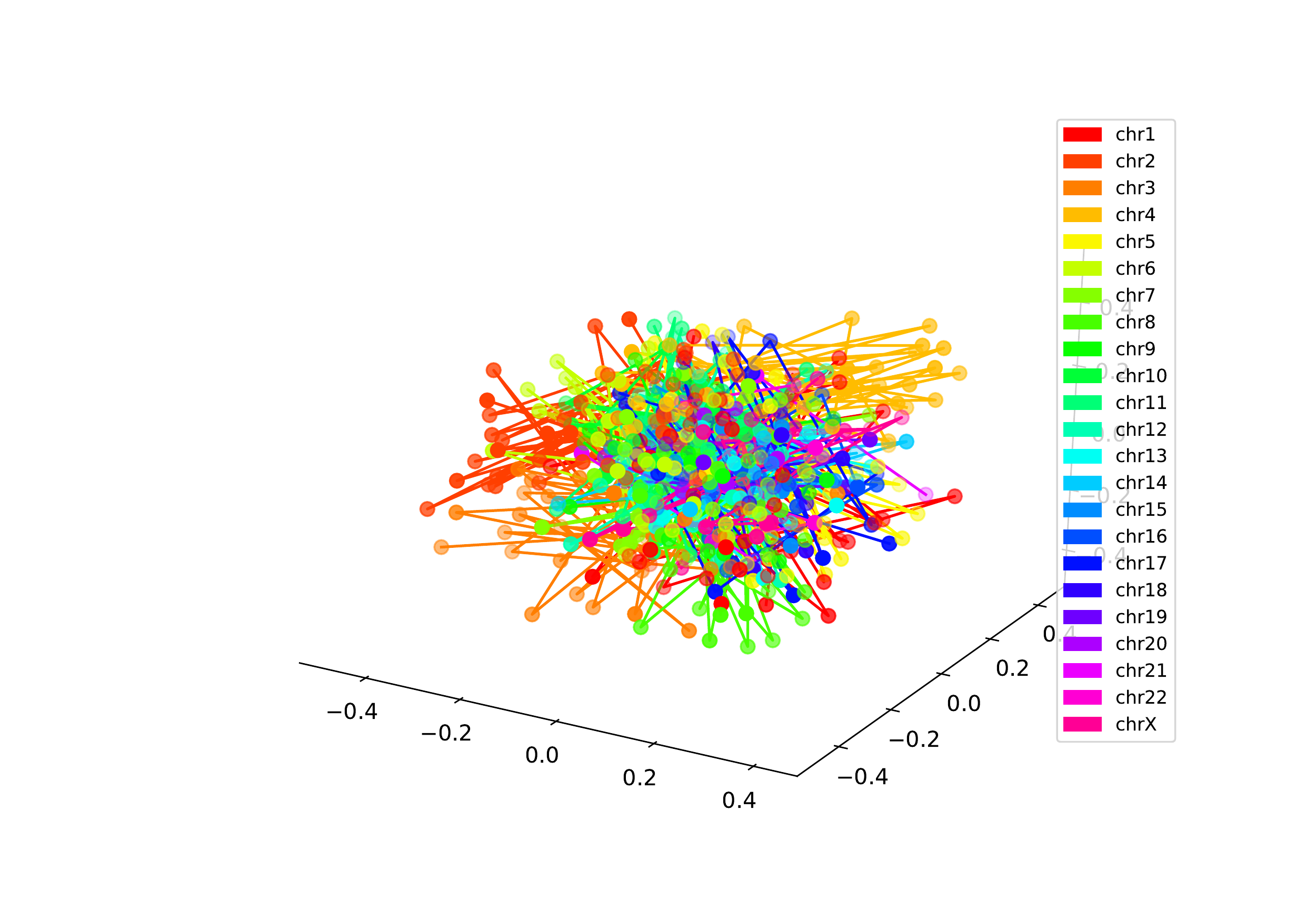}}
  ~
  \subfloat[]{\includegraphics[width=0.4\textwidth]{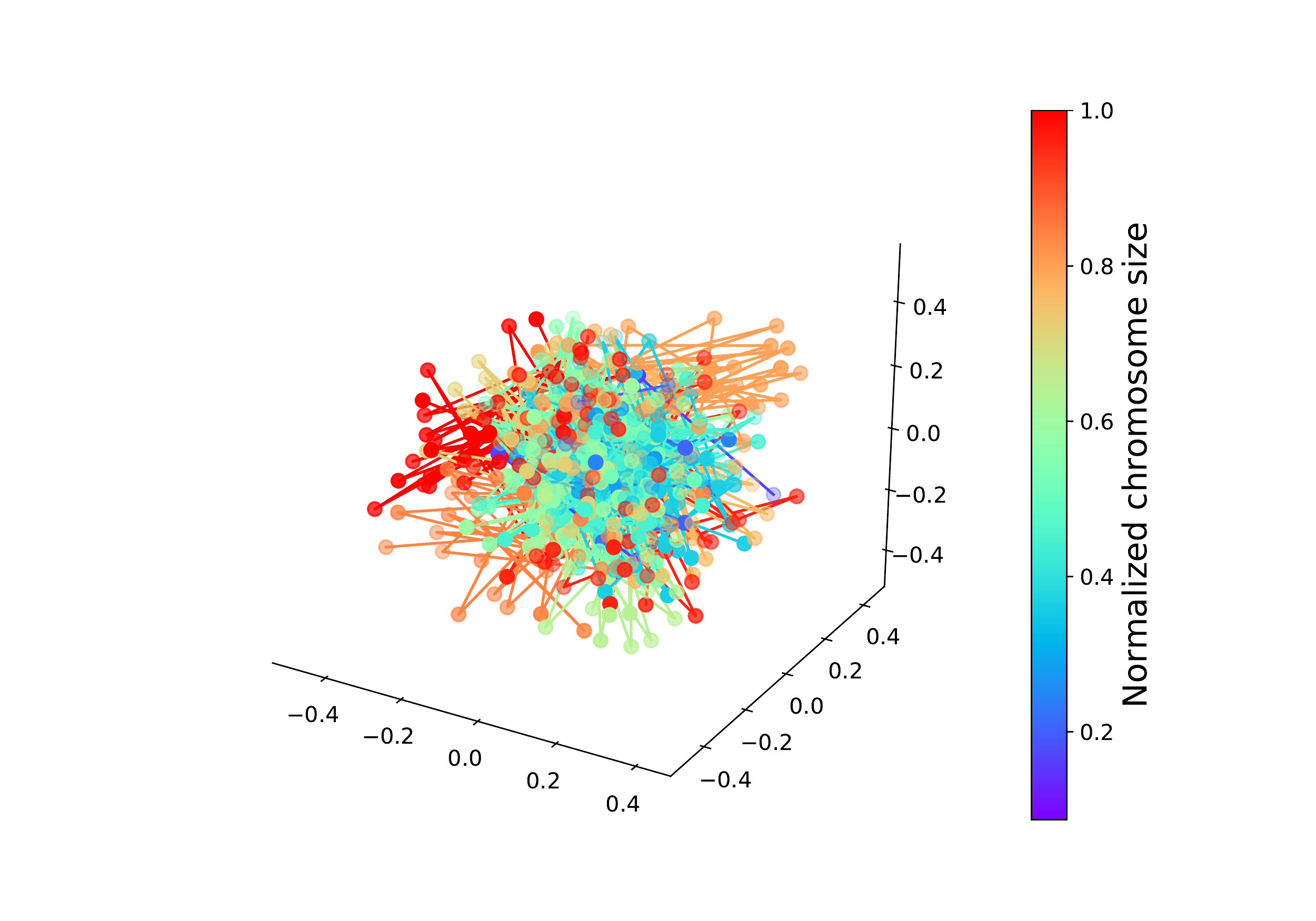}}
  \newline
  \subfloat[]{\includegraphics[width=0.38\textwidth]{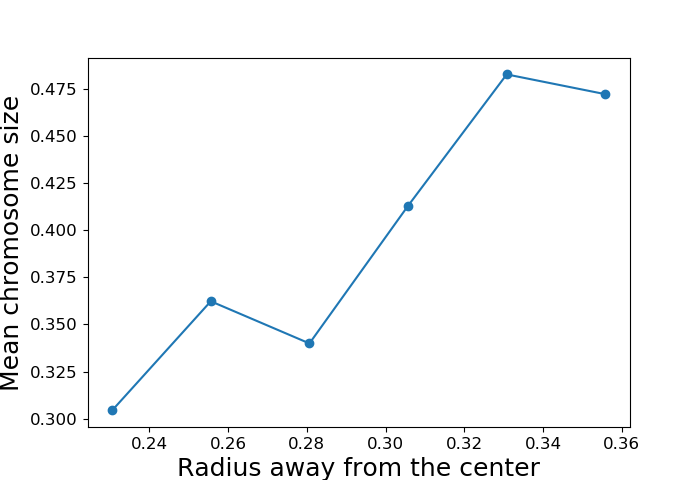}}
  ~
  \subfloat[]{\includegraphics[width=0.35\textwidth]{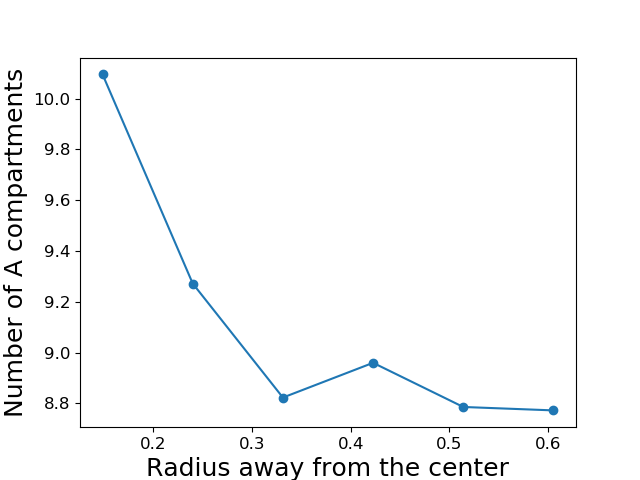}}  
    \caption{3D diploid genome reconstruction with a different parameter (0.7 instead of 0.65) for the distance between neighboring beads. (a) Estimated 3D positions of all chromosomes and their corresponding homologs with chromosomes colored according to chromosome number. (b) Whole diploid organization obtained via our method, colored by chromosome size. (c) Mean chromosome size as the distance from the center increases. (d) The number of A compartments as the distance from the center increases.}
    \label{fig:dist_neighbor_0.7}
\end{figure*}

We note that our observations are robust to the exact choice of the distance between neighboring beads. In \Cref{fig:dist_neighbor_0.7} we show the resulting 3D reconstruction as well as chromosome size and A compartment trends when using a parameter of 0.7 as the distance between neighboring beads.

\vspace{0.5cm}
\section{Comparison with ChromSDE}\label{appendix:real_data_chromsde}

\begin{figure*}[!h]
    \centering
  \subfloat[]{\includegraphics[width=0.32\textwidth]{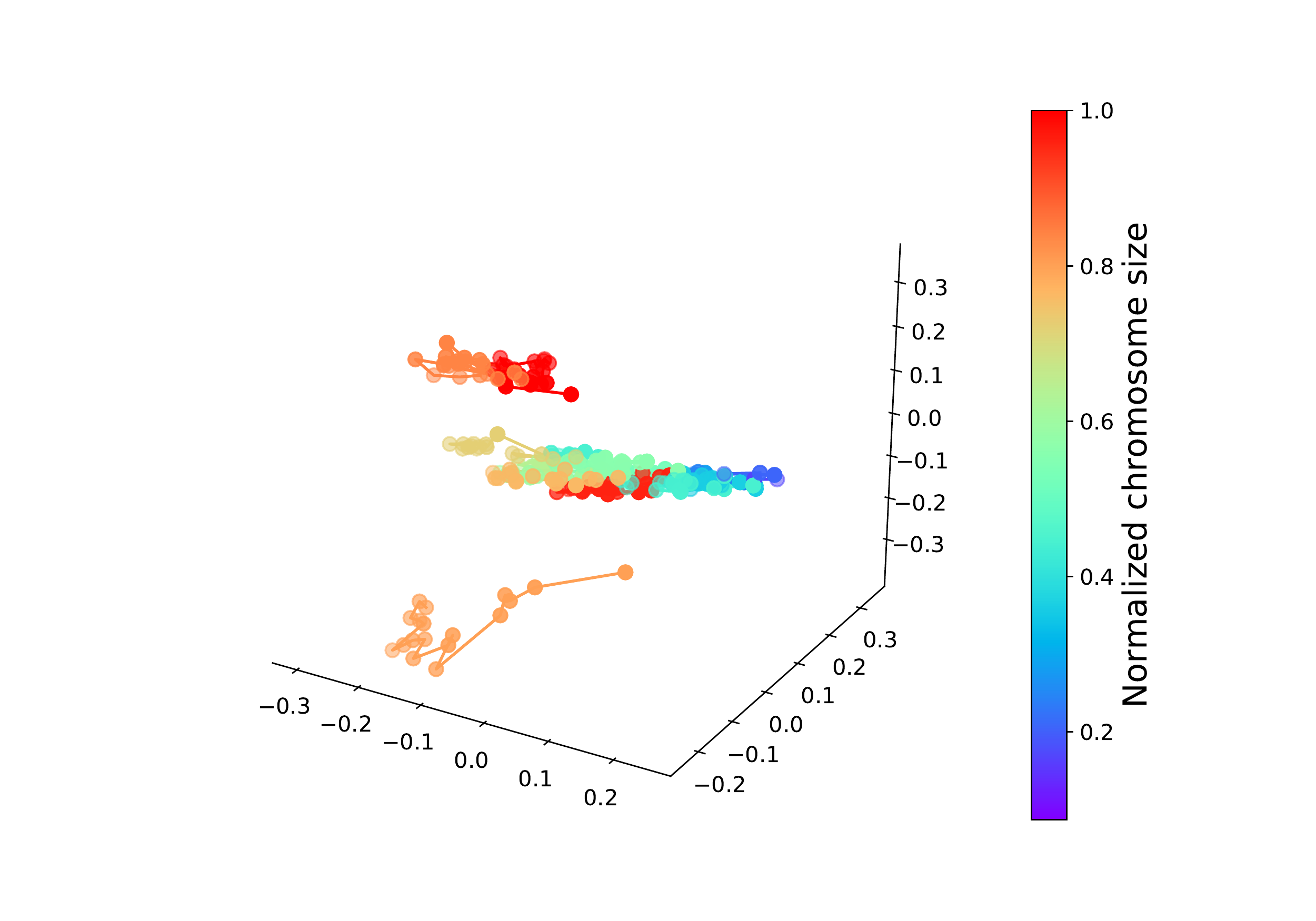}}
  ~
  \subfloat[]{\includegraphics[width=0.31\textwidth]{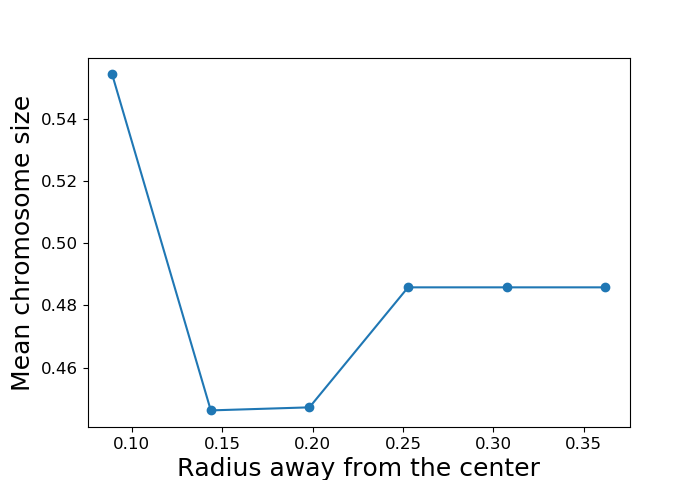}}
  ~    
  \subfloat[]{\includegraphics[width=0.3\textwidth]{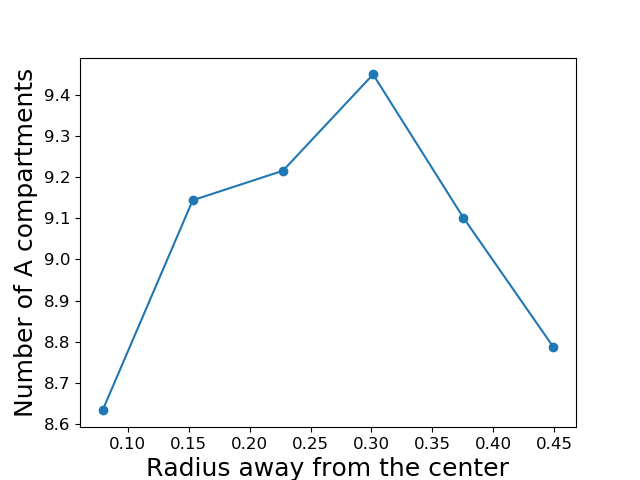}}
    \caption{3D diploid genome reconstruction with ChromSDE. ChromSDE was run using a distance matrix where the distances for $\|x_i - x_j\|^2, \|x_i - y_j\|^2, \|y_i - x_j\|^2$, and $\|y_i - y_j\|^2$ were set to $D_{ij}/4$. (a) Estimated 3D positions of all chromosomes and their corresponding homologs at 10Mb resolution colored by chromosome size. (b) Mean chromosome size as the distance from the center increases. (c) The number of A compartments as the distance from the center increases.}
    \label{fig:chromsde}
\end{figure*}

We compare our whole genome reconstruction to the reconstruction inferred by ChromSDE~\cite{chromsde}. Since ChromSDE does not account for the fact that the measured contact frequencies and corresponding observed distances are a sum of four different distances, i.e. $\|x_i - x_j\|^2, \|x_i - y_j\|^2, \|y_i - x_j\|^2$, and $\|y_i - y_j\|^2$, we converted frequencies to distances using $D_{ij}=F_{ij}^{-1/2}$ and used $D_{ij}/4$ for each of the four distances so that the diploid configuration of the genome could be computed. We assumed that homologous loci are far apart, as has been observed in imaging studies~\cite{bolzer2005three, nir2018walking}, and thus set $\|x_i - y_i\|^2 = \infty$. Given the described distance constraints, we solved the SDP for the Gram matrix and obtained the 3D coordinates using eigenvector decomposition, similar to our method. \Cref{fig:chromsde} shows the corresponding solution and quantification of the mean chromosome size and number of A compartments as the radius from the center increases. The computed 3D diploid genome configuration obtained via ChromSDE does not recapitulate that chromosome size increases with distance away from the center and that the number of A compartments decreases with distance away from the center.

\vspace{0.5cm}
\section{Analysis of 3D diploid genome reconstruction}\label{appendix:real_data_analysis}
\begin{figure*}[h]
\centering
\includegraphics[width=0.4\textwidth]{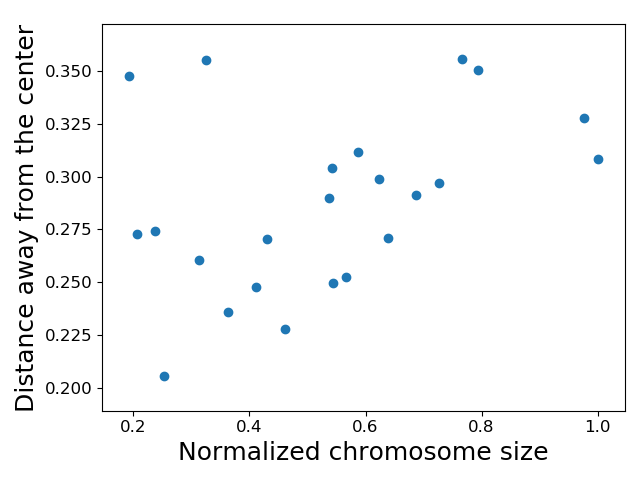}
\caption{Chromosome size (normalized by the size of the largest chromosome) versus the mean distance of the chromosome and its homolog away from the center.}
\label{fig:chromsize_realdata}
\end{figure*}

We provide further analysis of the 3D diploid genome reconstruction obtained using our algorithm from contact frequency data. \Cref{fig:chromsize_realdata} shows that chromosome size is correlated with the  distance of the chromosome to the center of the cell nucleus, whihc is in line known biological trends.

\vspace{0.5cm}
\section{Real data: impact of the tuning parameter $\rho$}\label{appendix:real_data_rho}

\begin{figure*}[!h]
\centering
\includegraphics[width=0.4\textwidth]{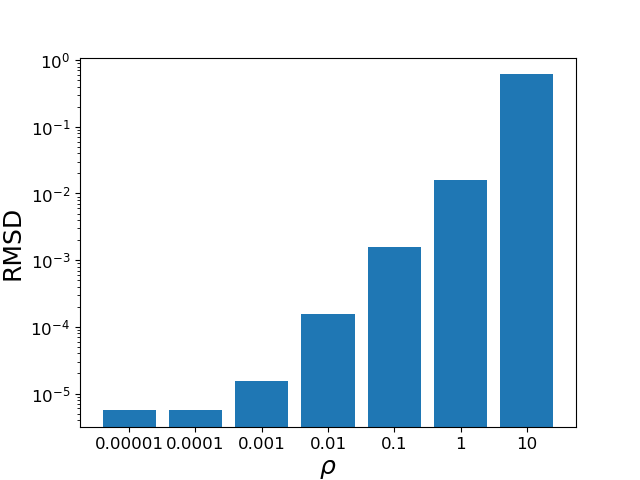}
\caption{The impact of $\rho$ in real data. Root-mean-square deviation (RMSD) between the 3D genome reconstruction computed with $\rho=0.000001$ and the  3D genome reconstructions computed with other choices of $\rho \in (0.00001, 0.0001, 0.001, 0.01, 0.1, 1, 10)$.}
\label{fig:impact_rho_real}
\end{figure*}

\begin{figure*}[!h]
    \centering
  \subfloat[]{\includegraphics[width=0.4\textwidth]{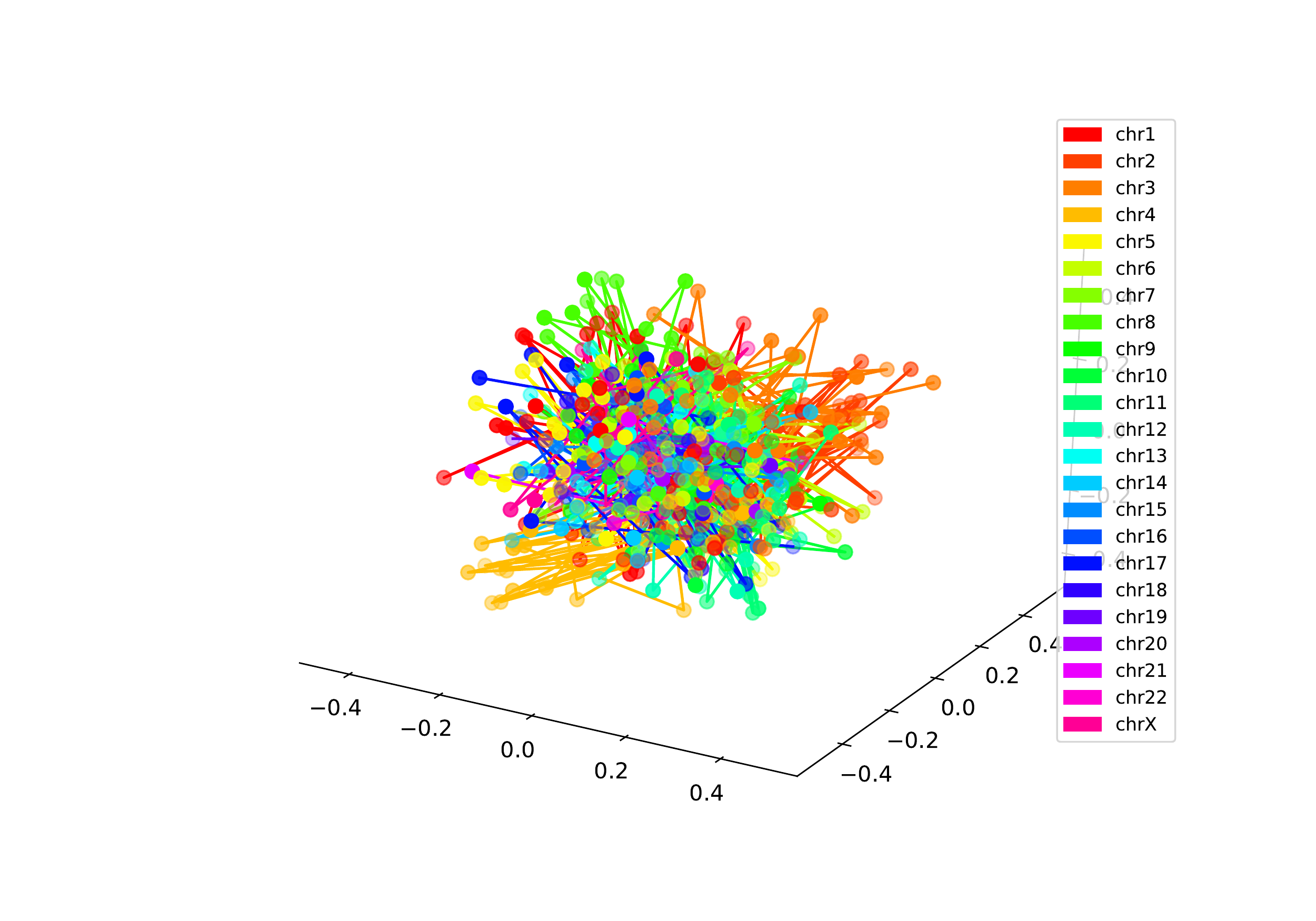}}
  ~
  \subfloat[]{\includegraphics[width=0.4\textwidth]{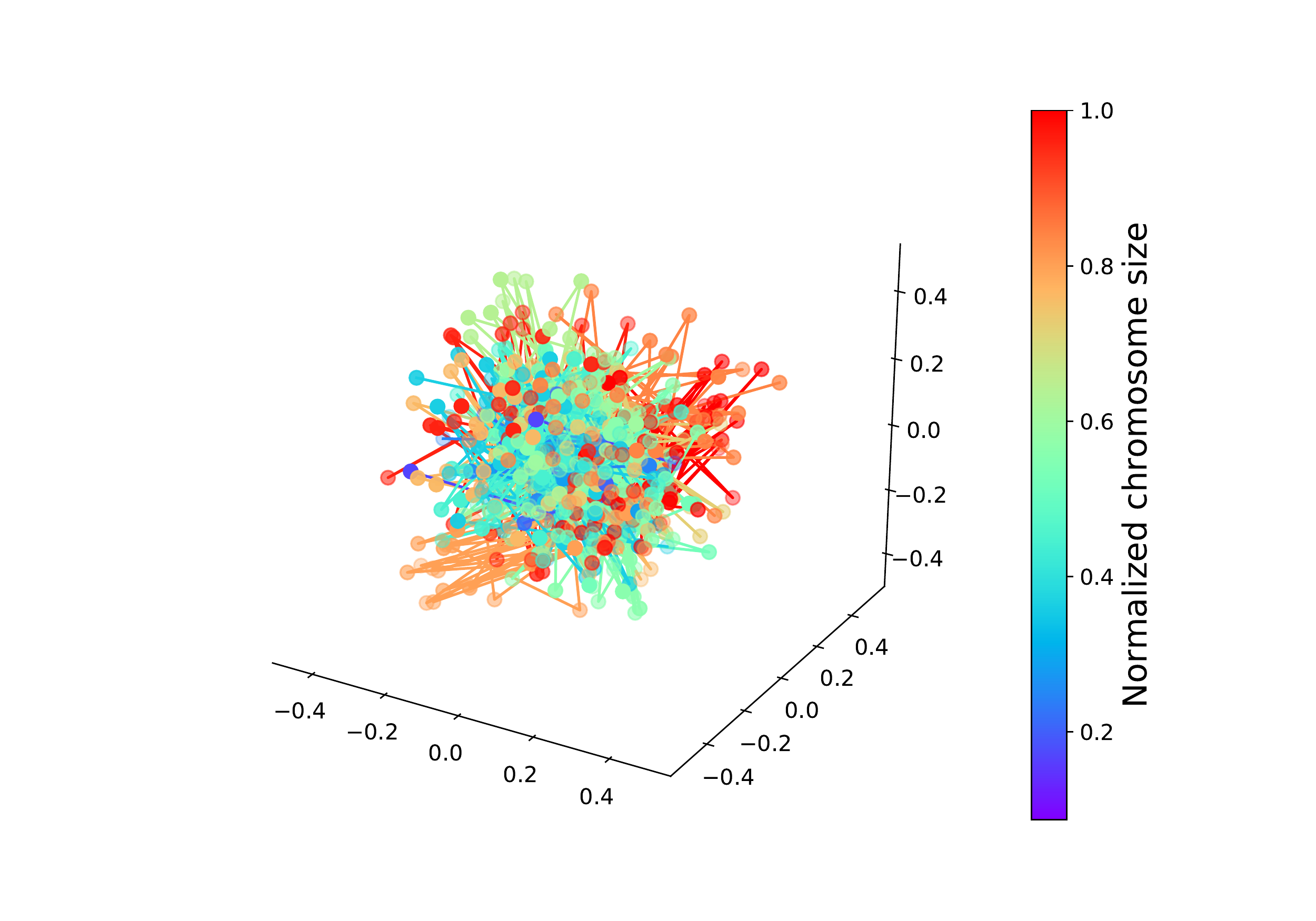}}
  \newline
  \subfloat[]{\includegraphics[width=0.35\textwidth]{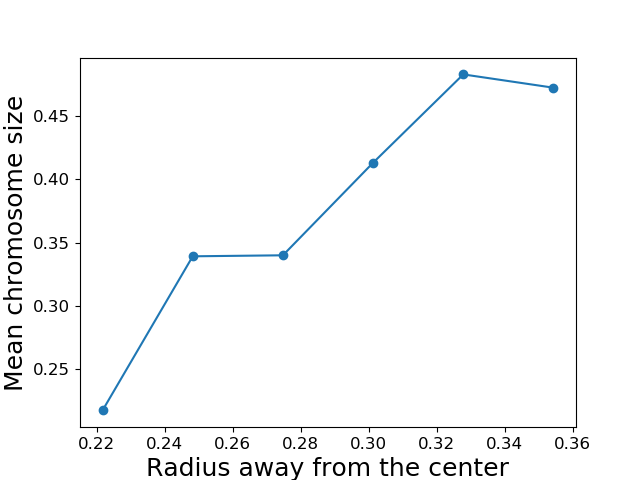}}
  ~
  \subfloat[]{\includegraphics[width=0.35\textwidth]{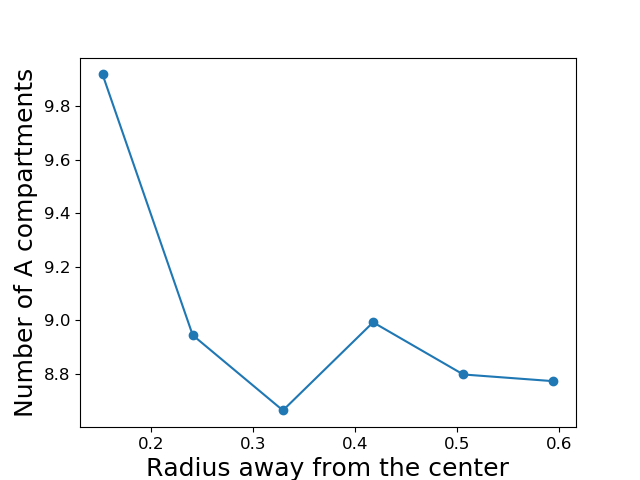}}
    \caption{3D diploid genome reconstruction with $\rho=10$. Estimated 3D positions of all chromosomes and their corresponding homologs at 10Mb resolution. Chromosomes are colored according to (a) chromosome number and (b) chromosome size. (c) Mean chromosome size as the distance from the center increases. (d) The number of A compartments as the distance from the center increases.}
    \label{fig:real_data_rho}
\end{figure*}

\Cref{fig:impact_rho_real} explores the impact of the tuning parameter $\rho$ 
%from \Cref{SDP_noisy} 
on the results of the real data analysis. We compute the RMSD between the 3D genome reconstruction computed with $\rho=0.000001$
%, shown in Figure~\ref{fig:real_data} 
and the 3D genome reconstructions computed with other choices of $\rho \in (0.00001, 0.0001, 0.001, 0.01, 0.1, 10, 1)$ to quantify how much the 3D structure changes when increasing $\rho$. As shown in \Cref{fig:impact_rho_real}, the RMSD is low across different choices of the tuning parameter. In \Cref{fig:real_data_rho}, we provide the 3D genome reconstruction and trends for mean number of A compartments and mean chromosome size versus the distance away from the center for the 3D reconstruction computed with $\rho=10$ since the RMSD was the highest for this choice of the tuning parameter. We observe that the trends remain the same also for this choice of $\rho$.

\section{Haploid distance matrices}\label{appendix:real_data_haploid}

\begin{figure*}[!h]
    \centering
  \subfloat[]{\label{fig:haploid_a}\includegraphics[width=0.33\textwidth]{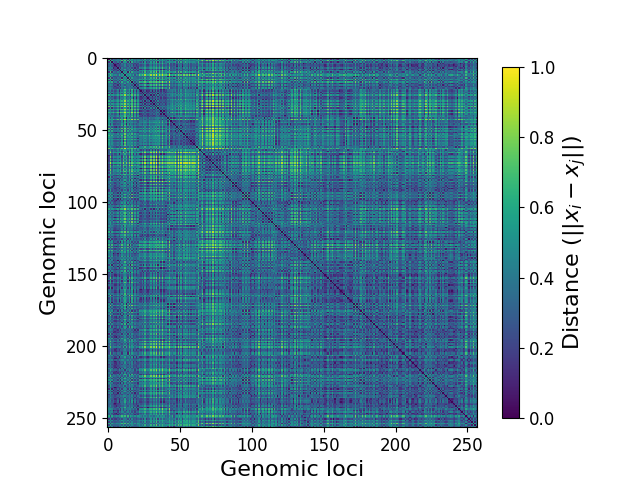}}
  \subfloat[]{\label{fig:haploid_b}\includegraphics[width=0.33\textwidth]{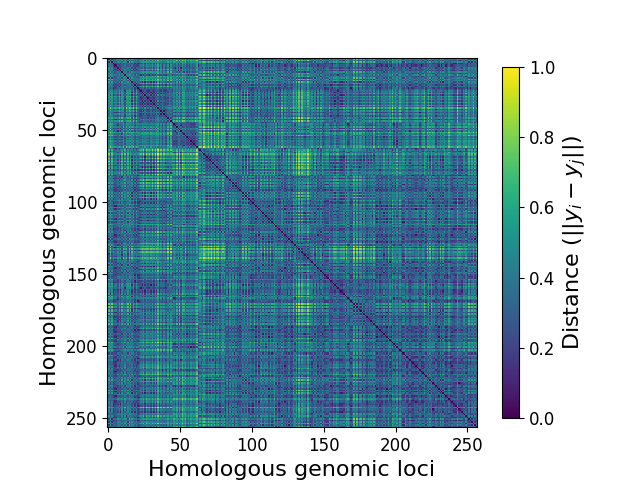}}
  \subfloat[]{\label{fig:haploid_c}\includegraphics[width=0.33\textwidth]{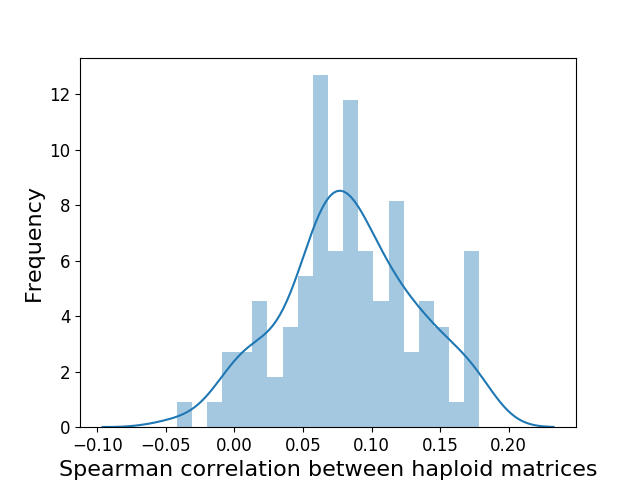}}
    \caption{ Haploid distance matrices.
 (a) Haploid distance matrix for points belonging to copy A and (b) its corresponding homologous haploid distance matrix for points belonging to copy B. (c) Spearman correlation between haploid distance matrices over 100 different assignments of each homolog to either copy A or copy B.}
    \label{fig:haploid}
\end{figure*}

To show that modeling the diploid aspect of the genome is critical and provides valuable information regarding the 3D organization of the genome, we randomly labeled each homolog of a particular chromosome to correspond to either copy A or copy B of the chromosome and
computed the Euclidean distances between all loci belonging to copy A, i.e.~$||x_i - x_j||$ 
to obtain one haploid distance matrix and the Euclidean distances between all loci belonging to copy B, i.e.~$||y_i - y_j||$
to obtain the second haploid distance matrix. \Cref{fig:haploid_a} and \Cref{fig:haploid_b} show the haploid distance matrices where the points $1, \dots, n$  are assigned to copy A and points $n+1, \dots, 2n$ are assigned to copy B. We were interested in comparing the two haploid distance matrices to see whether the two haploid matrices were the same or if modeling the diploid aspect of the genome also allowed us to learn about each homolog.
Inspection of the two matrices reveals that the distances are different in these two haploid matrices, suggesting that modeling the genome as a diploid structure is important and indeed provides additional information. We quantify the difference by computing the Spearman correlation between the distance matrices over 100 different samplings of assignments of chromosomes to either copy A or B. \Cref{fig:haploid_c} shows the histogram of the calculated Spearman correlations with mean Spearman correlation of 0.08.

\bibliographystyle{siamplain}
\bibliography{references}